\documentclass[iop]{emulateapj}

\usepackage{amsmath}
\usepackage{float}
\newcommand{\mv} {$M_V$}
\newcommand{\msun} {$M_{\odot}$}

\newcommand{\halpha} {H$\alpha$}
\newcommand{\hbeta} {H$\beta$}
\newcommand{\hgamma} {H$\gamma$}

\newcommand{\hepsilon} {H$\epsilon$}
\newcommand{\Te} {$T_{\rm eff}$}
\newcommand{\logg} {$\log g$}
\newcommand{\loghe} {$\log$ He/H}
\newcommand{\heii} {He {\sc ii} $\lambda$4686}
\newcommand{\hei} {He {\sc i} $\lambda$4471}
\newcommand{\heif} {He {\sc i} $\lambda$5877}

\slugcomment{Accepted for publication in the Astrophysical Journal}
\shortauthors{GIANNINAS, BERGERON \& RUIZ} 
\shorttitle{A SPECTROSCOPIC SURVEY \& ANALYSIS OF BRIGHT,
  HYDROGEN-RICH WHITE DWARFS}
\begin{document}

\title{A SPECTROSCOPIC SURVEY AND ANALYSIS OF BRIGHT, HYDROGEN-RICH
  WHITE DWARFS\footnote{Based on observations made with ESO Telescopes
    at the La Silla or Paranal Observatories under program ID
    078.D-0824(A).}}

\author{A. Gianninas$^{1,3}$, P. Bergeron$^{1}$, and
  M. T. Ruiz$^{2,3}$}

\affil{$^{1}$D\'epartement de Physique, Universit\'e de Montr\'eal,
  C.P.~6128, Succ.~Centre-Ville, Montr\'eal, Qu\'ebec H3C
  3J7, Canada;} 
\affil{$^{2}$Departamento de Astronom\'\i a, Universidad de Chile, 
  Casilla 36-D, Santiago, Chile}
\affil{$^{3}$Visiting astronomers at Las Campanas Observatory operated
  by Carnegie Institution of Washington}
\email{gianninas@astro.umontreal.ca, bergeron@astro.umontreal.ca,
  mtruiz@das.uchile.cl}

\begin{abstract}

  We have conducted a spectroscopic survey of over 1300 bright ($V
  \leq 17.5$), hydrogen-rich white dwarfs based largely on the last
  published version of the McCook \& Sion catalog. The complete
  results from our survey, including the spectroscopic analysis of
  over 1100 DA white dwarfs, are presented. High signal-to-noise ratio
  optical spectra were obtained for each star and were subsequently
  analyzed using our standard spectroscopic technique where the
  observed Balmer line profiles are compared to synthetic spectra
  computed from the latest generation of model atmospheres appropriate
  for these stars. First, we present the spectroscopic content of our
  sample, which includes many misclassifications as well as several
  DAB, DAZ, and magnetic white dwarfs. Next, we look at how the new
  Stark broadening profiles affect the determination of the
  atmospheric parameters. When necessary, specific models and analysis
  techniques are used to derive the most accurate atmospheric
  parameters possible. In particular, we employ M dwarf templates to
  obtain better estimates of the atmospheric parameters for those
  white dwarfs that are in DA+dM binary systems. Certain unique white
  dwarfs and double-degenerate binary systems are also analyzed in
  greater detail. We then examine the global properties of our sample
  including the mass distribution and their distribution as a function
  of temperature.  We then proceed to test the accuracy and robustness
  of our method by comparing our results to those of other surveys
  such as SPY and Sloan Digital Sky Survey. Finally, we revisit the ZZ
  Ceti instability strip and examine how the determination of its
  empirical boundaries is affected by the latest line profile
  calculations.

\end{abstract}


\section{INTRODUCTION}

For many years, the catalog of \citet[][hereafter MS99]{mccook99} was
the most important reference in the white dwarf community including a
total of 2249 spectroscopically identified white dwarf
stars. Similarly, the now regularly updated online
incarnation\footnote{http://www.astronomy.villanova.edu/WDCatalog/index.html}
is used by white dwarf researchers on a daily basis and has grown to
include nearly 13,000 entries as this publication goes to press. This
is due in large part to the increasingly large number of white dwarfs
discovered in the Sloan Digital Sky Survey (SDSS). The SDSS's huge
database of newly discovered white dwarfs has permitted the discovery
of previously unknown types of white dwarfs like the carbon atmosphere
``Hot DQ'' stars \citep{dufour08}, white dwarfs with gaseous disks
\citep{gansicke08}, and even white dwarfs with oxygen dominated
atmospheres \citep{gansicke10}. However, the fact remains that most of
the SDSS data lack the necessary signal-to-noise ratio (S/N) to
measure accurately the atmospheric parameters for each star
\citep{gianninas05}.  This is largely due to the rather faint white
dwarfs observed by the SDSS and the practice of having a set exposure
time for every field. Consequently, the fainter the star, the lower
the S/N of the data. On the other hand, white dwarfs from MS99 are
comparatively bright, especially if we restrict ourselves to stars
with $V \leq 17.5$. Historically, many of the white dwarfs from MS99
have not been particularly well studied despite their relative
brightness, which makes it easier to obtain high S/N spectroscopic
observations. Indeed, many of these white dwarfs have only a spectral
classification, based on photographic spectra, without any further
analysis ever having been performed, let alone with modern CCD
spectroscopy. Therefore, we believe that it is both important and
worthwhile to study these stars, the more so that it is much more
feasible to conduct follow-up observations (i.e., high-resolution
spectroscopy, high-speed photometry, etc.) for objects of interest, if
they are bright.

Indeed, some of the early results from our survey have already allowed
us to make a few discoveries of our own. The most notable of these is
likely the discovery of the unique nature of GD~362, originally
reported as a highly metallic and massive DAZ white dwarf
\citep{gianninas04}, it was later discovered that the main atmospheric
constituent is helium \citep{zuckerman07}. Even more important was the
discovery of an infrared excess due to the presence of a circumstellar
debris disk \citep{becklin05,kilic05} around this star. This lone
discovery essentially opened the door to a whole new area of white
dwarf research with many groups launching observational campaigns with
the specific goal of discovering debris disks around other cool,
metal-enriched white dwarfs. We have also successfully used our survey
to refine the empirical determination of the boundaries of the ZZ Ceti
instability strip and confirm its purity \citep{gianninas05}. We were
thus able to predict the variability of several new ZZ Ceti pulsators,
while also confirming the non-variability of other DA white dwarfs
near the ZZ Ceti instability strip \citep{gianninas06}.

Ours is not the only survey to have recently looked at a substantial
number of bright white dwarfs. The ESO SN Ia Progenitor Survey
\citep[SPY;][]{koester01} also observed about 800 bright white dwarfs,
of which many were selected from MS99. It is not surprising then that
there is a rather large overlap between the SPY sample and our own.
However, in the case of SPY, the aim of the survey was to obtain
multi-epoch, high-resolution spectra in order to detect the radial
velocity variations of double-degenerate binary systems. These could
then lead to the identification of possible SN Ia progenitors, which
were ultimately what they were looking to find. Results of the
analysis of the DA white dwarfs from SPY are summarized in
\citet{koester09}. \citet{finley97} also presented an analysis of 174
hot (\Te~$>$~25,000~K) DA white dwarfs selected from MS99 as well as
the {\it ROSAT} \citep{pounds93} and {\it Extreme-Ultraviolet
  Explorer} \citep[EUVE;][]{bowyer94} surveys. Similarly,
\citet{marsh97a,marsh97b} and \citet{vennes96,vennes97,vennes98} also
presented analyses of DA white dwarfs selected from {\it ROSAT} and
{\it EUVE}, while \citet{vennes02} analyzed over 200 DA white dwarfs
discovered in the 2dF QSO Redshift Survey \citep{smith05}. Other
important studies of large samples of DA white dwarfs include the
works of \citet[][hereafter LBH05]{lbh05} and \citet{limoges10} who
looked at the complete samples from the Palomar-Green
\citep[PG;][]{green86} and Kiso \citep[KUV;][]{noguchi80,kondo84}
surveys, respectively. These studies defined statistically significant
samples of white dwarfs in an effort to derive the luminosity function
of the disk white dwarf population.  Finally, an analysis of over 7000
DA white dwarfs from the SDSS Data Release 4 (DR4) was presented in
\citet{kepler07} with particular attention to the mass distribution of
the sample.  \citet{tremblay11a} performed a new analysis of the
aforementioned sample of DA white dwarfs from the SDSS and showed that
for the analysis of ensemble properties, such as the mass
distribution, the SDSS data are reliable down to S/N $\sim$
15. However, the precise determination of the atmospheric parameters
for individual stars remains problematic with such low S/N
observations, as in the case of the ZZ Ceti instability strip
\citep{gianninas05}.

In an effort to derive the most accurate atmospheric parameters
possible, we also want to incorporate all available improvements to
the model atmospheres that we use to study these white dwarfs. With
that in mind, we will be using models that include the improved
theoretical calculations of the Balmer line profiles, including
non-ideal effects, by \citet[][hereafter TB09]{TB09}. Seeing as our
spectroscopic technique relies on a detailed comparison between
observed and theoretical Balmer lines, it is crucial that we employ
the most accurate and up-to-date models at our disposal. In the same
vein, two sub-categories of white dwarfs contained within our sample
required particular attention in order to obtain accurate atmospheric
parameters. First, there are the hot DA and DAO stars exhibiting the
so-called Balmer line problem. A comprehensive and detailed analysis
of these stars has already been presented by \citet{gianninas10} and
those results will be included in our analysis. Second, we have DA+dM
binary systems whose Balmer line profiles are contaminated, to varying
degrees, by the spectrum of the dM secondary.

As a whole, we wish to present a detailed analysis of all the hydrogen
line white dwarfs from MS99 for which we have obtained optical
spectra. In Section 2, we will present the spectroscopic content of
our survey. Section 3 will deal with the model atmospheres used for
the analysis of our data, and we will present in Section 4 the
spectroscopic analysis of our entire sample, including particular
analyses for certain subsets of white dwarfs. Subsequently, in Section
5 we will examine the global properties of our sample including the
mass distribution and the mass distribution as a function of effective
temperature, and how these compare to other large surveys. We will
then revisit the PG luminosity function, and the empirical
determination of the ZZ Ceti instability strip. Finally, we end with
some concluding remarks in Section 6.

\section{SPECTROSCOPIC CONTENT}\label{sec:spec}

The goal of our survey was to observe all the white dwarfs listed as
DA in MS99, including all subtypes (DAB, DAH/P, DAO, DAZ, DA+dM), with
a visual magnitude $V \leq 17.5$. Along the way, several DA stars were
added to the sample through various other projects and collaborations
the Montr\'eal group was involved in. In the final reckoning, we had
defined a sample of 1503 white dwarfs for which we wanted to obtain
high S/N optical spectra. After nearly 8 years of observing, and a
total of 19 observing runs in both the northern and southern
hemispheres, we succeeded in securing spectra for 1360 of the 1503
white dwarfs we had selected. This translates to a 90\% completion
rate. Indeed, the majority of the missing spectra are those of the
least bright stars in our sample and objects observable only from the
southern hemisphere where we were unable to secure additional
telescope time.

In all, 726 spectra were obtained specifically for the purposes of
this survey since it began in 2003. Of those 726 spectra, 568 were
obtained between 2003 September and 2011 April at Steward
Observatory's 2.3~m telescope equipped with the Boller \& Chivens
spectrograph. The 4$\farcs$5 slit together with the 600~line~mm$^{-1}$
grating blazed at 3568~\AA\ in first order provide a spectral coverage
from about 3000 to 5250~\AA\ at a resolution of $\sim 6$~\AA\
(FWHM). An additional 18 spectra were obtained in 2006 April and 2007
January at the 1.6 m telescope of the Observatoire du Mont-M\'egantic
where the 600~line~mm$^{-1}$ grating provided a spectral coverage from
about 3100 to 7500~\AA\ at a similar resolution. Furthermore, a total
of 140 spectra from the southern hemisphere were secured during the
course of two observing runs in 2007 March and 2007 October,
respectively. First, 84 spectra were obtained at ESO's 3.6 m telescope
at La Silla, Chile with the ESO Faint Object Spectrograph and Camera
(v.2). The \#7 grism and a 1$\farcs$0 slit provided a spectral
coverage from about 3300 to 5200~\AA\ with a resolution of $\sim 6$
\AA\ (FWHM).  The remaining 56 southern spectra were obtained at
Carnegie Observatories' 2.5~m Ir\'en\'ee du Pont Telescope at Las
Campanas, Chile with the Boller \& Chivens spectrograph. The
1$\farcs$5 slit with the 600~line~mm$^{-1}$ grating blazed at
5000~\AA\ provided a spectral coverage from about 3500 to 6600~\AA\ at
a slightly better resolution of $\sim 3$~\AA\ (FWHM). There are also
169 spectra that were kindly provided by C. Moran (1999, private
communication). Of these, eight were obtained at the South African
Astronomical Observatory's 1.9~m Radcliffe Telescope using the Grating
Spectrograph. The remaining 161 spectra were obtained at the Isaac
Newton Telescope, on La Palma, with the Intermediate Dispersion
Spectrograph. In both cases, the spectra have a resolution of $\sim 3$
\AA\ (FWHM).  Finally, the sources of the spectra for 0950+139 and
1136+667 (two DAO white dwarfs) are described in \citet{gianninas10}.
For the remaining 464 stars, we have used archival spectra that have
been obtained throughout the years by the Montr\'eal group during the
course of various other projects (e.g., \citealt{bergeron92};
LBH05). These were acquired at Steward Observatory's 2.3~m telescope
with the same instrument and setup described above.

\begin{figure}[!t]
\includegraphics[scale=0.45,bb=20 117 592 604]{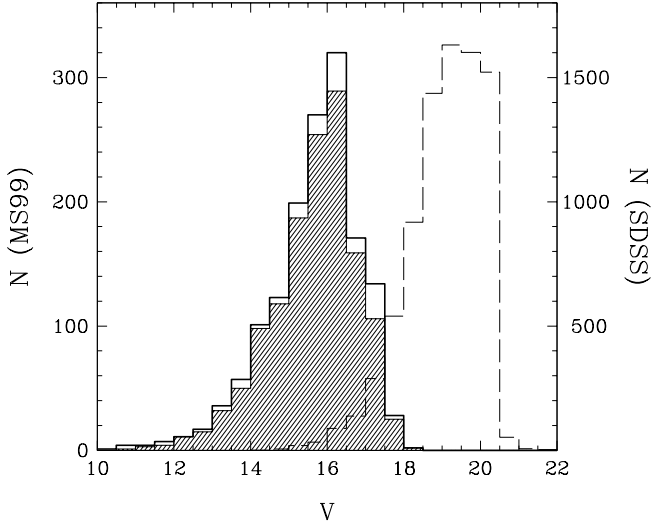}
\figcaption[f01.eps] {Distribution of visual magnitudes $V$ for our
  sample selected from MS99 (bold histogram) and for the white dwarfs
  for which we have obtained optical spectra (hatched histogram). In
  comparison, the distribution for the SDSS sample as of DR4
  \citep{eisenstein06} is also shown (dashed histogram). Note that the
  scale is different for the McCook \& Sion sample (left) and the SDSS
  sample (right) and that we use the SDSS $g$ magnitudes as equivalent
  to $V$. \label{fg:histoV}}
\end{figure}

\begin{figure}[!t]
\includegraphics[scale=0.45,bb=20 117 592 604]{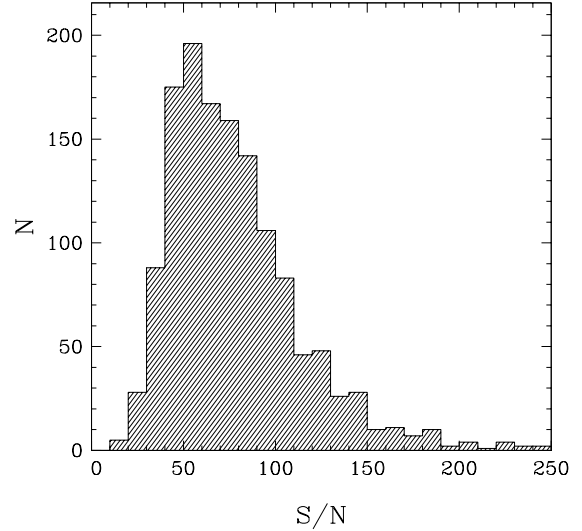}
\figcaption[f02.eps]{Distribution of S/N for all the stars in our
  spectroscopic survey. The distribution peaks at the 50-60 bin and
  more than 50\% of the spectra have S/N $\geq$
  70. \label{fg:histoSN}}
\end{figure}

We show in Figure~\ref{fg:histoV} the final distribution of visual
magnitudes $V$ for our sample of white dwarfs \citep[see Figure~1
of][as a comparison]{gianninas09}. We note that 10 white dwarfs have
no magnitude listed in the literature. The bold histogram shows the
distribution for the entire sample while the hatched histogram
represents the stars for which we have successfully obtained
spectra. We see that the distribution peaks near $V=16.5$ and that the
majority of our stars have $V \leq 16.5$. We contrast this result with
the sample of DA white dwarfs from the SDSS DR4
\citep{eisenstein06}. Although the SDSS sample contains over six times
more white dwarfs, the vast majority of these have $20.5 \geq g \geq
18.5$, which makes for a much fainter sample overall. The distribution
of S/N for our entire sample of 1360 optical spectra is displayed in
Figure~\ref{fg:histoSN}. The S/N\footnote{For the majority of our
  spectra, obtained at Steward Observatory's 2.3~m telescope, there
  are $\sim$3.2 pixels per resolution element but this value varies
  for the other instrument setups employed.} is determined by an
estimate of the rms noise per pixel in the continuum between $\sim$
4500 and $\sim$ 4700 \AA. Over 52\% of our sample have S/N $> 70$ and
almost 80\% above 50. Only the faintest objects in our sample have S/N
that is significantly lower. As such, we have achieved our goal of
obtaining high-quality data for the majority of the white dwarfs in
our survey.

\begin{figure*}[p]
\includegraphics[scale=0.85]{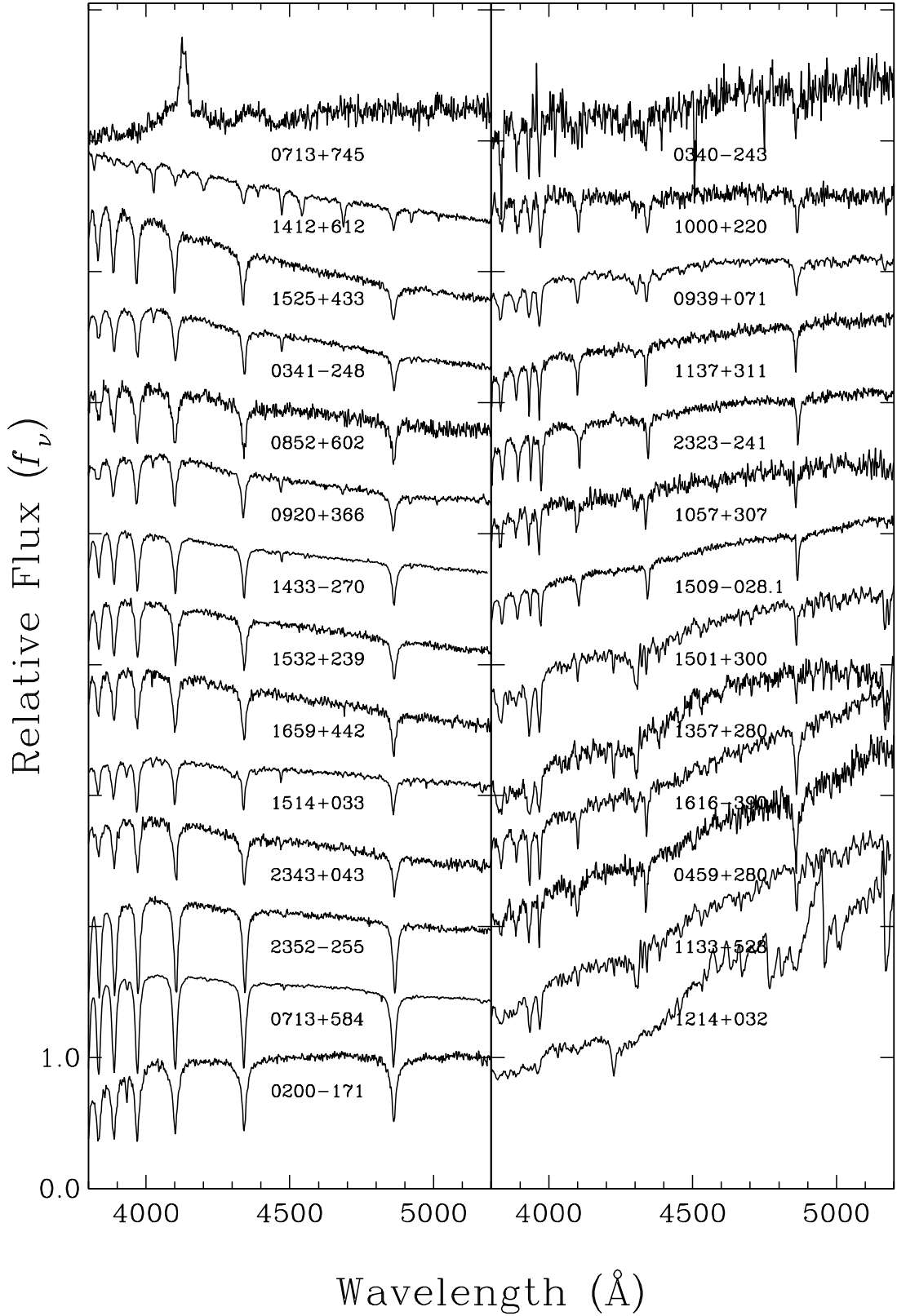}
\figcaption[f03.eps]{Optical spectra of stars misclassified as DA
  white dwarfs in MS99 and reported for the first time in this
  work. The objects are approximately ordered as a function of their
  slope from upper left to bottom right. The spectra have been
  normalized to a continuum set to unity and vertically shifted for
  clarity. 0713+745 (left panel, first object at the top) is a
  quasar. \label{fg:misclass}}
\end{figure*}

As previously stated, many of these stars have never been studied
using modern CCD spectroscopy. Consequently, it is not surprising to
discover that many of the spectral classifications are erroneous; many
of the objects listed in MS99 are not even white dwarfs. We list in
Table~\ref{tab:misclass} 68 objects that are incorrectly classified as
white dwarfs in MS99. We provide the spectral classification from MS99
as well as our revised spectroscopic classification for each object
along with the appropriate references. Some of these have been known
to be misclassified for some time but are still listed as white dwarfs
in the online version of MS99. Of the 68 stars listed in
Table~\ref{tab:misclass}, 20 were recently reported as being
misclassified objects in \citet{limoges10} and we refer the reader to
Figure~1 of Limoges \& Bergeron for the spectra of those stars. In
this paper, we report the erroneous classification of an additional 27
objects for the first time. The spectra of these 27 white dwarfs are
displayed in Figure~\ref{fg:misclass}. Many of these are sdB stars
that show strong Balmer lines that are not as broad as those of their
white dwarf cousins due to their lower surface gravity. However, it is
not too hard to see how these objects could easily have been mistaken
as white dwarfs, especially using photographic spectra. On the other
hand, it is harder to explain how several main-sequence stars were
just as easily misclassified. Indeed, even a quasar (0713+745) has
found its way into MS99. Furthermore, MS99 still lists stars like
BD+28~4211 (2148+286) and Feige~34 (1036+433), popular
spectrophotometric standards that are known hot subdwarfs. In addition
to the misclassified objects, several stars had two separate entries
in MS99 under two different names. Although virtually all of these
have since been corrected, we report here that at least two still
remain. The entries for 0525+271 and 0526+271 refer to the same star,
which also happens to be one of the misclassified objects in
Table~\ref{tab:misclass}, and 1100+604 and 1104+602 are also one in
the same. We also note that 1959+059 is GD~226, not GD~266 as listed
in MS99 \citep{giclas65}.

\begin{table*}\scriptsize
\caption{Misclassified Objects}
\begin{center}
\begin{tabular}{@{\extracolsep{1.25cm}}@{}lcccc@{}}
\hline
\hline
\noalign{\smallskip}
WD & Name & ST (MS99) & ST (This Work) & Notes \\
\noalign{\smallskip}
\hline
\noalign{\smallskip}
0021$-$234   &Ton S 155           &DAWK  &sdB    &1  \\
0031$-$274   &Ton S 163           &DA1   &sdB    &1  \\
0107$-$342   &GD 687              &DA3   &sdB    &1  \\
0109$-$264   &Ton S 201           &DA1   &sdB    &1  \\
0113$-$243   &GD 693              &DA6   &sdB    &2  \\
0154$-$071   &PB 8949             &DA1   &sdB    &1  \\
0200$-$171   &G272-B5B            &DA    &sdB    &3  \\  
0222+314     &KUV 02222+3124      &DA    &sdB    &2  \\
0240+341     &KUV 02409+3407      &DA    &MS(dF) &2  \\
0258+184     &PG 0258+184         &DA    &sdB    &1  \\
0340$-$243   &Ton S 372           &DAWK  &MS(dF) &3  \\
0341$-$248   &Ton S 374           &DA    &sdB    &3  \\
0459+280     &CTI 045934.1+280335 &DA    &MS(dF) &3  \\
0509+168     &KUV 05097+1649      &DA    &MS(dF) &4  \\
0510+163     &KUV 05101+1619      &DA    &MS(dF) &2  \\
0526+271     &KUV 05260+2711      &DA    &sdB    &2  \\
0529+261     &KUV 05296+2610      &DA    &sdB    &2  \\
0627+299     &KUV 06274+2958      &DA    &sdB    &2  \\
0628+314     &KUV 06289+3126      &DA    &sdB    &2  \\
0713+584     &GD 294              &DA4   &sdB    &3  \\
0713+745     &Mrk 380             &DA    &QSO    &3  \\
0720+304     &SA 51-822A          &DA2   &sdB    &5  \\
0752+412     &KUV 07528+4113      &DA?   &sdB    &2  \\
0852+602     &PG 0852+602         &DA2   &sdB    &3  \\
0920+366     &CBS 98              &DA1   &sdB    &3  \\
0927+388     &KUV 09272+3854      &DA    &sdB    &2  \\
0930+376     &KUV 09306+3740      &DA    &sdB    &2  \\
0932+396     &KUV 09327+3937      &DA    &sdB    &2  \\
0937+395     &KUV 09372+3933      &DA    &sdB    &2  \\
0939+071     &PG 0939+072         &DA7   &MS(dF) &3  \\
0943+371     &KUV 09436+3709      &DA    &sdB    &2  \\
0944$-$090   &GD 104              &DA1   &sdO    &6  \\
0946+381     &KUV 09467+3809      &DA    &sdB    &2  \\
1000+220     &Ton 1145            &DA6   &MS(dF) &3  \\
1008$-$179   &EC 10081$-$1757     &DA1   &sdO    &7  \\
1036+433     &Feige 34            &DA    &sdO    &8  \\
1057+307     &CSO 64              &DA5   &MS(dF) &3  \\
1121+145     &PG 1121+145         &DA1   &sdB    &9  \\
1133$-$528   &BPM 21065           &DA5   &MS(dK) &3  \\
1137+311     &CSO 105             &DA5   &MS(dF) &3  \\
1207$-$032   &PG 1207$-$033       &DA4   &sdB    &1  \\
1214+032     &LP 554-63           &DA    &dM     &3  \\
1256+286     &KUV 12562+2839      &DA?   &sdB    &2  \\
1304+313     &PB 3322             &DA    &sdB    &2  \\
1357+280     &CTI 135700.6+280448 &DA    &MS(dF) &3  \\
1412+612     &HS 1412+6115        &DAH   &sdO    &3  \\
1412$-$049   &PG 1412$-$049       &DA    &DA+dM  &10 \\ 
1433$-$270   &BPS CS 22874-57     &DA    &sdB    &3  \\
1501+300     &\ldots              &DA    &MS(dF) &3  \\
1509$-$028.1 &LP 622-13           &DA:   &MS(dF) &3  \\
1514+033     &PG 1514+034         &DAWK  &sdB    &3  \\
1524+438     &CBS 246             &DABZ4 &sdO    &11 \\
1525+433     &GD 344              &DA    &sdB    &3  \\
1532+239     &Ton 241             &DA    &sdB    &3  \\
1544+009     &LB 898              &DA1   &sdO    &12 \\
1603+175     &KUV 16032+1735      &DAH?  &sdO+dK &2  \\
1611+390     &KUV 16118+3906      &DA    &sdB    &2  \\
1616$-$390   &Ton 264             &DA4   &MS(dF) &3  \\
1659+442     &PG 1659+442         &DA    &sdB    &3  \\
2122+157     &PG 2122+157         &DA    &sdB    &1  \\
2148+286     &BD+28 4211          &DA    &sdO    &13 \\
2204+071     &PG 2204+071         &DA    &sdOB   &10 \\
2309+258     &KUV 23099+2548      &DA    &sdB    &2  \\
2323$-$241   &G275-B16A           &DA    &MS(dF) &3  \\
2329$-$291   &Ton S 102           &DAWK  &sdB    &1  \\
2333$-$002   &PB 5462             &DA2   &sdO    &14 \\
2343+043     &PB 5529             &DA    &sdB    &3  \\
2352$-$255   &G275-B17B           &DA    &sdB    &3  \\
\noalign{\smallskip}
\hline
\noalign{\smallskip}
\multicolumn{5}{@{}p{14.5cm}@{}}{{\bf References.} 
(1) \citealt{lisker05};
(2) \citealt{limoges10};
(3) this work;
(4) \citealt{kawka04};
(5) \citealt{finley97};
(6) \citealt{stroeer07};
(7) \citealt{christlieb01};
(8) \citealt{farihi05};
(9) \citealt{green86};
(10) LBH05;
(11) \citealt{eisenstein06};
(12) \citealt{catalan08};
(13) \citealt{kilkenny88};
(14) \citealt{hugelmeyer06}.} \\
\label{tab:misclass}
\end{tabular}
\end{center}
\end{table*}

\begin{figure*}[p]
\includegraphics[scale=0.85]{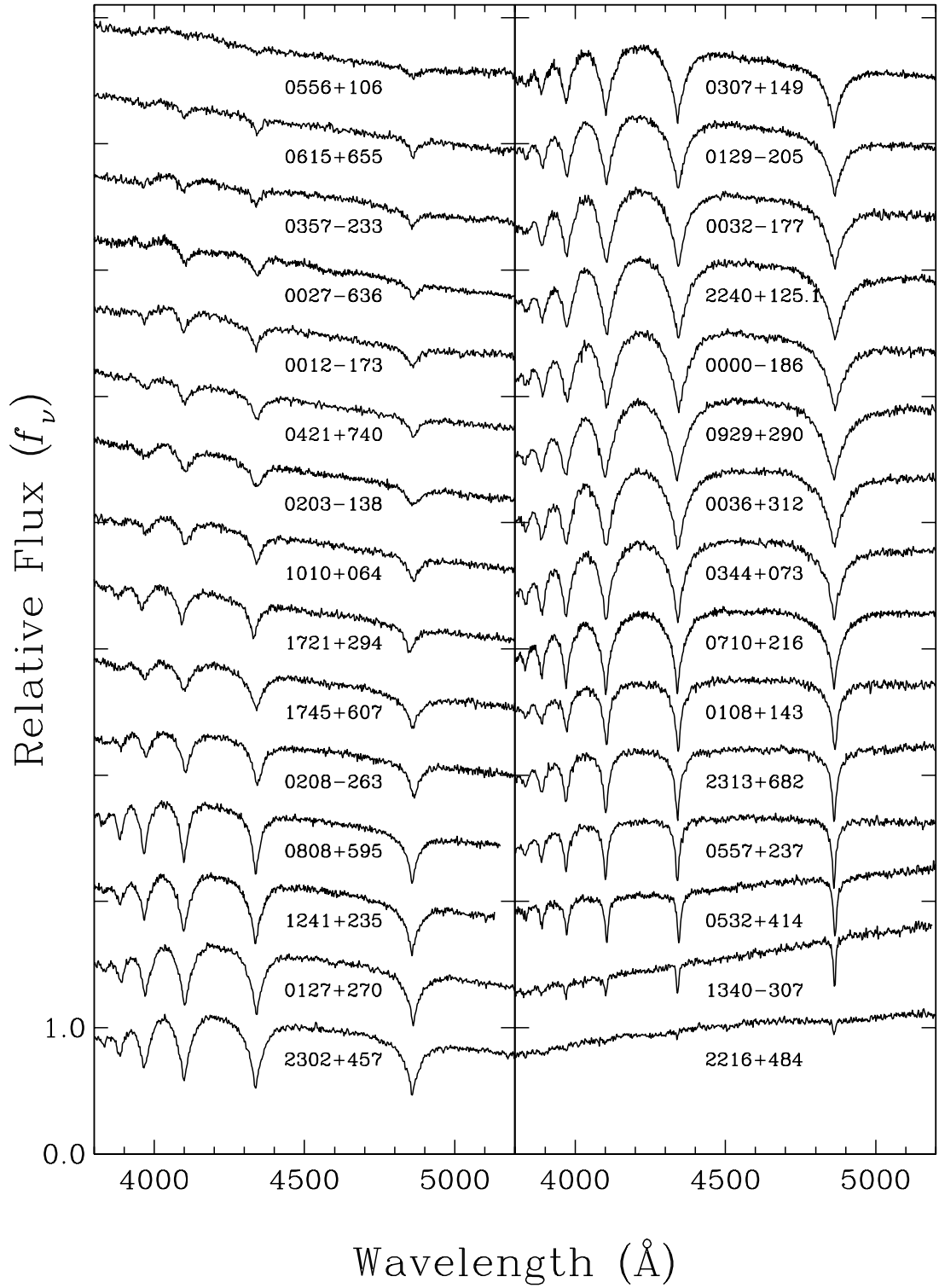}
\figcaption[f04.eps]{Optical spectra for a subsample of the DA white
  dwarfs from our survey of MS99. The spectra are normalized at 4500
  \AA\ and shifted vertically for clarity. The effective temperature
  decreases from upper left to bottom right. All the spectra have 75
  $\geq$ S/N $\geq$ 65. \label{fg:DA}}
\end{figure*}

\begin{figure*}[p]
\includegraphics[scale=0.85]{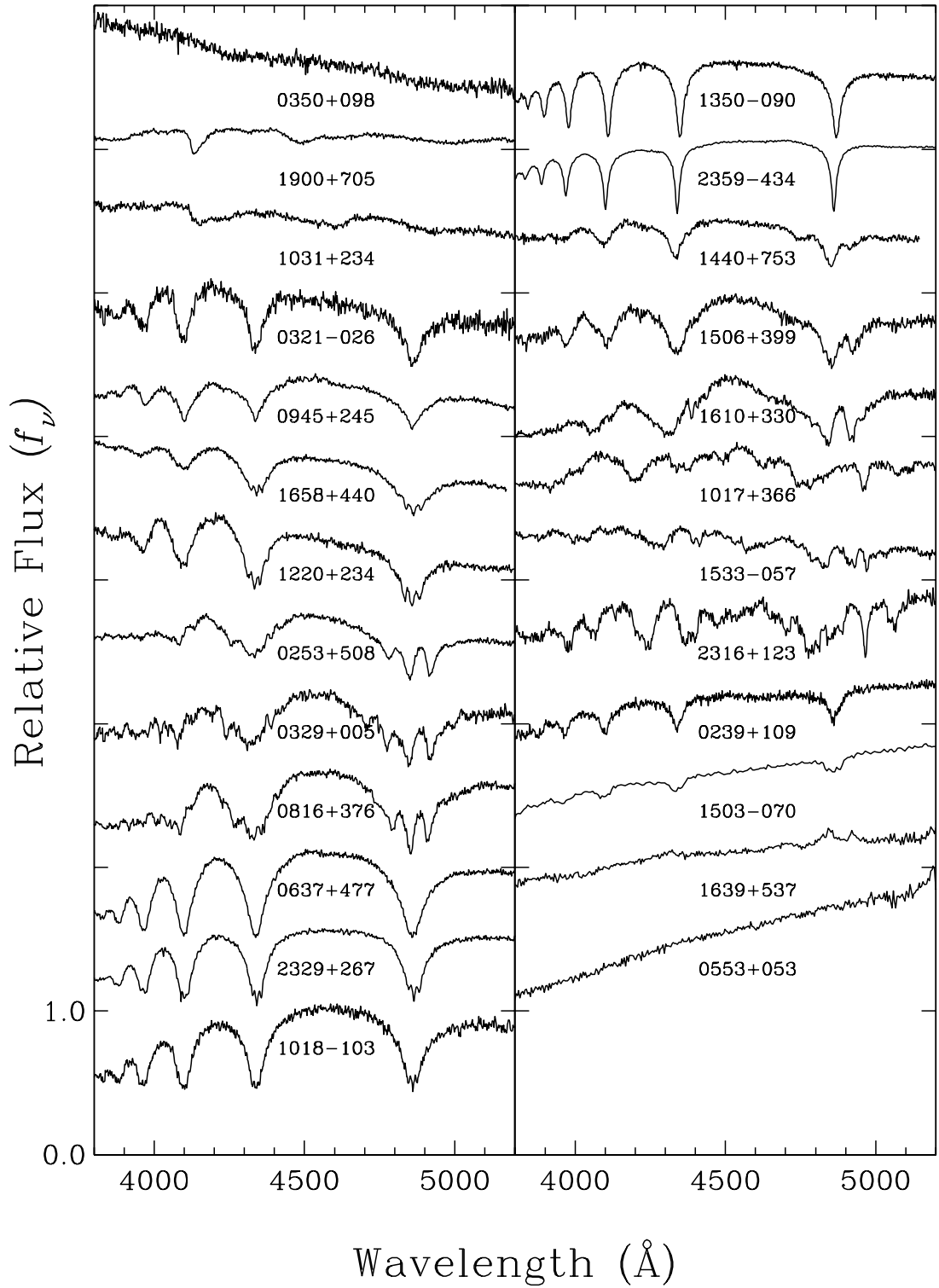}
\figcaption[f05.eps]{Optical spectra for 25 magnetic (or suspected
  magnetic) DA white dwarfs from our survey of MS99. The objects are
  approximately ordered as a function of their slope from upper left
  to bottom right. The spectra are normalized at 4500 \AA\ and shifted
  vertically for clarity. \label{fg:MAG}}
\end{figure*}

\begin{figure*}[p]
\includegraphics[scale=0.85]{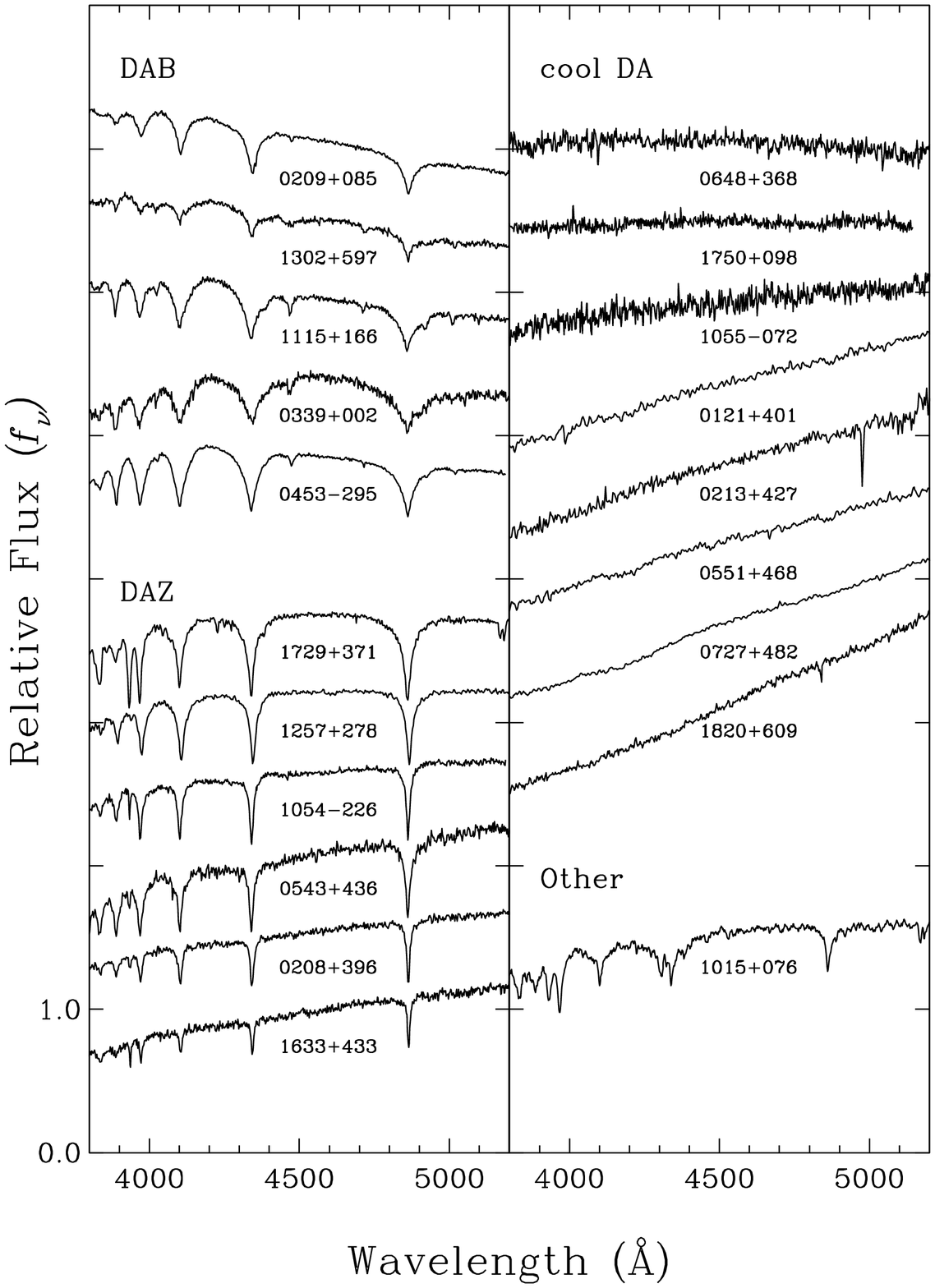}
\figcaption[f06.eps]{Optical spectra for 20 white dwarfs from our
  survey of miscellaneous spectral types. The objects are
  approximately ordered as a function of their slope from upper left
  to bottom right. The spectra are normalized at 4600 \AA\ and shifted
  vertically for clarity. \label{fg:misc}}
\end{figure*}

In Figure~\ref{fg:DA}, we show a series of representative DA spectra
in order of decreasing effective temperature, all of which have $75
\geq$~S/N~$\geq 65$. We have chosen this range of S/N values to
emphasize the mean quality of our data, which is quite high overall.
We refer the reader to \citet[][see Figure~1]{gianninas10} for a
detailed presentation of the DAO white dwarfs contained within our
sample. Spectra for the 25 magnetic white dwarfs in our sample are
displayed in Figure~\ref{fg:MAG} as a function of their continuum
slope. In many of these we see the classical Zeeman splitting of the
hydrogen lines for somewhat lower magnetic field strengths on the
order of $\sim$1 to $\sim$10 MG. On the other hand, stars with much
stronger fields, on the order of tens to hundreds of MG, are rendered
almost unrecognizable as DA stars. Of note in Figure~\ref{fg:MAG} are
possibly four new magnetic white dwarfs. With regards to 1018$-$103
and 1610+330, there is little doubt as to their magnetic nature. On
the other hand, the cases of 0321$-$026 and 0350+098 are more
problematic. In the case of 0321$-$026, the Balmer lines predicted by
our best-fit model are too deep and we obtain a somewhat high-\logg\
value (see Section \ref{sec:mag}). This is often the case with
magnetic white dwarfs as the Zeeman splitting broadens the observed
line profiles leading to a higher \logg\ value. We suggest that
0321$-$026 might harbor a weak magnetic field, and a spectrum with
higher S/N could reveal weak Zeeman splitting such as that seen in
0637+477. Second, 0350+098 was detected by {\it ROSAT}
\citep{fleming96} and we therefore conclude that it must be a
relatively hot star, and the slope would seem to corroborate that
conclusion. Unfortunately, nothing else is known about this
object. The absence of any spectral features could be an indication of
a strong magnetic field that has completely diluted any absorption
lines that would otherwise be present in the spectrum of
0350+098. Spectropolarimetric observations would be needed to confirm
the magnetic nature of this star. We also note that three of the
magnetic white dwarfs are members of double-degenerate binary
systems. Namely, 0945+245 \citep[LB~11146,][]{liebert93}, 1440+753
\citep[RE~J1440+750,][]{vennes99}, and 1506+229
\citep[CBS~229,][]{gianninas09}.

Finally, in Figure~\ref{fg:misc} we display spectra for white dwarfs
of other miscellaneous spectral types included in our sample. This
includes five DAB stars, six DAZ white dwarfs, and eight cool
DA. Among the DAB white dwarfs we find HS~0209+0832 (0209+085) and
GD~323 (1302+597), which have been shown to be spectroscopically
variable \citep[][respectively]{heber97,pereira05}, while the other
three DAB stars (0339+002, 0453$-$295, and 1115+166) are actually
DA+DB binary systems
\citep[][respectively]{limoges10,wesemael94,bergeron02}. Regarding the
DAZ white dwarfs, our sample includes the first DAZ ever discovered,
G74-7 \citep[0208+396,][]{lacombe83} as well as GD~362
\citep[1729+371,][]{gianninas04}. Although, as previously mentioned,
GD~362 has since been shown to have a helium-dominated atmosphere
\citep{zuckerman07}, a discovery made possible by the detection of a
very faint helium line at 5877 \AA. However, the dominant absorption
lines in the spectrum, which define the spectral class of an object,
remain the hydrogen and metal lines, as such the spectral type should
at most be updated to DAZB. In addition, we report for the first time
that 0543+436 and 1054$-$226 are also members of the DAZ class. In
particular, 1054$-$226 was initially discovered by
\citet{subasavage07} but their discovery spectrum lacked the necessary
S/N to detect the H and K lines of calcium. In addition, there are
eight white dwarfs whose spectra are completely devoid of absorption
lines in the observed wavelength range. However, with the exception of
0648+368 and 1055$-$072, the other six cool DA stars have detections
of \halpha\ \citep{BLR01,silvestri01}, hence their DA spectral
type. We also note that 0727+482 is actually a double-degenerate
system \citep{strand76}. Finally, we also include the peculiar
spectrum obtained for 1015+076. It seems that 1015+076 is a DA white
dwarf whose spectrum is completely dominated by an unresolved,
background G dwarf \citep{farihi05}. Indeed, the SPY spectrum analyzed
in \citet{voss06} shows a perfectly normal DA white dwarf with \Te\
$\sim$ 25,600~K and \logg\ $\sim$ 7.70.

\begin{figure*}[!t]
\includegraphics[scale=0.65,angle=-90,bb=47 16 596 784]{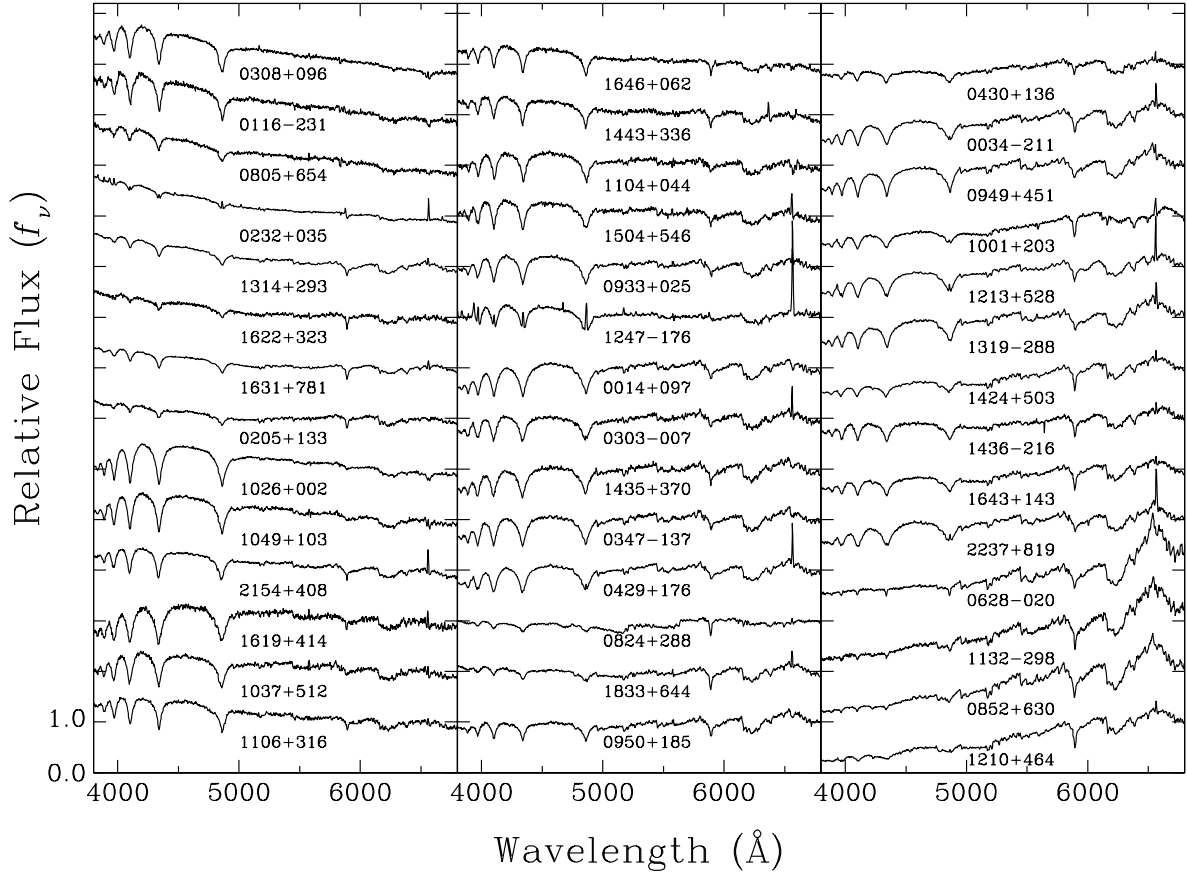}
\figcaption[f07.eps]{Optical spectra for 42 DA+dM binary systems
  covering the range 3500--7000 \AA. The spectra are normalized to
  unity at 6500 \AA\ and shifted vertically for
  clarity. \label{fg:specDAdM}}
\end{figure*}

Our spectroscopic sample also contains a total of 47 DA+dM binary
systems, all of which have been previously identified. In order to
obtain accurate atmospheric parameters for the DA white dwarfs in
these binary systems, we need to remove the contamination from the dM
secondary. With that in mind, we have obtained follow-up spectra for
all the DA+dM systems in our sample, with the exception of five
(0208$-$153, 0309$-$275, 0419$-$487, 1541$-$381, 1717$-$345), covering
a larger spectral range from 3500 to 7000~\AA. These spectra were
obtained at Steward Observatory's 2.3~m telescope using the same
instrument and setup as described above except that here we used the
400 line mm$^{-1}$ grating blazed at 4800~\AA\ to achieve the desired
wavelength coverage. The one drawback was that the use of this
particular grating degrades our spectral resolution to $\sim$ 9~\AA\
(FWHM) from our usual $\sim$ 6~\AA. However, this slightly reduced
spectral resolution is still sufficient for our purposes. We present
in Figure~\ref{fg:specDAdM} what we will refer to as our {\it red}
spectra for 42 of the 47 DA+dM binaries in our sample. These will be
coupled with our usual optical spectra (3000 -- 5250~\AA), which we
will call our {\it blue} spectra, for the analysis of the DA+dM
systems. Several of the tell tale features found in the spectra of
DA+dM binaries can be seen in Figure~\ref{fg:specDAdM}.  We see in
these spectra how the contribution from the M dwarf becomes more
significant at longer wavelengths. More specifically, we see the TiO
absorption band near $\sim$ 4950~\AA\ that produces the ``kink'' in
the red wing of \hbeta.  We also note the presence of the strong Na D
line near 5895~\AA\ as well as emission, from the M dwarf, in the core
of several Balmer lines, especially at \halpha.

As a summary, we present in Table~\ref{tab:spectypes} a breakdown of
the white dwarfs in our sample listing the number of stars in each
spectral class.

\clearpage

\section{MODEL ATMOSPHERES}

\subsection{Pure Hydrogen Atmosphere Models}

For the analysis of the vast majority of white dwarfs in our sample,
we have employed model atmospheres and synthetic spectra appropriate
for DA white dwarfs. These are described at length in LBH05, and
references therein. Briefly, these are pure hydrogen, plane-parallel
model atmospheres where non-local thermodynamic equilibrium (NLTE)
effects are explicitly taken into account above \Te\ = 20,000~K, and
energy transport by convection is included in cooler models following
the ML2/$\alpha$ = 0.8 prescription of the mixing-length theory
\citep[see][]{tremblay10}. The one major upgrade from the models
described in LBH05 is that we now use the new and improved Stark
broadening profiles of TB09 that account for nonideal effects directly
in the line profile calculation. However, since we will wish to
compare our results with those of other large surveys, such as SPY, we
have also fit our sample using models computed with the older
broadening profiles of \citet{lemke97}, and consequently with the
ML2/$\alpha$ = 0.6 version of the mixing-length theory
\citep{bergeron95}, since they were still in use at the time that the
SPY analysis was performed. We can thus compare the independent
results on a level playing field. Beyond that, performing our analysis
with both sets of hydrogen line profiles will also allow us to explore
the implications of these improved models on the analysis of DA white
dwarfs.

\subsection{Hydrogen Atmosphere Models with CNO}

Although the pure hydrogen models described above are sufficient in
most cases, some white dwarfs present unique challenges that require
somewhat more elaborate models. This is the case with the hot DAO
white dwarfs as well as the hot DA stars that exhibit the so-called
Balmer line problem (DA+BP), which manifests itself as an inability to
fit all the Balmer lines simultaneously with consistent atmospheric
parameters. Though the models used to properly fit the spectra of
these stars are presented and described at length in
\citet{gianninas10}, for the benefit of the reader, we briefly explain
them here. The above-mentioned models have all been computed using the
TLUSTY atmosphere code and the accompanying spectrum synthesis code
SYNSPEC, both developed by \citet{hubeny95}. The models are computed
in NLTE and include carbon, nitrogen, and oxygen (CNO) at solar
abundances \citep{asplund05}. As \citet{werner96} demonstrated, the
inclusion of CNO reduces the temperature in the upper layers of the
atmosphere, which in turn produces deeper cores for the hydrogen
Balmer lines, thus overcoming the Balmer line problem. As clearly
stated by \citet{gianninas10}, CNO acts as a proxy for all metals that
may be present in the atmosphere. As such, the inclusion of CNO at
solar abundances should in no way be considered as a determination of
the abundance of CNO in these stars. The model atmospheres used for
the DAO and DA+BP stars are identical, save for the inclusion of
helium in the DAO models in order to reproduce the \heii\ line.

\begin{figure*}[!t]
\includegraphics[scale=0.85]{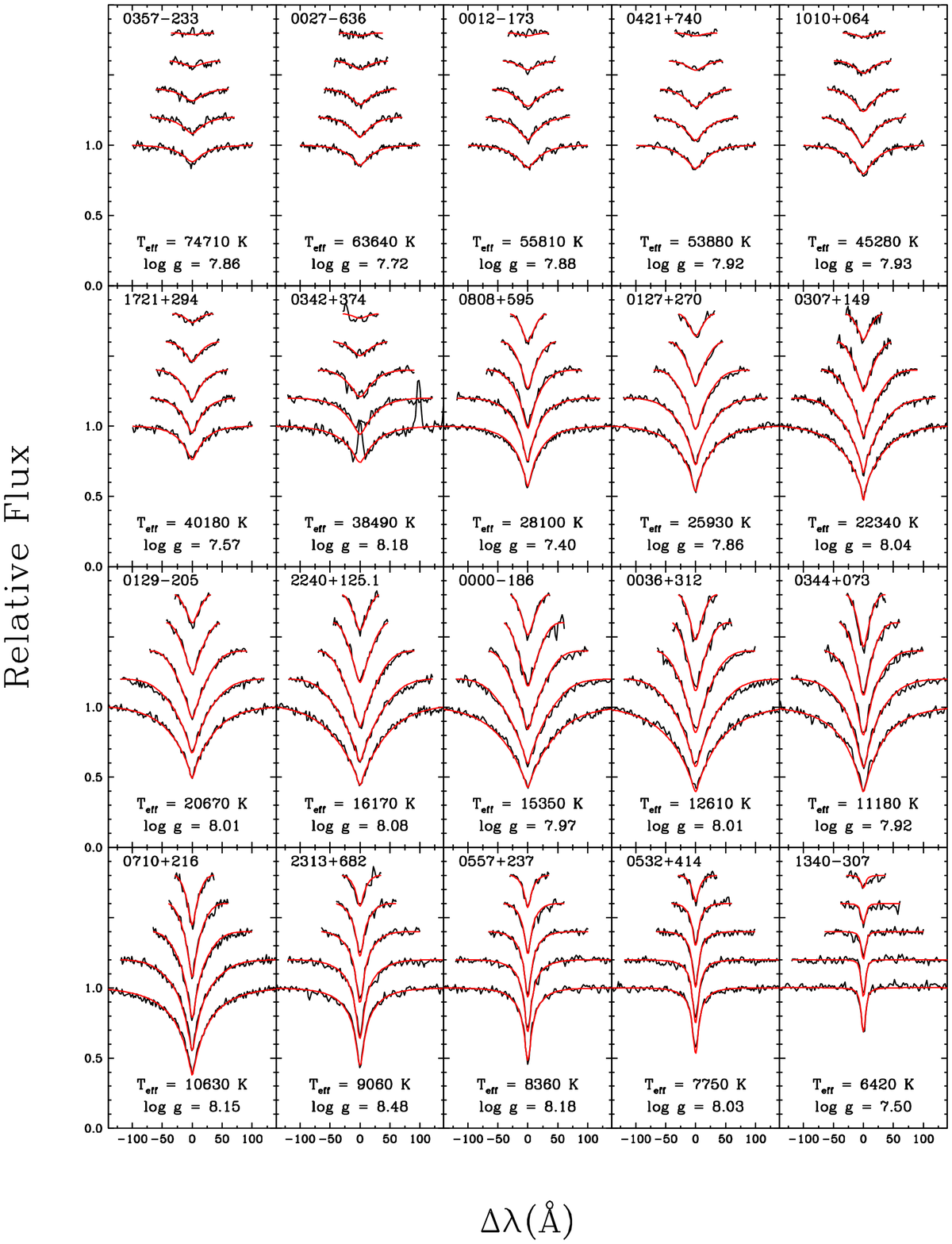}
\figcaption[f08.eps]{Model fits (red) to the individual Balmer line
  profiles (black) of a sample of DA white dwarfs taken from
  Figure~\ref{fg:DA} in order of decreasing effective temperature. The
  lines range from \hbeta\ (bottom) to H8 (top), each offset by a
  factor of 0.2. The best-fit values of \Te\ and \logg\ are indicated
  at the bottom of each panel. \label{fg:fits_DA}}
\end{figure*}

\subsection{Helium Atmosphere Models}\label{sec:helium}

In Figure~\ref{fg:misc}, we presented three DA+DB binary systems that
are included in our sample. In order to properly fit the spectra of
these stars, we will need to combine our pure hydrogen models with
pure helium models. The helium models we have employed are described
in \citet{bergeron11}, which are based on the updated grid of white
dwarf models described in TB09 in which the improved Stark profiles of
neutral helium of \citet{beauchamp97} have been incorporated. These
models are comparable to those described by \citet{beauchamp95} and
\citet{beauchamp96}, with the exception that at low temperatures (\Te\
$<$ 10,800~K), the free-free absorption coefficient of the negative
helium ion of \citet{john94} is now used. The models are in local
thermodynamic equilibrium (LTE) and include convective energy
transport within the mixing-length theory. As in the analysis of
\citet{beauchamp99}, the parameterization described as ML2/$\alpha$ =
1.25 is implemented.

\begin{table}\scriptsize
\caption{Breakdown of Spectral Types}
\begin{center}
\begin{tabular}{@{}lccccccr@{}}
\hline
\hline
\noalign{\smallskip}
DA & DAO & DAZ & DAB & DAmag & DA+dM & Cool DA & Total\\
\noalign{\smallskip}
\hline
\noalign{\smallskip}
1171 & 30 & 6 & 5 & 25 & 47 & 8 & 1292 \\
\noalign{\smallskip}
\hline
\label{tab:spectypes}
\end{tabular}
\end{center}
\end{table}

\section{SPECTROSCOPIC ANALYSIS}

\subsection{Fitting Technique}

The method used for fitting the observations relies on the so-called
spectroscopic technique developed by \citet{bergeron92}, which has
been refined by \citet{bergeron95} and more recently by LBH05, and
includes the Balmer lines from \hbeta\ to H8.  The first step is to
normalize the flux from an individual line, in both observed and model
spectra, to a continuum set to unity at a fixed distance from the line
center. The comparison with model spectra, which are convolved with
the appropriate Gaussian instrumental profile (3, 6, 9, and 12 \AA),
is then carried out in terms of these line shapes only. The most
sensitive aspect of this fitting technique is to define the continuum
of the observed spectra. The approach is slightly different depending
on the temperature range in question. For stars in the interval
16,000~K $\gtrsim$ \Te\ $\gtrsim$ 9000~K, where the Balmer lines are
at their strongest, the normalization procedure involves the use of
several pseudo-Gaussian profiles \citep[][and references
therein]{bergeron95}. This procedure is quite reliable as the sum of
the pseudo-Gaussian profiles represents a good approximation to the
observed Balmer lines. Outside of this temperature range, the method
becomes more unstable as the continuum between the Balmer lines
becomes essentially linear. Consequently, for stars with \Te\ $>$
16,000~K and \Te\ $<$ 9000~K, we rely on our synthetic spectra to
reproduce the observed spectrum, including a wavelength shift, as well
as several order terms in $\lambda$ (up to $\lambda^{6}$) using the
nonlinear least-squares method of Levenberg-Marquardt
\citep{press86}. The normal points are then fixed at the points
defined by this smooth model fit. Note that the values of \Te\ and
\logg\ at this stage are meaningless since too many fitting parameters
are used, and the model just serves as a smooth fitting function to
define the continuum of the observed spectrum. Once the lines are
normalized to a continuum set to unity, we use our grid of model
spectra to determine \Te\ and \logg\ in terms of these normalized
profiles only. Our minimization technique again relies on the
nonlinear least-squares method of Levenberg-Marquardt, which is based
on a steepest descent method. Sample fits of 20 DA white dwarfs,
covering almost the entire temperature range of our sample are
displayed in Figure~\ref{fg:fits_DA}. We see that the combination of
high S/N spectra with the procedures described above allow us to
achieve a proper normalization in each case. Hence, we are able to
obtain an excellent agreement between the observed and predicted line
profiles for nearly every star. The spectrum of one star, 0342+374,
features emission lines from its associated planetary nebula. These
emission lines can interfere with our normalization procedure but also
with our fitting technique and so we simply exclude the affected
spectral ranges from both the normalization and fitting procedure.

\begin{figure}[!t]
\includegraphics[scale=0.50,bb=20 42 592 749]{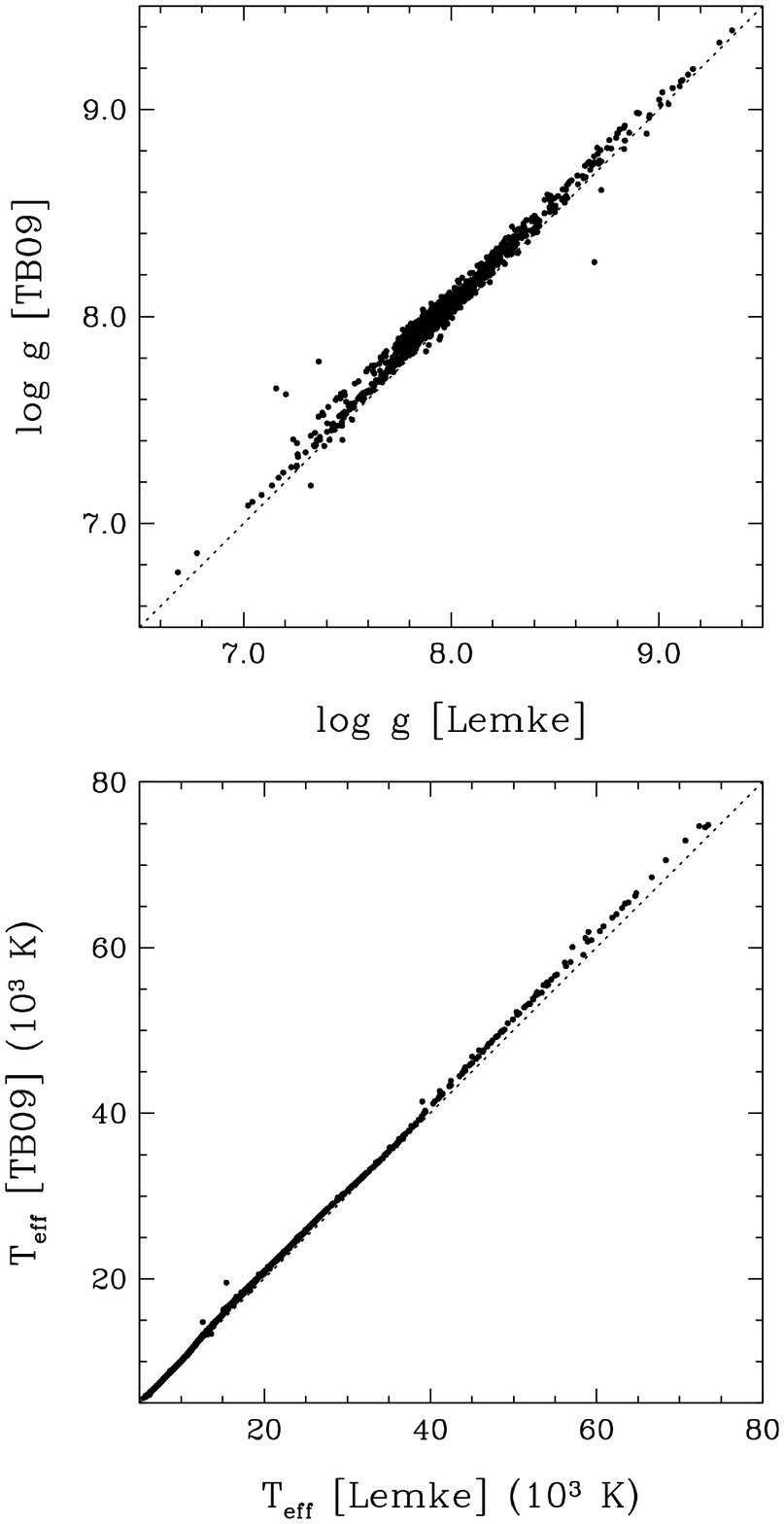}
\figcaption[f09.eps]{Comparison of \logg\ (top) and \Te\ (bottom)
  values derived from fits using the improved models of TB09 and the
  profiles of \citet{lemke97} for 1151 DA white dwarfs. The dotted
  lines represent the 1:1 correlation. \label{fg:comptg}}
\end{figure}

\subsection{Effects of New Stark Profile Calculations}

As we have mentioned, in this study we will be passing from the use of
the old \citet{lemke97} profiles to the new TB09 Stark broadening
profiles that include non-ideal effects directly in the line profile
calculation. In doing so, it is important to understand what effects
these new profiles will have on our atmospheric parameter
determinations. TB09 explored some of these differences (see their
Figure~12) using the PG sample of LBH05, but this limited them to a
particular range of \Te\ whereas we can examine the effects along
virtually the entire white dwarf cooling sequence.

In Figure~\ref{fg:comptg}, we present the comparison between the
atmospheric parameters measured using the older \citet{lemke97} and
the new TB09 Stark profiles for the 1151 DA stars in our sample. In
the lower panel, we see that the correlation between the old and new
values of \Te\ is very tight although the overall trend is to higher
temperatures. This effect is somewhat more evident at higher \Te. In
the upper panel of Figure~\ref{fg:comptg}, we compare the old and new
values of surface gravity. Unlike with the values of \Te\, we note
here a much more obvious shift toward higher measurements of \logg\
when using the new TB09 Stark broadening profiles. These results are
consistent with those presented in TB09.

\begin{figure}[!t]
\includegraphics[scale=0.425,bb=20 167 592 639]{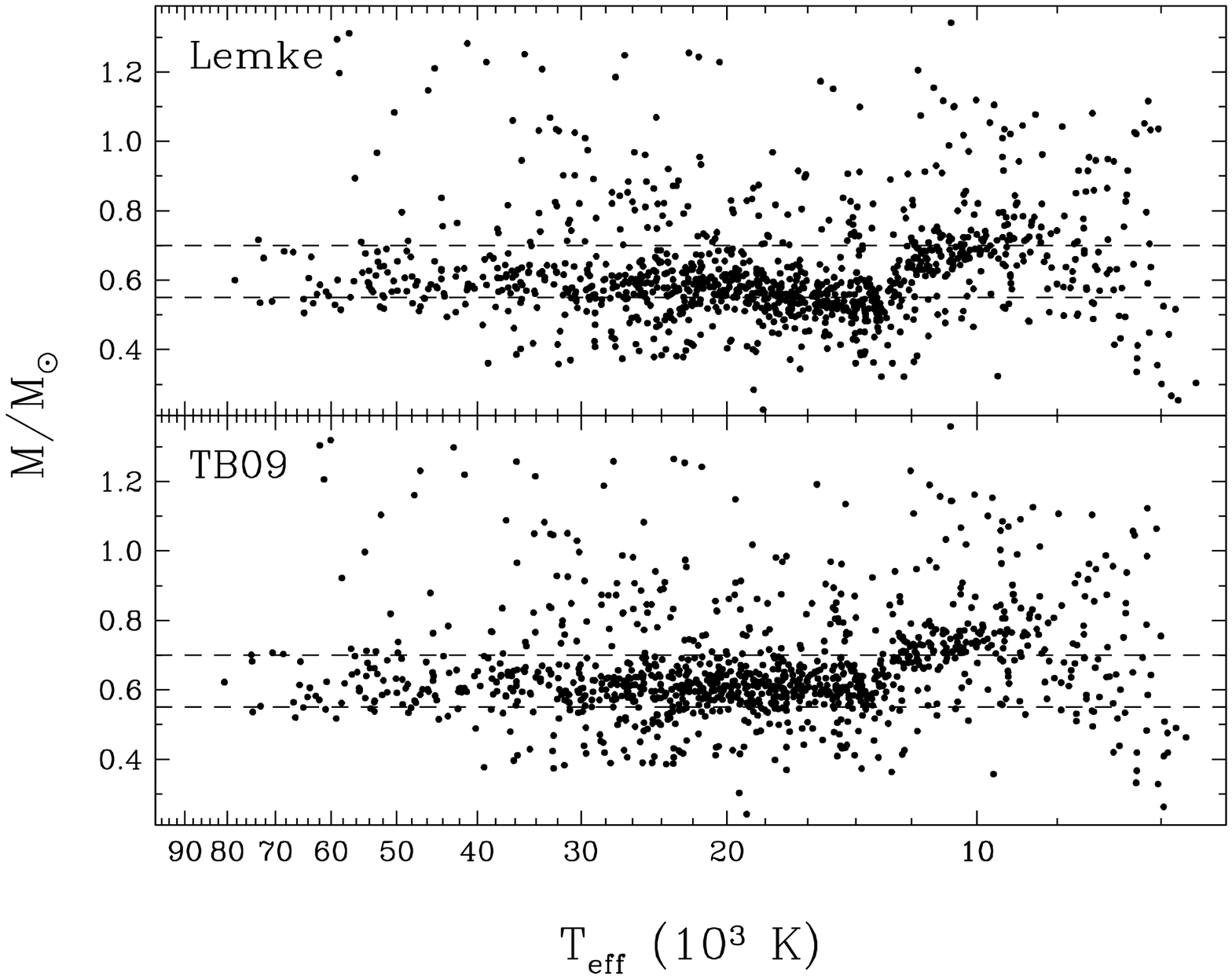}
\figcaption[f10.eps]{Mass vs. \Te\ distribution for 1151 DA white
  dwarfs in our sample using the Stark broadening profiles of
  Lemke~(1997;~top) and the improved calculations of TB09
  (bottom). Lines of constant mass at 0.55 \msun\ and 0.7 \msun\ are
  shown as a reference. \label{fg:comp_mass}}
\end{figure}

\begin{figure}[!h]
\includegraphics[scale=0.45,bb=20 117 592 679]{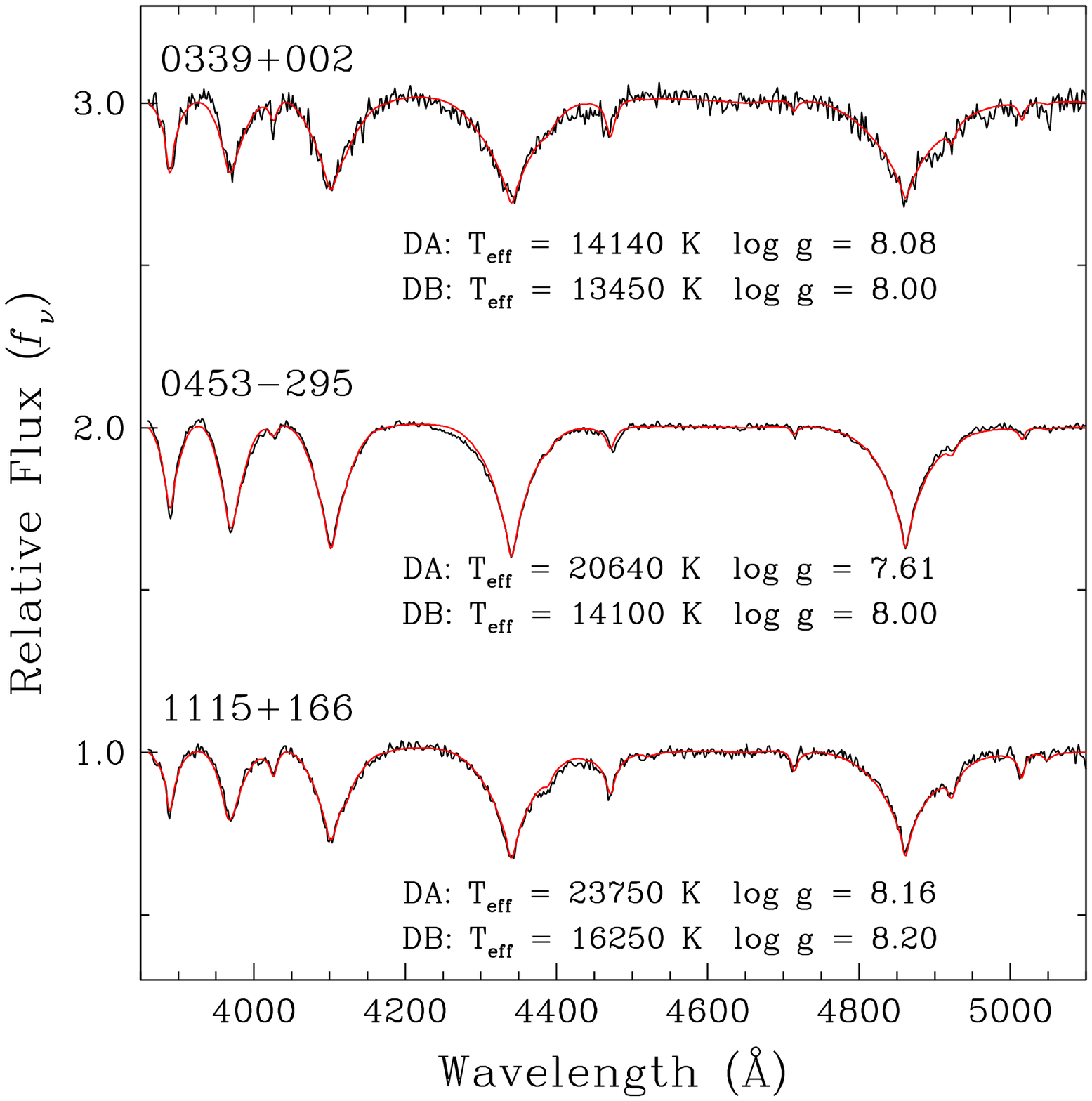}
\figcaption[f11.eps]{Model fits (red) to the optical spectra (black)
  for the three DA+DB binary systems in our sample. The atmospheric
  parameters for each solution are given in the figure. Both the
  observed and theoretical spectra are normalized to a continuum set
  to unity and the spectra are vertically shifted for
  clarity. \label{fg:DADB}}
\end{figure}

In Figure~\ref{fg:comp_mass}, we examine how these differences in the
measured atmospheric parameters translate to the determination of the
mass distribution as a function of \Te. Using the old profiles, we see
in the top panel of Figure~\ref{fg:comp_mass} that besides the
high-\logg\ problem at lower temperatures \citep[][and Section
\ref{sec:mass} below]{tremblay10}, the mass distribution dips near
13,000~K. On the other hand, the new profiles seem to even out the
mass distribution with the bulk of the stars showing masses between
0.55 \msun\ and 0.7 \msun\ before the high-\logg\ problem takes hold
at lower \Te. As such, the new profiles permit an improved
determination of the mass distribution as a function of \Te\ over the
older models that employ Lemke's Stark broadening
profiles. Specifically, the TB09 profiles produce a relatively
constant mass distribution down to about 13,000~K where the
high-\logg\ problem begins to manifest itself.

In the following sections, we will focus on the spectroscopic analysis
of all the objects and systems that require particular
attention. These include white dwarfs of spectral types DAB, DAZ,
DA+dM binary systems, magnetic white dwarfs, and a selection of other
unique objects that demanded a more detailed analysis. The motivation
here is to ensure that we determine the most accurate atmospheric
parameters possible for each star. With that in mind, we also note
that from this point forward, we will be using the new Stark profiles
of TB09 exclusively.

\subsection{DAB White Dwarfs}\label{sec:DAB}

In Figure~\ref{fg:misc}, we displayed five white dwarfs classified as
DAB stars. In other words, besides the hydrogen Balmer lines, the
spectra of these stars also feature lines due to neutral helium. For
three of these objects (0339+002, 0453$-$295, 1115+166), previous
analyses have shown that fits assuming a single star do not produce
satisfactory results. In contrast, fits performed under the assumption
that the spectrum is that of a DA+DB double-degenerate binary system
have proven successful
\citep[][respectively]{limoges10,wesemael94,bergeron02}.  With that in
mind, we proceed to fit the spectra of these systems by combining our
usual pure hydrogen models with the pure helium models described in
Section~\ref{sec:helium}. When fitting DA+DB model spectra, the total
flux of the system is obtained from the sum of the monochromatic
Eddington fluxes of the individual components, weighted by their
respective radius. The stellar radii are obtained from evolutionary
models similar to those described in \citet{fontaine01} but with C/O
cores, $q({\rm He})\equiv \log M_{\rm He}/M_{\star}=10^{-2}$ and
$q({\rm H})=10^{-4}$, which are representative of hydrogen atmosphere
white dwarfs, and $q({\rm He})=10^{-2}$ and $q({\rm H})=10^{-10}$,
which are representative of helium atmosphere white dwarfs.

The fitting technique for these DA+DB systems first requires that we
normalize both the observed and synthetic spectra to a continuum set
to unity. The calculation of $\chi^2$ is then carried out in terms of
these normalized line profiles only. The atmospheric parameters, \Te\
and \logg, for the DA white dwarf and \Te\ for the DB white dwarf are
considered free parameters. We set \logg\ = 8.0 for the DB components
in 0339+002 and 0453$-$295 since the helium lines are rather weak and
are not especially sensitive to the surface gravity. The helium lines
are comparatively stronger in 1115+166 so for the fit of this star, we
allow \logg\ to vary.

\begin{figure}[!t]
\includegraphics[scale=0.80,angle=-90,bb=162 216 516 784]{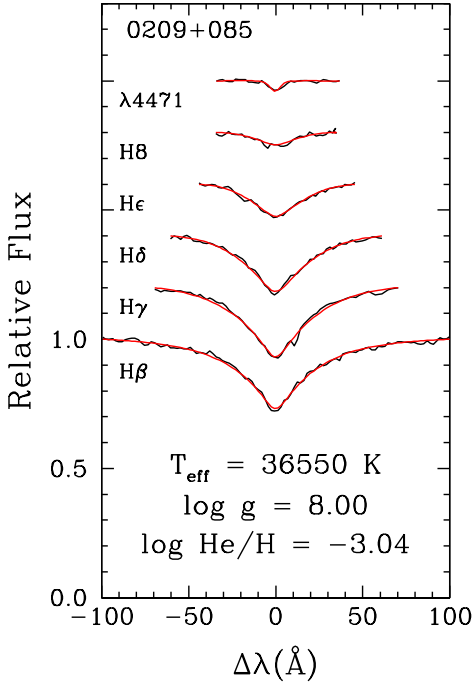}
\figcaption[f12.eps]{Model fits (red) to the hydrogen Balmer lines and
  the \hei\ line in the observed optical spectrum (black) of the DAB
  white dwarf HS~0209+0832. The lines range from \hbeta\ (bottom) to
  H8 in addition to \hei\ (top), each offset by a factor of 0.2. The
  best-fit values of \Te, \logg, and log He/H are indicated at the
  bottom of the figure. \label{fg:DAB}}
\end{figure}

Our best fits to the spectra of 0339+002, 0453$-$295, and 1115+166 are
shown in Figure~\ref{fg:DADB}. We obtain excellent fits for all three
systems with the fit for 1115+166 being exceptionally good. For
0339+002, we obtain higher values of \Te\ and \logg\ for the DA
component as compared to the results of \citet{limoges10} with a
comparable value of \Te\ for the DB component. This is not an
altogether surprising result for the DA star since we are using the
new profiles of TB09 in our analysis. With regards to 0453$-$295, we
obtain a substantially higher \Te\ and \logg\ for the DA component as
compared to \citet{wesemael94} and also a higher \Te\ for the DB white
dwarf. However, the models used in their analysis are considerably
outdated and they also performed a four parameter fit allowing the
surface gravity of the DB star to vary. Finally, we get somewhat
higher values of \Te\ and \logg\ than \citet{bergeron02} for the DA
component but identical parameters for the DB.

The case of HS~0209+0832 is different from the above three systems as
it is a true DAB white dwarf, as first reported by
\citet{jordan93}. In other words, it is a single star with both
hydrogen and neutral helium lines present in its optical
spectrum. Furthermore, like GD~323, this star presents neutral helium
lines that are variable, in particular the \hei\ line
\citep{heber97}. However, contrary to GD~323, \citet{wolff00} were
able to explain the variability as resulting from the accretion of
pockets of interstellar matter of varying densities leading to
different accretion rates over the course of time. We show in
Figure~\ref{fg:DAB} the best fit to our optical spectrum of
HS~0209+0832.  Besides the usual Balmer lines, we also fit the \hei\
line in order to measure the helium abundance. To perform the fit, we
have used the grid of homogeneously mixed H/He models described by
\citet{gianninas10} for DAO white dwarfs, with some
modifications. Namely, we have extended the grid down to \Te\ =
30,000~K. However, we do not include CNO in the models as there is no
evidence of the Balmer line problem in HS~0209+0832.

Our new determinations (\Te\ = 36,550, \logg\ = 8.00) are quite
comparable to the previous measurements of \citet[][\Te = 36,100,
\logg\ = 7.91]{heber97} and \citet[][\Te = 35,500, \logg\ =
7.90]{wolff00} with our higher values likely due to the use of the new
TB09 Stark broadening profiles. Conversely, we determine a much lower
helium abundance of \loghe\ = $-$3.04 as compared to the values
compiled in Table 2 of \citet{wolff00} that hover around \loghe\ $\sim
-2.0$ with the lowest value at \loghe\ = $-$2.32 \citep{heber97}. It
is important to note that our spectrum was obtained in 2002 December,
more than three years after the last observation listed in Table 2 of
\citet{wolff00}. Furthermore, Wolff et~al. state that ``A lower rate
would reduce the abundance within a few months due to the short
diffusion timescale''. Hence, the most logical explanation for our
considerably lower helium abundance determination is that in the
intervening time, HS~0209+0832 traversed a region of space where the
helium density was lower, leading to a reduced accretion rate.

Finally, we do not attempt to analyze GD~323. Indeed, to date there
has not been a model that has been able to achieve a satisfactory fit
to the optical spectrum of GD~323 despite many attempts throughout the
years. We refer the reader to \citet{pereira05} for a comprehensive
summary of the various models that have been elaborated over the years
in order to explain both the optical spectrum of GD~323, and the
observed spectral variability.

\begin{figure*}[!t]
\includegraphics[scale=0.85,bb=20 287 592 509]{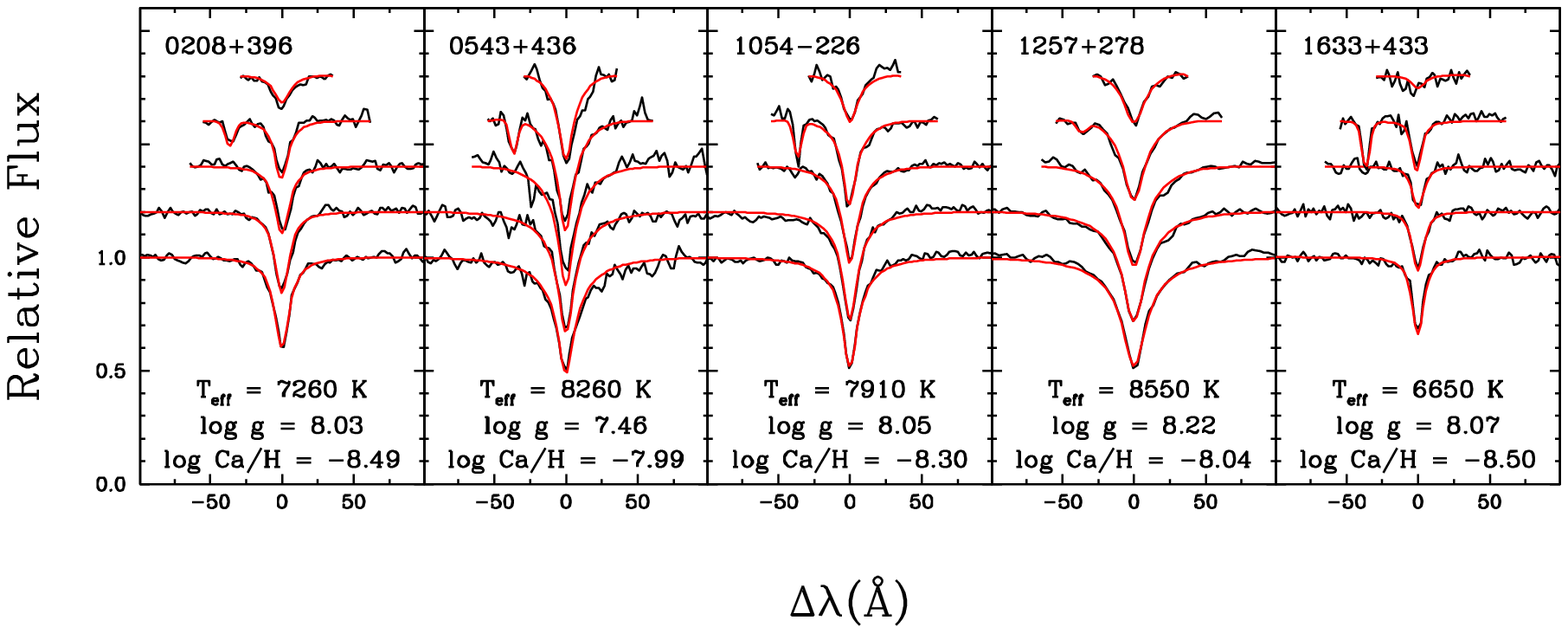}
\figcaption[f13.eps]{Model fits (red) to the hydrogen Balmer lines and
  the Ca H \& K lines in the observed optical spectra (black) of five
  DAZ white dwarfs. The lines range from \hbeta\ (bottom) to H8 (top),
  each offset by a factor of 0.2. We note that the Ca H line is
  blended with \hepsilon. The best-fit values of \Te, \logg, and log
  Ca/H are indicated at the bottom of each panel. \label{fg:DAZ}}
\includegraphics[scale=0.85,bb=20 287 592 529]{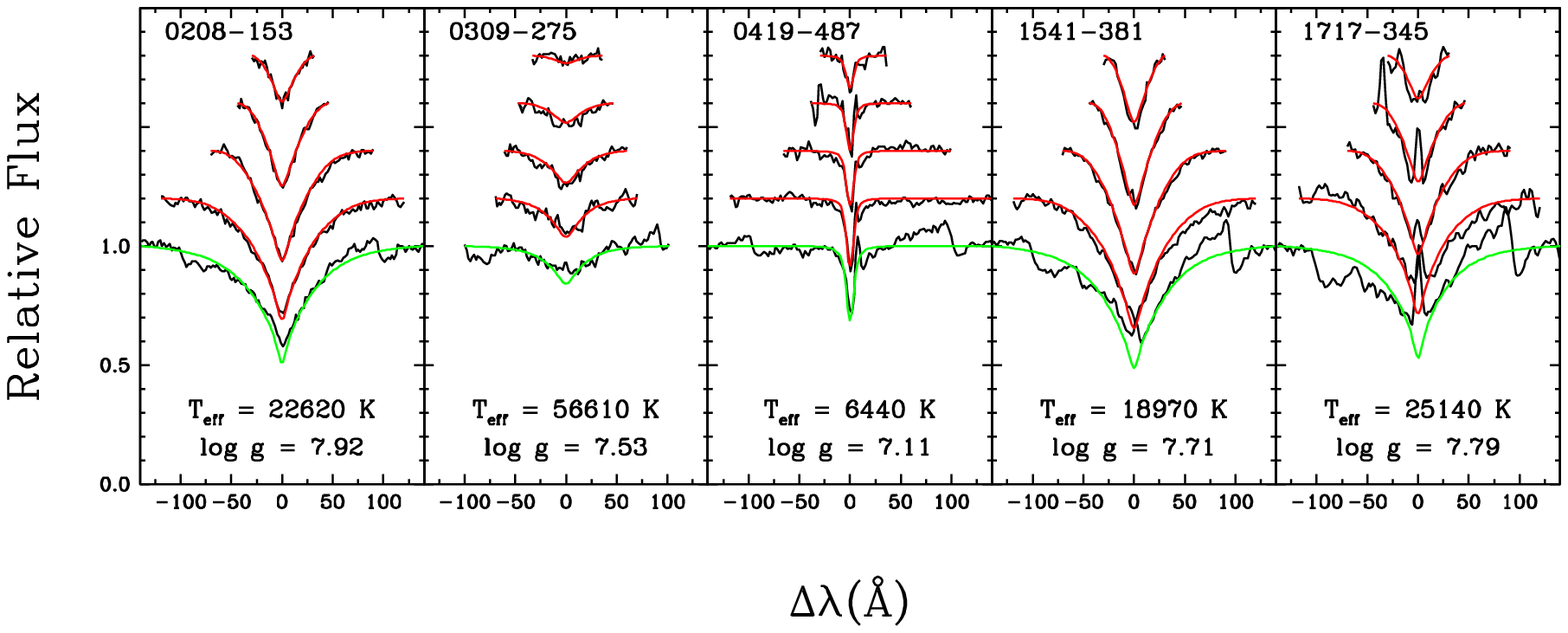}
\figcaption[f14.eps]{Same as Figure~\ref{fg:fits_DA} for five DA+dM
  binary systems for which we do not have red spectra. In these cases,
  \hbeta\ (green) is omitted from the fitting procedure as the only
  contaminated spectral line. \label{fg:DAdM2}}
\includegraphics[scale=0.85,bb=20 287 592 529]{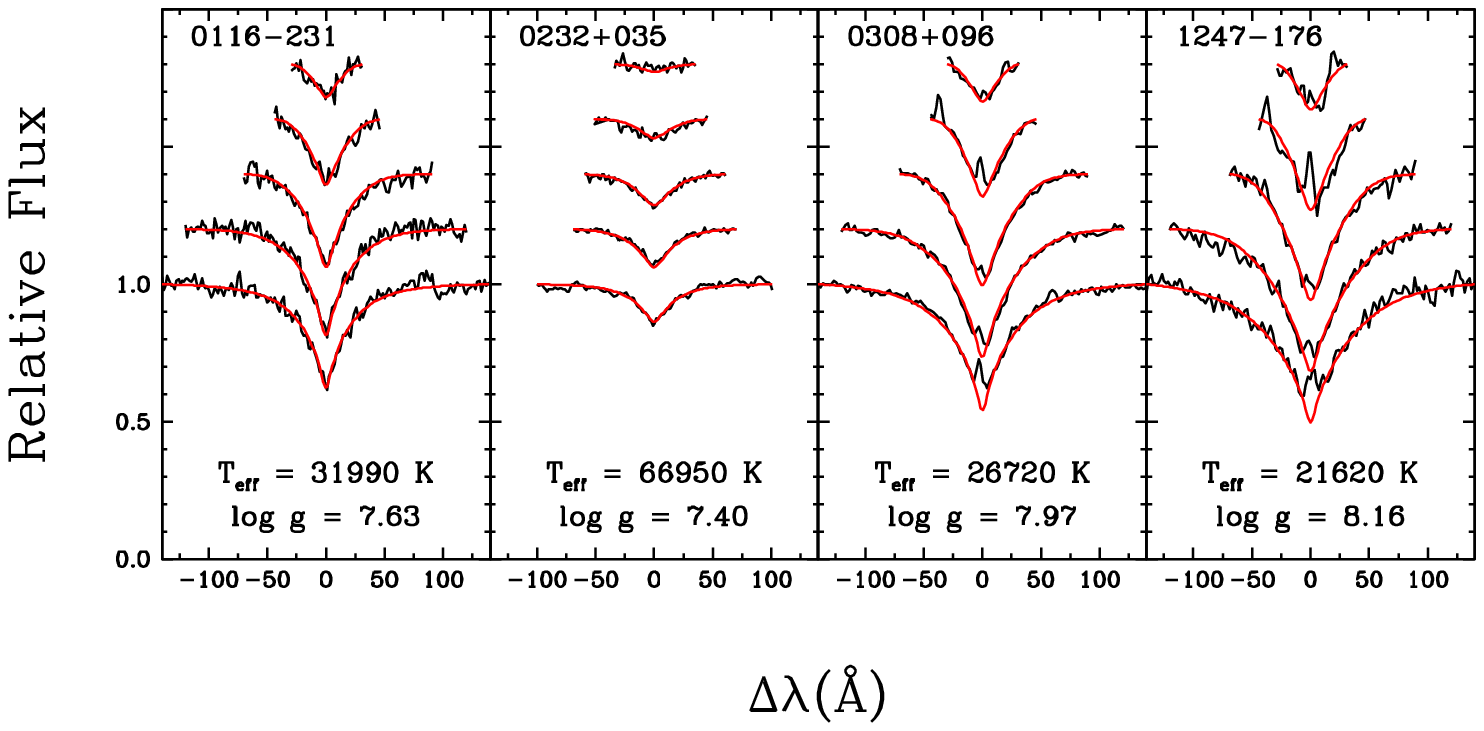}
\figcaption[f15.eps]{Same as Figure~\ref{fg:fits_DA} for four
  additional DA+dM binary systems where little to no contamination is
  detected in their blue spectra. \label{fg:DAdM1}}
\end{figure*}

\subsection{DAZ White Dwarfs}

Our sample contains the six DAZ white dwarfs shown in
Figure~\ref{fg:misc}. With the exception of GD~362, we fit the
remaining five DAZ white dwarfs to obtain \Te, \logg, as well as the
calcium abundance. In order to do so, we computed a small grid of
synthetic spectra based on our grid of pure hydrogen atmospheres and
added calcium in the calculation of the synthetic spectrum only, as
described in \citet{billeres97}. The inclusion of calcium in the
calculation of the synthetic spectrum alone, and not in the model
atmosphere itself, is justified since the trace amounts of calcium
will have no effect whatsoever on the thermodynamic structure of the
atmosphere \citep[see][]{gianninas04}. Our grid covers \Te\ from 6000
to 9000~K in steps of 500~K, \logg\ from 7.0 to 9.5 in steps of 0.5
dex and $\log$ Ca/H from $-$7.0 to $-$9.5 in steps of 0.5
dex. Although our standard spectroscopic technique works well even
without the inclusion of calcium in the synthetic spectra, the fact is
that the calcium H line (at 3968~\AA) is blended with \hepsilon\ (at
3970~\AA). Since the upper Balmer lines are especially sensitive to
the surface gravity, any failure to properly model the line profile
would lead to a less precise measurement of \logg. Therefore, we are
compelled to include the calcium lines in our spectra to avoid this
potential source of uncertainty.

Our best fits of the five DAZ stars using the above grid are displayed
in Figure~\ref{fg:DAZ}. We have employed essentially the same fitting
procedure used for our regular DA white dwarfs but here we have three
free parameters: \Te, \logg, and the calcium abundance. Consequently,
we have extended the fitting region around \hepsilon\ blueward in
order to properly include the Ca K line in the fit. We can see that we
are able to reproduce the calcium lines rather well in all
cases. Furthermore, we can compare the atmospheric parameters of the
three known DAZ stars, G74-7 (0208+396), LTT~13742 (1257+278), and
G180-43 (1633+433), to previous determinations. Specifically,
\citet{billeres97} obtained \Te\ = 7260~K, \logg\ = 8.03, and log Ca/H
= $-$8.80 for G74-7, and \citet{zuckerman03} obtained \Te\ = 8481~K,
\logg\ = 7.90, and log Ca/H = $-$8.07 for LTT~13742, and \Te\ =
6569~K, \logg\ = 8.08, and log Ca/H = $-$8.63 for G180-43. These
values are quite comparable to the values we measure here. Indeed,
some variation is expected considering our models contain several
improvements over the ones used in the previous analyses. With respect
to the two newly identified DAZ stars, G96-53 (0543+436) and LP 849-31
(1054$-$226), their atmospheric parameters seem to suggest that they
are rather typical of DAZ white dwarfs within the same temperature
range \citep[see Figure~5 of][]{zuckerman03}. 

\subsection{DA+dM Binary Systems}

As stated in Section 2, we have a total of 47 DA+dM binary systems in
our sample. For nine of those systems, we employ our standard fitting
technique, as described above. The fits for these nine systems are
displayed in Figures \ref{fg:DAdM2} and \ref{fg:DAdM1}. Of these nine,
five are the DA+dM systems, listed earlier, for which we were unable
to secure red spectra (see Section~\ref{sec:spec}). In these cases, we
simply omit \hbeta, the only contaminated Balmer line, from our
fitting procedure and fit only the Balmer lines from \hgamma\ to
H8. We see in Figure~\ref{fg:DAdM2} that we are able to achieve a more
than satisfactory fit to the remaining four Balmer lines. On the other
hand, the remaining four DA+dM systems, shown in
Figure~\ref{fg:DAdM1}, display little to no contamination from their
companion in their blue spectra and we obtain excellent fits to their
observed line profiles without the need to omit any lines. First, the
red spectrum of 0116$-$231 shows virtually no contamination from a
companion, as we can see in the left panel of
Figure~\ref{fg:specDAdM}. Similarly, the red spectrum of 0308+096 is
also devoid of contamination except for emission at \halpha. In
contrast, the blue spectrum 0232+035 seems devoid of any contamination
while its red spectrum has emission lines in the cores of all the
Balmer lines. Finally, the blue spectrum of 1247$-$176 shows weak
emission for some of the Balmer lines while we see strong emission in
the core of all the Balmer lines in its red spectrum (see the middle
panel of Figure~\ref{fg:specDAdM}). In the first two cases, there is
insufficient contamination in the blue spectra to warrant using the
procedure described below to subtract the contamination due to the
secondary. In the latter two cases, the red spectrum is considerably
more contaminated than the blue spectrum. As such, our procedure would
likely only make matters worse. Furthermore, these seemingly
contradictory detections and non-detections of emission lines from the
M dwarf secondary in the latter two cases are likely due to the fact
that the red and blue spectra were obtained at different phases of the
binary orbit. Finally, as described above for 0342+374, we exclude the
center of each Balmer line from the fitting procedure for 0308+096,
1247$-$176, and 1717$-$345, as they are contaminated by emission lines
from their M dwarf companions.

\begin{figure}[!t]
\includegraphics[scale=0.45,bb=20 167 592 654]{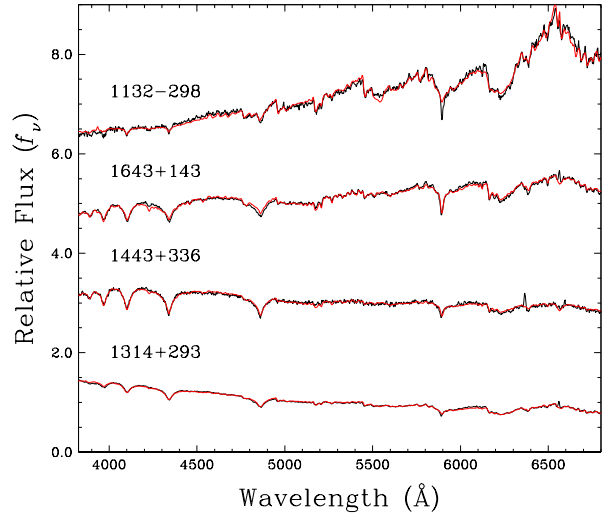}
\figcaption[f16.eps]{Best fits of the observed red spectra (black) for
  four of our DA+dM binary systems using our combination of synthetic
  DA spectra and dM spectral templates (red). The spectra are
  normalized to unity at 5100~\AA\ and shifted vertically for
  clarity. \label{fg:DAdMfit}}
\end{figure}

For the remaining 38 systems, we will take advantage of the red
spectra we obtained (see Figure~\ref{fg:specDAdM}) in order to remove
the contamination due to the presence of the M dwarf. There have been
several studies in the last few years that have analyzed DA+dM systems
from the SDSS \citep{silvestri06,heller09,rebassa10}. One important
distinction between these studies and our own is that in the above
cases the authors were interested in recovering both the spectrum of
the white dwarf and the spectrum of the M dwarf in an effort to study
the system as a whole and obtain atmospheric parameters for both
stars. In contrast, we are only interested in obtaining a more
accurate measurement of the parameters for the DA star.

\begin{figure*}[!t]
\includegraphics[scale=0.65,angle=-90]{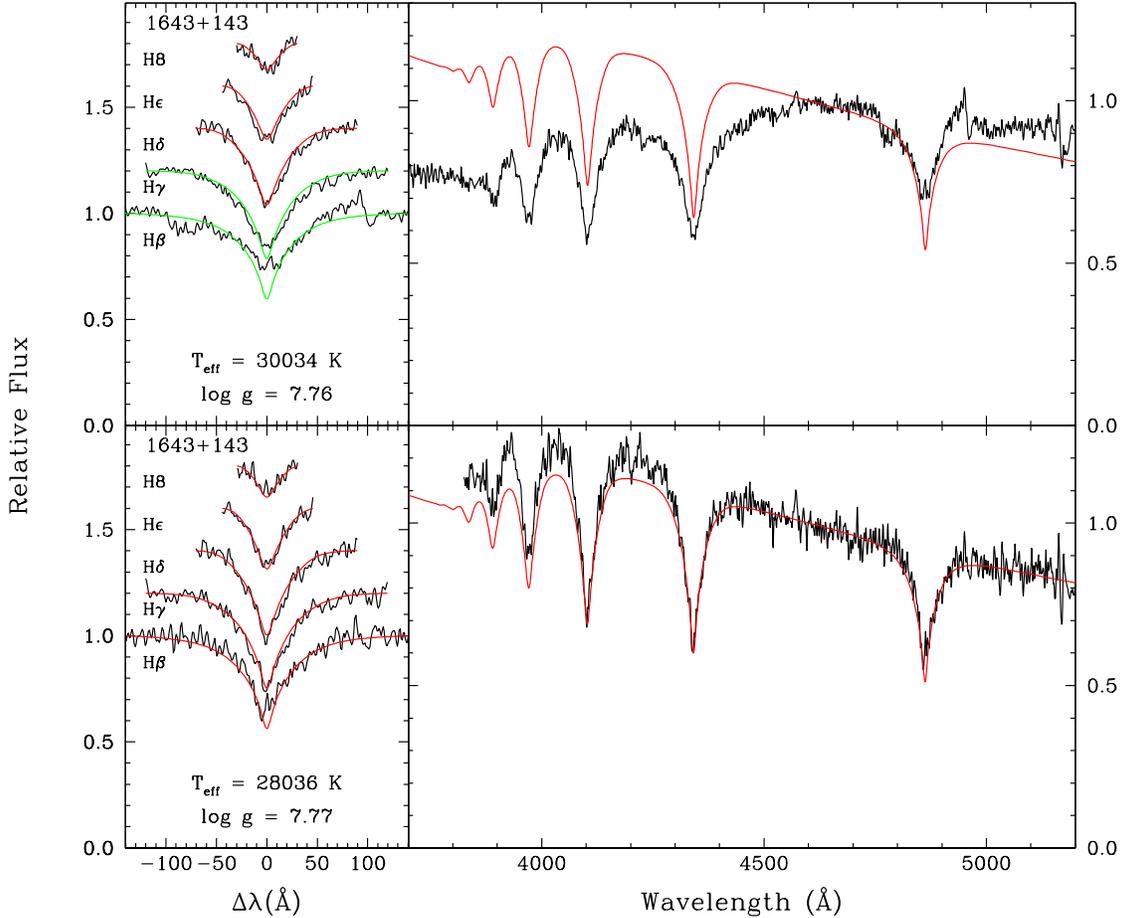}
\figcaption[f17.eps]{Comparison of the spectroscopic solutions for
  1643+143 before (top) and after (bottom) subtraction of the dM
  component. The left panels show the fits (red) to the observed
  Balmer lines (black) from \hbeta\ to H8 where fits in green indicate
  spectral lines that are not taken into account in the fitting
  procedure. The right panels compare the observed spectrum (black)
  with a synthetic spectrum (red) interpolated at the values of
  \Te\ and \logg\ obtained from the spectroscopic fit, both are
  normalized to unity at 4600 \AA. Note the scales for the relative
  flux are different in the left and right panels. \label{fg:WD1643}}
\end{figure*}

Our fitting procedure for these systems combines the synthetic spectra
computed from our pure hydrogen model atmospheres with the M dwarf
spectral templates compiled and presented by \citet{bochanski07}. We
attempt to fit a function of the form

\begin{align*}
F_{\rm obs} &= [F_{\rm DA}(T_{\rm eff},\log g) + a_{1}\cdot 
F_{\rm dM}({\rm Sp.~type})] \nonumber \\
&\quad {}\times(a_{2} + a_{3}\lambda + a_{4}\lambda^{2} + a_{5}\lambda^{3}) 
\end{align*}

\noindent to the observed spectrum. $F_{\rm DA}$ is the flux due to
the DA white dwarf and depends on \Te\ and \logg\ while $F_{\rm dM}$,
the flux from the dM component, is a function solely of the M dwarf
spectral type, from M0 to M9. The remaining free parameters are
$a_{1}$, which is the relative contribution of the synthetic DA
spectrum and the M dwarf spectral template, $a_{2}$, a scaling factor
between the composite synthetic spectrum and the observed spectrum,
and the coefficients $a_{3-5}$ of a third-order polynomial that is
meant to account for errors in the flux calibration. In total, there
are eight free parameters when we fit the composite spectra. This
means that the values of \Te, \logg, and spectral type that we obtain
are meaningless since there are too many parameters used in the
fitting process. However, as we mentioned above, our only interest is
to obtain the necessary function to subtract from the composite
spectrum in order to remove the dM contamination. We show in
Figure~\ref{fg:DAdMfit} four examples of the fits we obtain to our
composite DA+dM spectra. We see that we are able to achieve some very
good fits to the observed spectra. It is interesting to note the
varying degrees of contamination from one system to the next. For
example, 1443+336 shows only light contamination in contrast to
1132$-$298 where the white dwarf is nearly drowned out by the flux
from its M dwarf companion. Indeed, in the case of 1132$-$298, and
several other systems, when letting all the free parameters listed
above vary, the fitting procedure produced rather poor results. In
cases such as those, we forced \Te\ and \logg\ to the values obtained
from fitting our blue spectra of the same objects, similar to the fits
presented in Figure~\ref{fg:DAdM2}, and this allowed us to achieve
satisfactory results as evidenced by 1132$-$298.

Once the spectroscopic fit to the composite spectrum is obtained, we
subtract from the spectroscopic solution the contribution of the DA
component [$F_{\rm DA}$(\Te, \logg)]. In other words, we subtract a
synthetic DA spectrum interpolated to the values of \Te, and \logg, as
determined by the fit to the composite spectrum, from the composite
spectroscopic solution. The residual spectrum generated by this
procedure represents the contribution of the M dwarf only. This
residual spectrum is then interpolated onto the wavelength grid of the
blue spectrum of the same object. Finally, we subtract the residual
spectrum from the blue spectrum removing the contamination from the
companion.

The next step is to fit the corrected blue spectrum with our standard
fitting technique. We show in Figure~\ref{fg:WD1643} an example of the
fits for 1643+143 before and after the correction has been applied.
We see quite a dramatic change here. Indeed, where we had to omit the
fits to \hbeta\ and \hgamma, we are now able to fit them quite well
with the contamination removed. The \logg\ value is virtually
identical whereas we determine a \Te\ that is over 2000~K
lower. Finally, we note the greatly improved agreement between the
slope of the observed spectrum and the spectroscopic solution.

\begin{figure}[!t]
\includegraphics[scale=0.55,bb=70 117 592 679]{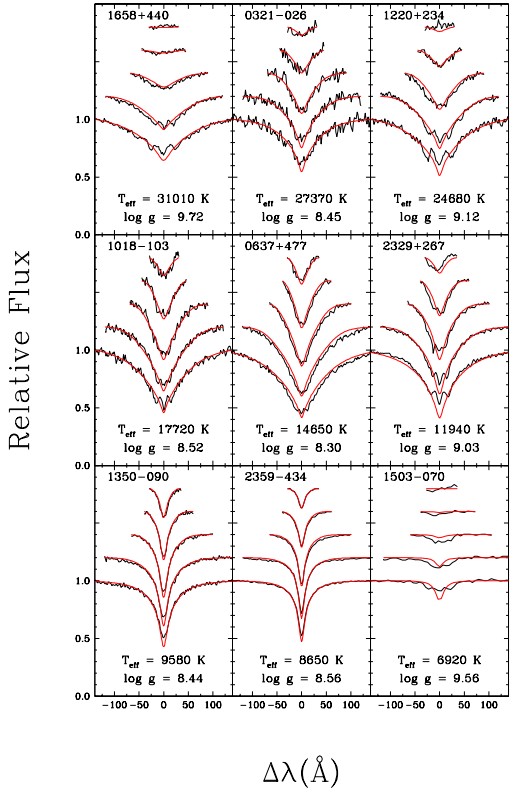}
\figcaption[f18.eps]{Same as Figure~\ref{fg:fits_DA} for nine weakly
  magnetic DA white dwarfs. \label{fg:fitsMAG}}
\end{figure}

\begin{table}[!t]\scriptsize
\caption{Parameters for Magnetic White Dwarfs}
\begin{center}
\begin{tabular}{@{}llcrcc@{}}
\hline
\hline
\noalign{\smallskip}
WD & \multicolumn{1}{c}{Name} & $B_p$ & \multicolumn{1}{c}{\Te} & \logg & Notes \\
   &                          &  (MG) & \multicolumn{1}{c}{(K)} &       &       \\
\noalign{\smallskip}
\hline
\noalign{\smallskip}
0253+508    &KPD 0253+5052  &   17   & 15,000 &   8.00 & 1 \\
0329+005    &KUV 03292+0035 &   12.1 & 26,500 & \ldots & 2 \\
0553+053    &LTT 17891      &   20   &   5500 &   8.00 & 3 \\
0816+376    &GD 90          &    9   & 14,000 &   8.00 & 4,5 \\
0945+245    &LB 11146a      &  670   & 14,500 &   8.50 & 6,7,8 \\
1017+366    &Ton 1206       &   65   & 16,000 & \ldots & 9 \\
1031+234    &Ton 527        & 1000   & 15,000 &   8.50 & 10 \\
1440+753    &RE J1440+750   &   15   & 42,000 &   8.80 & 6,11 \\
1506+399    &CBS 229        &   20   & 17,000 & \ldots & 6,12,13 \\
1533$-$057  &PG 1533$-$057  &   31   & 20,000 &   8.50 & 1,14 \\
1639+537    &GD 356         &   13   &   7510 &   8.14 & 15,16\\
1900+705    &LFT 1446       &  320   & 16,000 &   8.58 & 15 \\
2316+123    &KUV 23162+1220 &   45   & 11,000 &   8.00 & 3 \\
\noalign{\smallskip}
\hline
\noalign{\smallskip}
\multicolumn{6}{@{}p{8.5cm}@{}}{{\bf References.} (1) \citealt{achilleos89}; 
(2) \citealt{gansicke02};
(3) \citealt{putney95}; 
(4) \citealt{putney97};
(5) \citealt{martin84};
(6) Double degenerate;
(7) \citealt{liebert93};
(8) \citealt{glenn94};
(9) \citealt{saffer89};
(10) \citealt{schmidt86};
(11) \citealt{vennes99};
(12) \citealt{gianninas09};
(13) \citealt{vanlandingham05};
(14) LBH05;
(15) \citealt{BLR01};
(16) \citealt{ferrario97}.} \\
\label{tab:mag}    
\end{tabular}
\end{center}
\end{table}

\subsection{Magnetic and Cool White Dwarfs}\label{sec:mag}

In Figure~\ref{fg:MAG}, we presented the spectra of 25 magnetic, or
suspected to be magnetic, white dwarfs that are found in our
sample. For most of these white dwarfs, our standard spectroscopic
technique is inadequate since the Balmer line profiles are severely
distorted and/or completely destroyed by a strong magnetic
field. However, some of the more weakly magnetic stars can be fit with
our spectroscopic method. We note that we do not exclude any part of
the observed line profiles from our fitting procedure. Consequently,
the values of \Te\ and \logg\ derived from these fits can only be
considered, at best, as rough estimates of the atmospheric
parameters. In Figure~\ref{fg:fitsMAG}, we present the fits for nine
weakly magnetic white dwarfs. For 1658+440, 1220+234, 1018$-$103,
0637+477, 2329+267, and 1503$-$070, the Zeeman splitting of the Balmer
lines is fairly obvious. In the cases of 1350$-$090 and 2359$-$434,
the spectroscopic solution predicts the Balmer lines to be deeper than
observed, this is also evidence of a weak magnetic field, which is
independently confirmed through spectropolarimetric measurements
\citep[][respectively]{schmidt94,aznar04}. Finally, the reason why we
classified 0321$-$026 as magnetic becomes evident here as the Balmer
lines are again predicted to be too strong. We note that all nine
stars have \logg\ values which are substantially higher than the
canonical value of \logg~=~8.0, this is not in the least surprising
since it is known that magnetic white dwarfs have masses which are
higher than the average for normal DA stars
\citep{wickramasinghe00}. When trying to fit these magnetic stars with
non-magnetic models, this phenomenon is exacerbated by the Zeeman
splitting which broadens the lines even further causing our
spectroscopic solutions to require even higher \logg\ values. There
are no better examples of this effect than 1658+440 and 1503$-$070
whose \logg\ values have actually been extrapolated outside of our
model grid.

The 16 remaining magnetic white dwarfs are simply beyond the scope of
this work. The same can be said of the cool white dwarfs shown in the
right panel of Figure~\ref{fg:misc}. For this reason, these white
dwarfs will not be further analyzed here. However, in the interest of
completeness, we have searched throughout the literature in order to
compile atmospheric parameter determinations, and other derived
quantities, for these white dwarfs. First, Table~\ref{tab:mag} lists
the properties for 13 of the 16 remaining magnetic white dwarfs,
including the magnetic field strength (assuming a magnetic dipole),
\Te, and \logg, where available, along with the references to the
original analyses. No information could be found for 0350+098, and
1610+330 is identified as a magnetic white dwarf for the first time in
this work. Finally, 0239+109 is analyzed at length in the next
section. Similarly, Table~\ref{tab:cool} lists the atmospheric
parameters for all the cool white dwarfs in our sample. Specifically,
we indicate \Te, \logg, mass, and the absolute visual magnitude along
with the necessary references. We also include in Table~\ref{tab:cool}
the parameters for GD~362 from \citet{zuckerman07} as we have not
computed appropriate helium models, which include the necessary
metals, for a proper re-analysis of this particular object.

\begin{table}[!t]\scriptsize
\caption{Parameters for Cool White Dwarfs}
\begin{center}
\begin{tabular}{@{}llrcccc@{}}
\hline
\hline
\noalign{\smallskip}
WD & \multicolumn{1}{c}{Name} & \multicolumn{1}{c}{\Te} & \logg & $M$/\msun & \mv & Notes \\ 
   &                          & \multicolumn{1}{c}{(K)} &       &           &     &       \\
\noalign{\smallskip}
\hline
\noalign{\smallskip}
0121+401    &G133-8    &   5340 & 7.90 & 0.52 & 14.65 & 1 \\
0213+427    &G134-22   &   5600 & 8.12 & 0.66 & 14.72 & 1 \\
0551+468    &LP 159-32 &   5380 & 8.01 & 0.59 & 14.78 & 1 \\
0648+368    &GD 78     &   5700 &\ldots&\ldots&\ldots & 2 \\
0727+482A   &G107-70A  &   5020 & 7.92 & 0.53 & 15.03 & 1,3 \\
0727+482B   &G107-70B  &   5000 & 8.12 & 0.66 & 15.33 & 1,3 \\
1055$-$072  &LFT 753   &   7420 & 8.42 & 0.85 & 13.91 & 1 \\
1750+098    &G140-B2B  &   9527 &\ldots& 1.17 &\ldots & 4 \\
1820+609    &G227-28   &   4780 & 7.83 & 0.48 & 15.16 & 1 \\
\noalign{\smallskip}
\hline
\noalign{\smallskip}
\multicolumn{7}{c}{DAZ} \\
\noalign{\smallskip}
\hline
\noalign{\smallskip}
1729+371    &GD 362    & 10,540 & 8.24 &\ldots&\ldots & 5 \\
\noalign{\smallskip}
\hline
\noalign{\smallskip}
\multicolumn{7}{@{}p{8.5cm}@{}}{{\bf References.} (1) \citealt{BLR01}; 
(2) \citealt{angel81};
(3) Double degenerate;
(4) \citealt{silvestri01};
(5) \citealt{zuckerman07}.} \\
\label{tab:cool}
\end{tabular}
\end{center}
\end{table}

We note that from here on, we shall not take into consideration the
values compiled in the two tables described above in our analysis of
the global properties of our spectroscopic sample. We wish to include
only white dwarfs whose atmospheric parameters were determined
directly by us, as we endeavor to maintain the homogeneity of our
analysis.

\subsection{Peculiar Objects}\label{sec:pec}

\begin{figure}[!t]
\includegraphics[scale=0.60,bb=95 192 592 604]{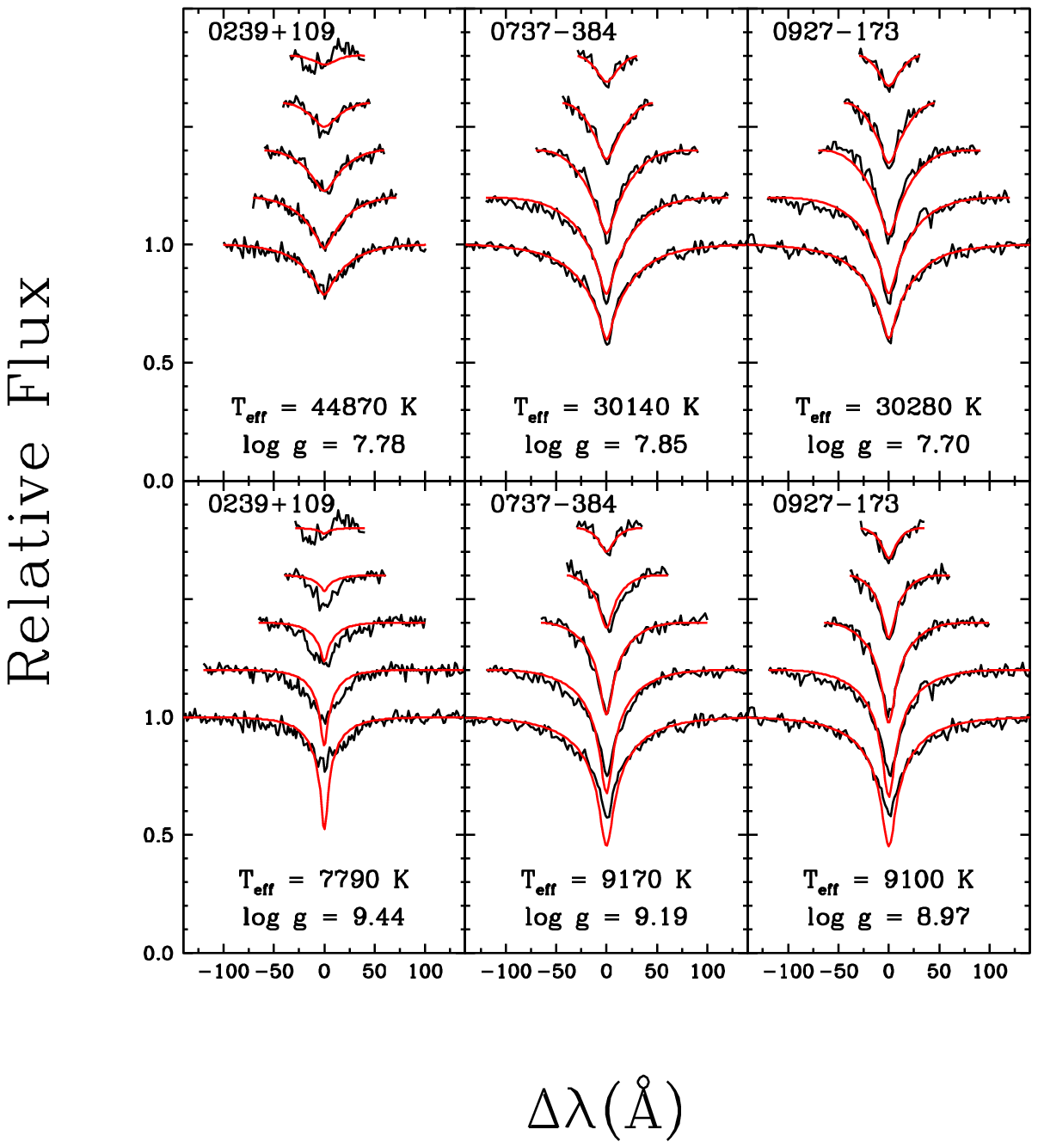}
\figcaption[f19.eps]{Model fits (red) to the observed Balmer lines
  (black) for the three peculiar white dwarfs in our sample using the
  hot seed (top) and the cold seed (bottom) for the determination of
  the spectroscopic solution. The atmospheric parameters for each
  solution are given in the figure. Both the observed and theoretical
  spectra are normalized to a continuum set to unity and the lines
  range from \hbeta\ (bottom) to H8 (top), each offset by a factor of
  0.2. \label{fg:DD_panel}}
\includegraphics[scale=0.625,bb=105 167 592 639]{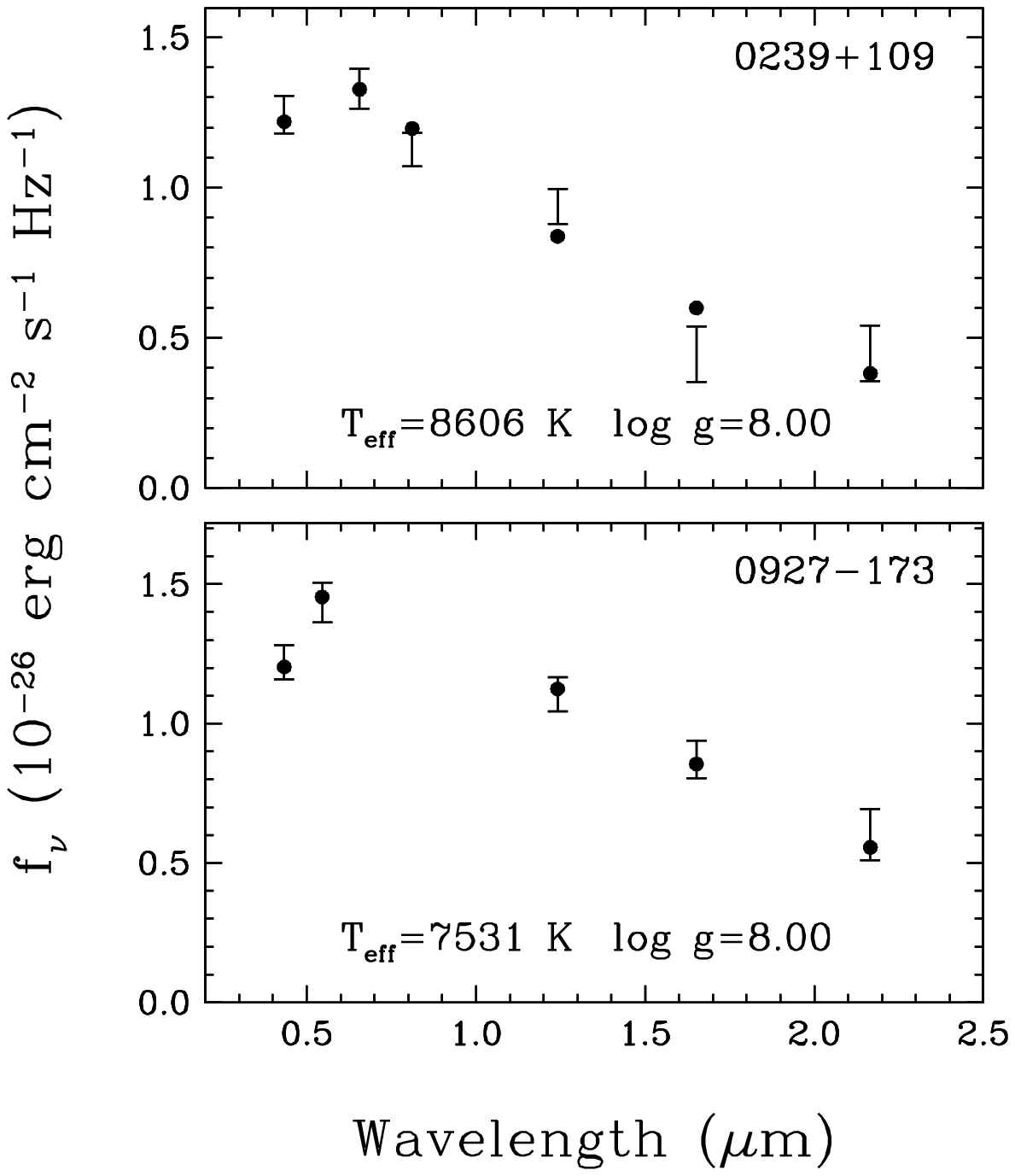}
\figcaption[f20.eps]{Fits to the energy distribution of 0239+109 and
  0927$-$173 with pure hydrogen models. The $BVI$ ($BV$ only for
  0927$-$173) and $JHK_{S}$ photometric observations are represented
  by error bars, while the model fluxes are shown as filled
  circles. The atmospheric parameters corresponding to the photometric
  solution are given in the figure. \label{fg:fitphoto}}
\end{figure}

In the following section, we will take a closer look at three white
dwarfs (0239+109, 0737$-$384, and 0927$-$173) whose analysis demanded
particular attention, in order to be certain we are obtaining the
correct values for their atmospheric parameters. In all three cases,
although the hot solution seemed to be the correct choice, based on
our Balmer line fits, the slopes of the corresponding spectroscopic
solutions proved incompatible with those of the observed spectra. The
fits for the three stars in question are presented in
Figure~\ref{fg:DD_panel}. We see that for all three stars, the hot
solutions (top panels) seem to match the observed spectra rather well,
whereas the cold solutions are completely incompatible with the
data. We now take a closer look at each object in turn.

\subsubsection{0239+109 (G4-34)}\label{sec:0239}

The star 0239+109 is actually one of the 25 magnetic white dwarfs we
presented in Figure~\ref{fg:MAG}, however we will see that this case
is not quite so simple. This star was analyzed in \citet{bergeron90a}
and they had similar difficulties in properly fitting the Balmer-line
profiles (see their Figure~1). To be more exact, the slope of their
observed spectrum clearly suggested a cooler spectroscopic solution
despite the fact that the hot solution produced the better fit to the
Balmer-line profiles. As a result, their conclusion was that 0239+109
must in fact be an unresolved DA+DC binary system. More recently, this
star was observed as part of the SPY survey and \citet{koester09}
classified it instead as a magnetic white dwarf. Indeed, the SPY
spectrum shown in Figure~7.3 of \citet{voss06} shows what seems to be
Zeeman splitting of both \halpha\ and \hbeta. Our spectrum of 0239+109
is one of the many provided to us by C. Moran, and as a result, we
have no information regarding the observations (airmass,
etc.). However, as we mentioned earlier, when trying to fit the
Balmer-line profiles of this star, we encounter the same incoherence
noted by \citet{bergeron90a}, i.e., the hot solution produces the
better fit to the observed Balmer lines, but the slope of the
spectroscopic solution is incompatible with that of our observed
spectrum, which suggests a much cooler temperature. Furthermore, we
have in our archives the older spectrum used in the
\citet{bergeron90a} analysis and obtained by \citet{greenstein86}. The
slopes of the two independent observations are in perfect agreement so
we are confident that there is nothing wrong with the data.

\begin{figure}[!t]
\includegraphics[scale=0.45,bb=20 202 592 604]{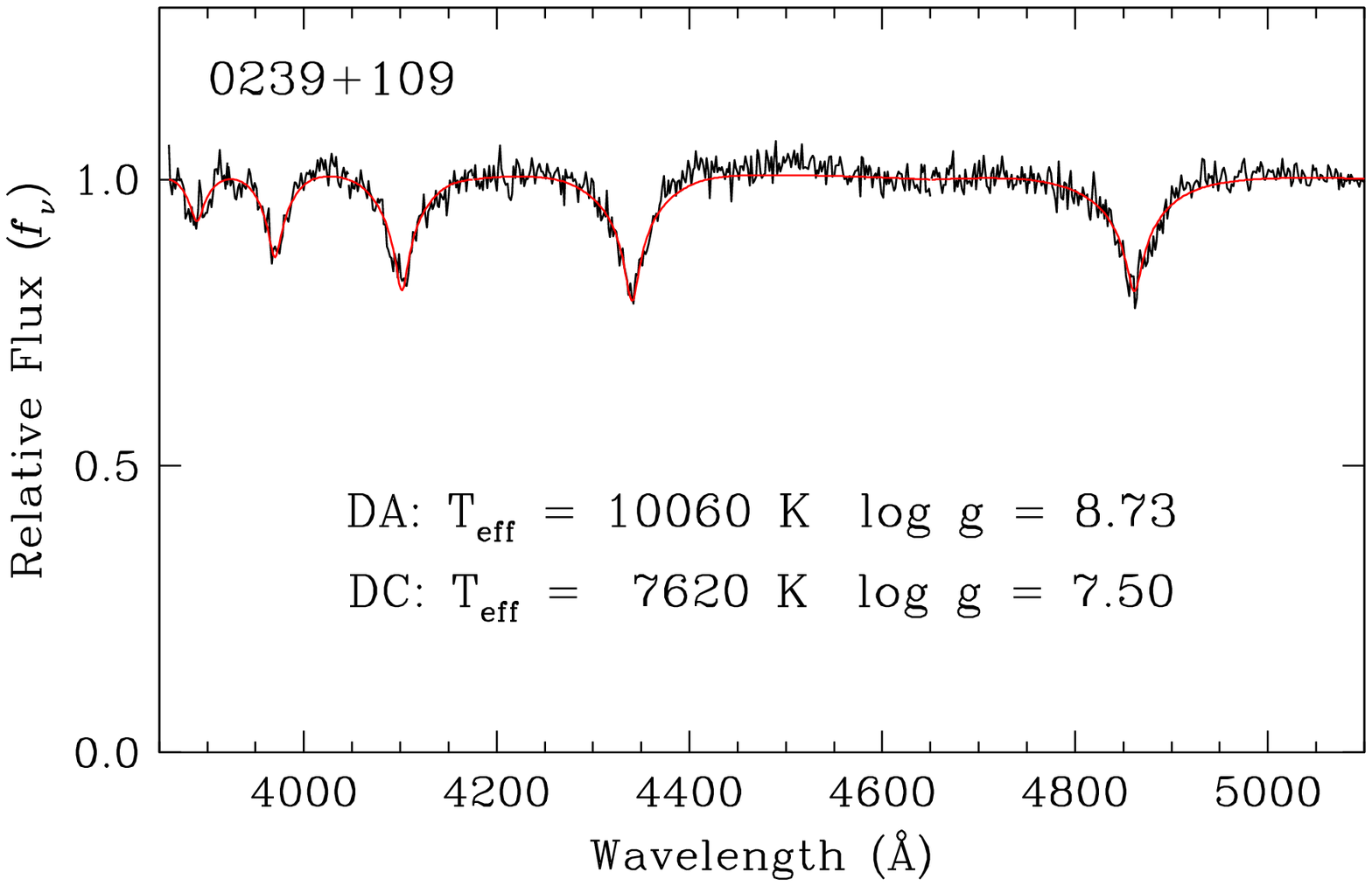}
\figcaption[f21.eps]{Model fit (red) to the optical spectrum (black)
  for the DA+DC binary 0239+109. The atmospheric parameters
  corresponding to the spectroscopic solution are given in the
  figure. Both the observed and theoretical spectra are normalized to
  a continuum set to unity. \label{fg:fitDADC_0239}}
\includegraphics[scale=0.45,bb=18 219 592 668]{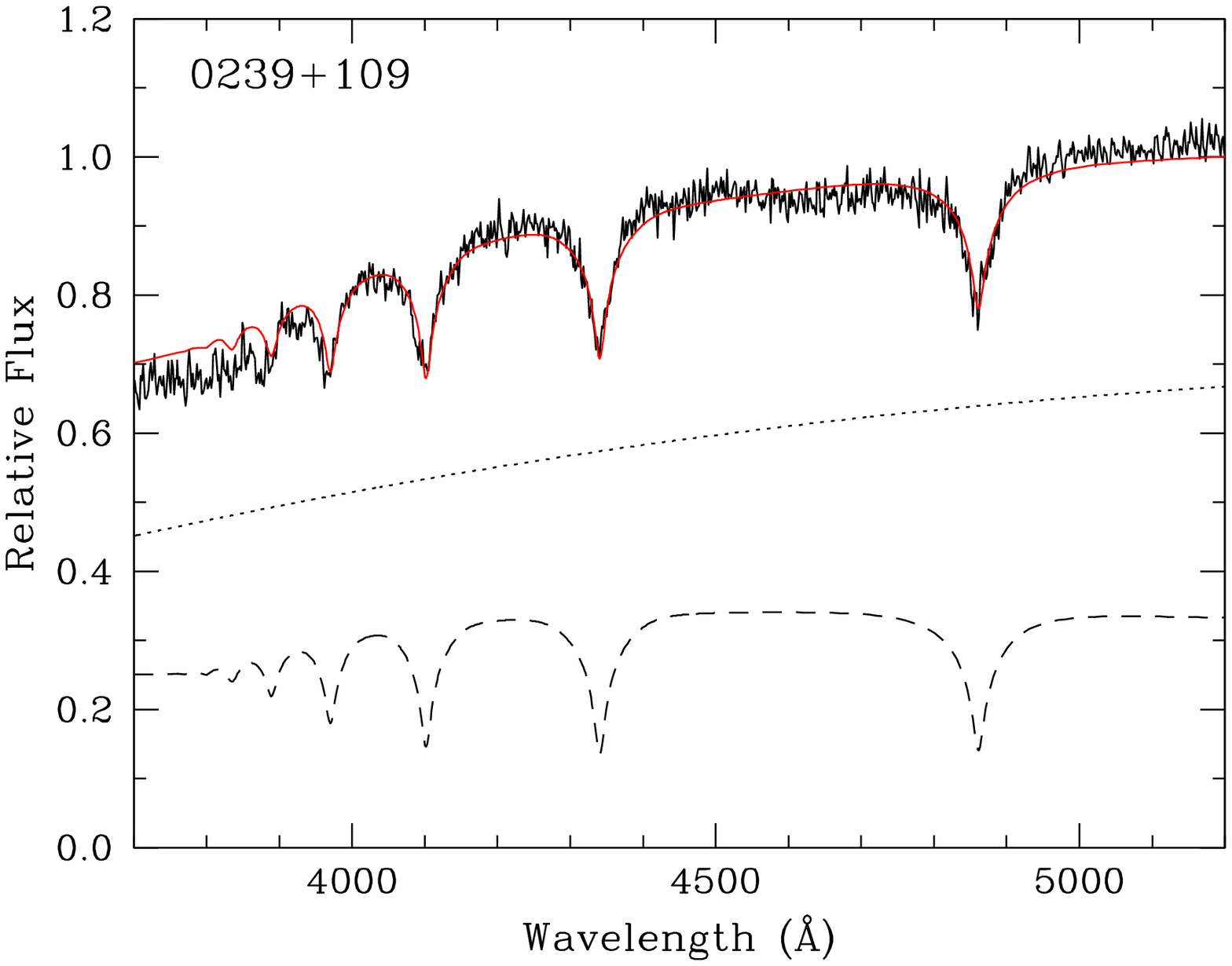}
\figcaption[f22.eps]{Relative energy distributions for our best
  composite DA+DC fit displayed in Figure~\ref{fg:fitDADC_0239}. The
  dashed line represents the contribution of the DA component while
  the dotted line represents the contribution of the DC component,
  both properly weighted by their respective radius. The red line
  corresponds to the total monochromatic flux of the composite system
  superimposed on our spectrum of 0239+109 (black line), which has
  been scaled to the flux of the composite model at 4600
  \AA. \label{fg:plotDADC_0239}}
\end{figure}

In order to get an independent estimate of \Te, we exploit the
available photometry and fit the overall optical-near-infrared
spectral energy distribution combining the $BRI$ photometry from the
USNO-B Catalog \citep{monet03} with the available Two Micron All Sky
Survey \citep[2MASS;][]{cutri03} $JHK_{s}$ magnitudes. Synthetic
colors are obtained using the procedure outlined in \citet{holberg06}
based on the Vega fluxes taken from \citet{bohlin04}. The method used
to fit the photometric data is described in \citet{BLR01}. Since we
have no parallax measurement, we assume \logg~=~8.0. The best fit to
the photometry is presented in the top panel of
Figure~\ref{fg:fitphoto}. Although the photometric fit is not perfect,
overall we see that this must be a cool object with
\Te~$\sim8600$~K. To reconcile the discrepancy between the slopes of
the hot solution and that of our optical spectrum, taking into account
the \Te\ value from the photometric fit, the only possibility is that
there are two stars instead of one. Hence, we believe that in the end,
both \citet{bergeron90a} and \citet{koester09} were right. In other
words, the system is an unresolved double-degenerate binary composed
of a magnetic DA star and a cooler DC white dwarf, a system analogous
to G62-46 \citep{bergeron93}. In order to get a more accurate measure
of the atmospheric parameters for 0239+109, we use a procedure
identical to the one employed for the DA+DB binary systems but here we
assume a DA and a DC component. Figure~\ref{fg:fitDADC_0239} shows our
best fit to the composite spectrum and our results are very similar to
those obtained by \citet{bergeron90a}. As a final check, we plot in
Figure~\ref{fg:plotDADC_0239} the model flux for the combined DA+DC
spectroscopic solution and compare the slope with that of our optical
spectrum of 0239+109 and see that the agreement is nearly perfect. We
note, however, that despite the excellent agreement between our
solution and the data, the magnetic nature of the DA component means
that our measured values of \Te\ and \logg\ for this star should be
considered approximate.

\begin{figure}[!t]
\includegraphics[scale=0.50,bb=70 207 592 579]{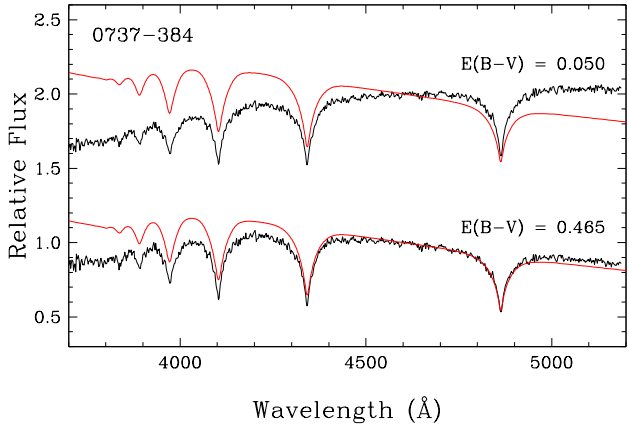}
\figcaption[f23.eps]{Comparison of the observed spectrum (black) with
  the synthetic spectrum (red) corresponding to the hot solution
  presented in Figure~\ref{fg:DD_panel} for 0737$-$384. The observed
  spectra have been corrected for interstellar reddening with values
  of $E(B-V)$ as indicated in the figure. Both the observed and
  synthetic spectra are normalized at 4600 \AA\ and offset vertically
  for clarity. \label{fg:0737}}
\end{figure}

\subsubsection{0737$-$384 (NGC 2451-6)}

This white dwarf is a supposed member of the open cluster NGC~2451 as
first reported by \citet{koester85}. However, this claim was later
challenged by \citet{platais01} based on the proper motion of the star
as compared to the proper motion of the cluster. The analysis of
\citet{koester85} yielded \Te\ = 31,000~$\pm$~3000 K, which would seem
to concur with our determination assuming the hot solution is indeed
the correct one. Unfortunately, the slope of the hot solution does not
match the slope of our observed spectrum, which suggests a much cooler
temperature for this star. How can this clear discrepancy between the
slopes of our observed spectrum and that of the hot solution be
explained? First, we verified our observing logs to see if our
observations were obtained at high airmass. This could lead to a
significant loss of flux in the blue portion of the spectrum, due to
atmospheric extinction, if the instrument is not properly rotated to
match the parallactic angle. As it turns out, 0737$-$384 was observed
at an airmass of $z$~=~1.08 so any such effects would be minimal and
we discount that possibility. Next, we attempted to fit 0737$-$384 as
a DA+DC system, but the slope of the combined model does not match the
slope of the observed spectrum, as was the case in the previous
section with 0239+109, thus a different solution is required to
explain the discrepancy. In the analysis of \citet{koester85}, a value
of $E(B-V)$~=~0.05 is adopted for the reddening toward NGC~2451 based
on several earlier studies (see references therein). Hence, we
speculated that our spectrum might actually be suffering from
interstellar reddening. We used the above value of interstellar
reddening to correct our observed spectrum, following the prescription
of \citet{seaton79}, but the effect is negligible as we see in the top
panel of Figure~\ref{fg:0737}. However, when querying the NASA/IPAC
Infrared Science Archive's Dust Extinction
maps\footnote{http://irsa.ipac.caltech.edu/applications/DUST/, derived
  using the data and techniques from \citet{schlegel98}.} in the
direction of NGC~2451, we get a value of $E(B-V)$~=~0.684. This is
significantly larger than the previously determined
values. Alternatively, if we enter the coordinates for 0737$-$384, we
obtain $E(B-V)$~=~0.465. This would seem to suggest that the
extinction in the direction fo the cluster is quite patchy. For our
purposes, we adopt the value of $E(B-V)$~=~0.465. When we apply this
new value of reddening to our data we see, in the bottom of
Figure~\ref{fg:0737}, that the slopes are now in much better
agreement. We therefore conclude that the hot solution for 0737$-$384
is indeed the correct one and the discrepancy between the slopes of
the observed spectrum and the spectroscopic solution are due to
interstellar reddening. The spectroscopic fit to the corrected
spectrum yields \Te~=~30,270 K and \logg~=~7.83.

\subsubsection{0927$-$173 (LP 787-49)}

\begin{figure}[!t]
\includegraphics[scale=0.45,bb=20 202 592 604]{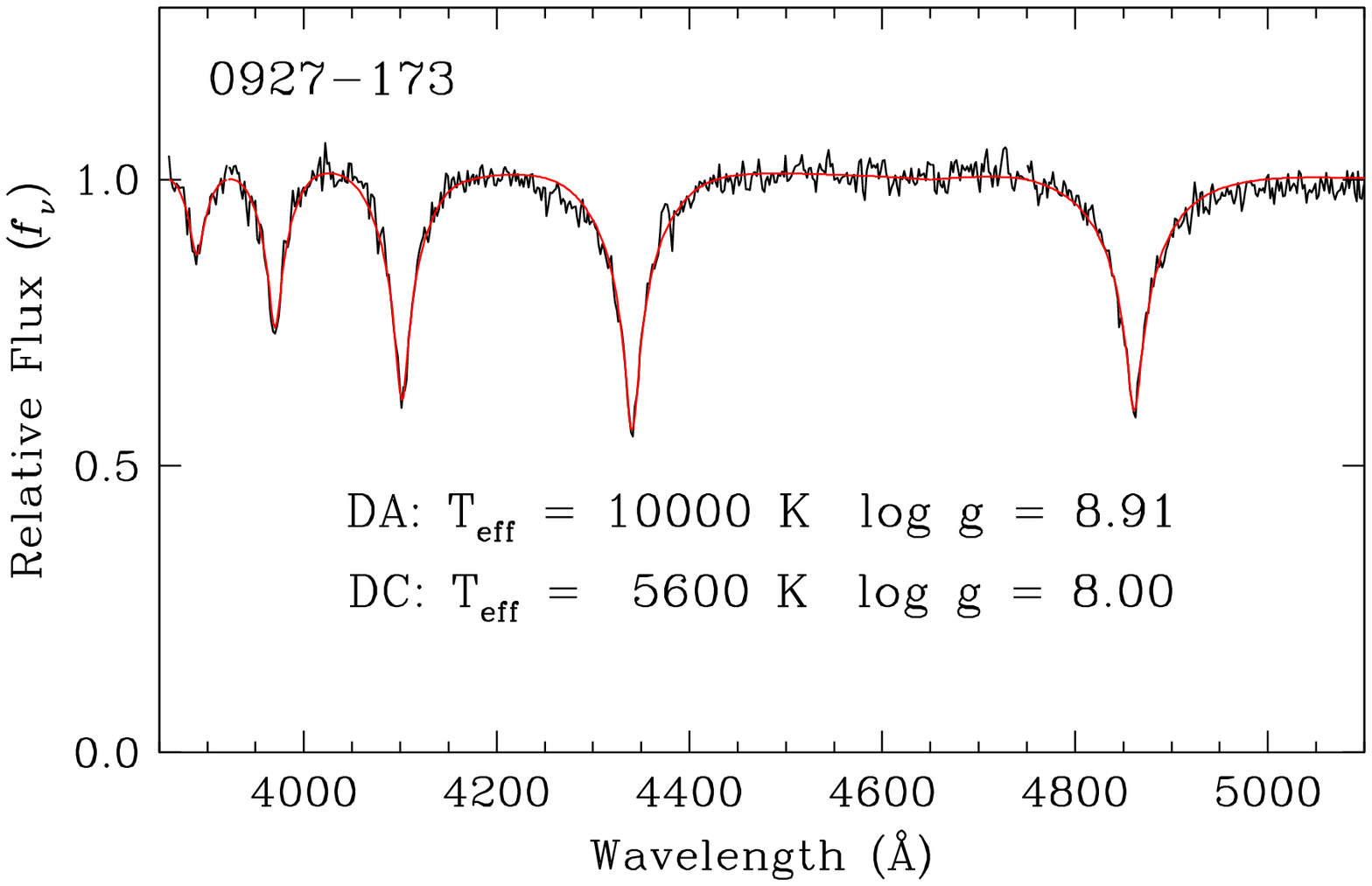}
\figcaption[f24.eps]{Model fit (red) to the optical spectrum (black)
  for the DA+DC binary 0927$-$173. The atmospheric parameters
  corresponding to the spectroscopic solution are given in the
  figure. Both the observed and theoretical spectra are normalized to
  a continuum set to unity. \label{fg:fitDADC_0927}}
\includegraphics[scale=0.45,bb=18 219 592 668]{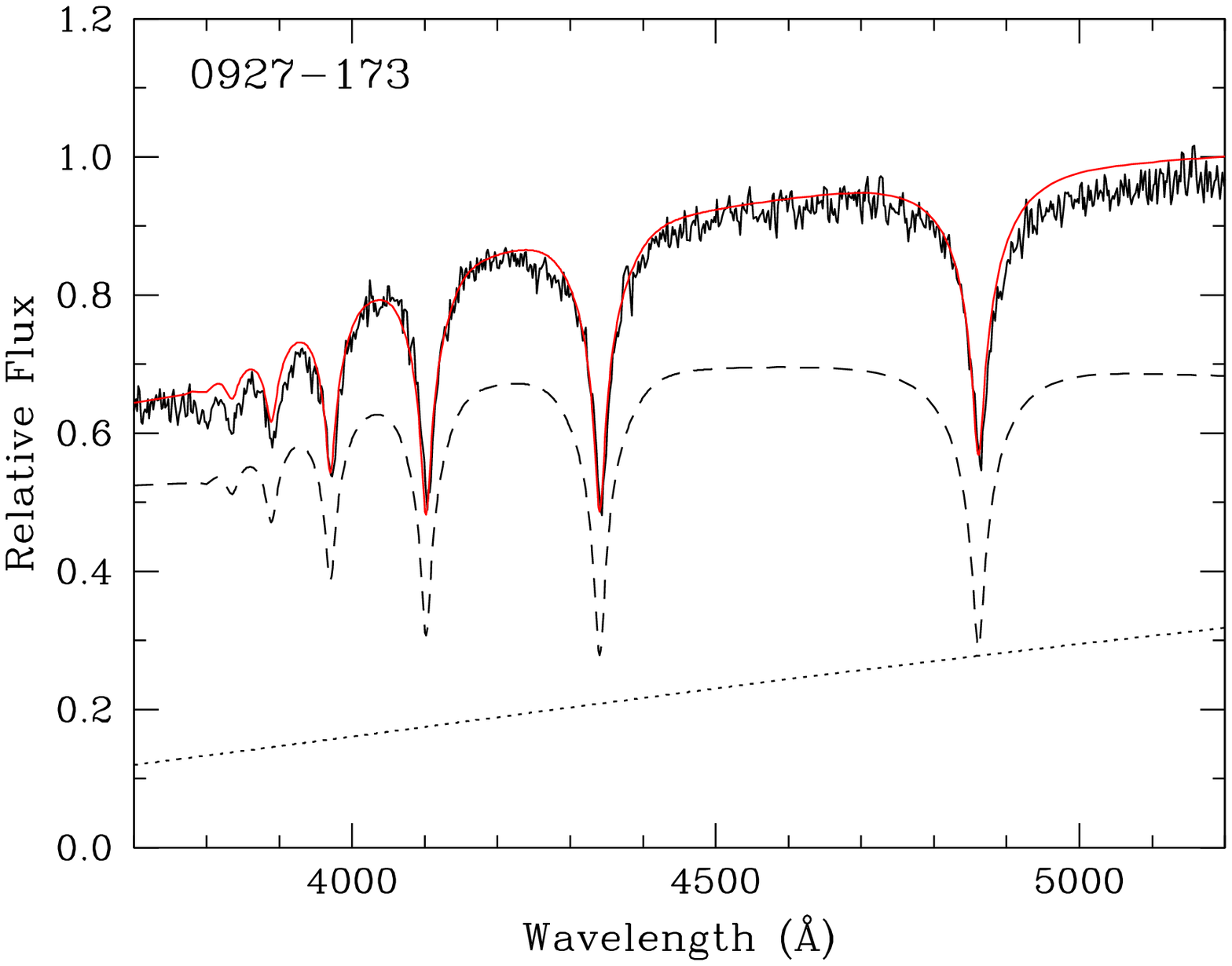}
\figcaption[f25.eps]{Relative energy distributions for our best
  composite DA+DC fit displayed in Figure~\ref{fg:fitDADC_0927}. The
  dashed line represents the contribution of the DA component while
  the dotted line represents the contribution of the DC component,
  both properly weighted by their respective radius. The red line
  corresponds to the total monochromatic flux of the composite system
  surperimposed on our spectrum of 0927$-$173 (black line), which has
  been scaled to the flux of the composite model at 4600
  \AA. \label{fg:plotDADC_0927}}
\end{figure}

\begin{table*}[!t]\scriptsize
\caption{Atmospheric Parameters of DA White Dwarfs from MS99}
\begin{center}
\begin{tabular}{@{}lllrccrrc@{}}
\hline
\hline
\noalign{\smallskip}
WD & \multicolumn{1}{c}{Name} & \multicolumn{1}{c}{ST} & \multicolumn{1}{c}{\Te} & \logg & $M$/\msun & \multicolumn{1}{c}{\mv} & \multicolumn{1}{c}{$D$} & Notes \\
   &                          &                        & \multicolumn{1}{c}{(K)} &       &           &                         & \multicolumn{1}{c}{(pc)} & \\
\noalign{\smallskip}
\hline
\noalign{\smallskip}
0000+171      & PG 0000+172     & DA2.4  & 21,130 \,\ (325) & 8.00 (0.05) & 0.63 (0.03) & 10.68 &  108 & \\
0000$-$186    & GD 575          & DA3.3  & 15,350 \,\ (265) & 7.97 (0.05) & 0.60 (0.03) & 11.19 &  105 & \\
0001+433      & RE J0003+433    & DA1.1  & 46,850  (1244) & 9.05 (0.10) & 1.23 (0.04) & 11.24 &  118 & \\
0004+061      & PHL 670         & DA2.1  & 24,400 \,\ (386) & 8.51 (0.05) & 0.94 (0.03) & 11.23 &  100 & \\
0004+330      & GD 2            & DA1.0  & 49,980 \,\ (898) & 7.77 (0.06) & 0.59 (0.02) &  8.79 &  103 & \\
0005$-$163    & G158-132        & DA3.4  & 14,920 \,\ (252) & 7.93 (0.05) & 0.57 (0.03) & 11.18 &  104 & 1 \\
0008+424      & LP 192-41       & DA7.0  & 7200   \,\ (107) & 8.12 (0.07) & 0.66 (0.05) & 13.64 &   21 &  \\
0009+501      & G217-37         & DA7.6  & 6620   \,\ (103) & 8.40 (0.09) & 0.85 (0.06) & 14.42 &   10 & 1 \\
0009$-$058    & G158-39         & DA4.8  & 10,560 \,\ (156) & 8.22 (0.06) & 0.74 (0.04) & 12.34 &   54 & \\
0010+280      & PG 0010+281     & DA1.9  & 27,220 \,\ (407) & 7.87 (0.05) & 0.57 (0.02) &  9.99 &  152 & \\
\noalign{\smallskip}
\hline
\noalign{\smallskip}
\multicolumn{9}{@{}p{12.75cm}@{}}{{\bf Notes.}
(1) Photometrically constant;
(2) DA+dM binary;
(3) ZZ Ceti;
(4) Balmer line problem;
(5) based on non-magnetic models;
(6) DA+DC binary;
(7) DA+DB binary;
(8) Suffers from interstellar reddening. 
(Table 5 is available in its entirety in the electronic edition of the
{\it Astrophysical Journal}. A portion is shown here for guidance
regarding its form and content.)} \\
\label{tab:MS99}
\end{tabular}
\end{center}
\end{table*}

As in the previous case, our first instinct was to verify if there
might be a problem with our observations of 0927$-$173. As it turns
out, 0927$-$173 was observed at an airmass of $z$~=~1.54, which puts it
at much greater risk of suffering from atmospheric
extinction. However, other objects observed on the same night and at
higher airmass ($z \geq$~1.80) do not manifest the same problem and
the slopes of their optical spectra are in perfect agreement with
those of their spectroscopic solutions. Once again, we are confident
that there is nothing wrong with the data itself. Furthermore,
previous determinations of \Te\ for this object suggest that it is a
rather cool DA star. Indeed, \citet{kilkenny97} find \Te~=~7000~K
while \citet{zuckerman03} find \Te~=~8000~K, in both cases the
temperature is determined solely from a measurement of
$(B-V)$. Combining the photometry from \citet{kilkenny97} with the
available measurements of $JHK_{\rm s}$ from 2MASS, we performed a
photometric fit using the same procedure described in
Section~\ref{sec:0239}. The results of this fit are displayed in
the bottom panel of Figure~\ref{fg:fitphoto}. Assuming \logg~=~8.0, our fit to the
available photometry yields \Te~=~7530 K, in excellent agreement with
the previous determinations, and we conclude that the cold solution
must be the correct one. But how do we reconcile this temperature with
the poor fit to the Balmer lines with the cold solution? The most
likely explanation is that 0927$-$173 is a DA+DC binary system where
the continuum flux of the DC component dilutes the observed Balmer
lines leading to the results seen in Figure~\ref{fg:DD_panel}. In
order to fit the composite spectrum, we employ exactly the same method
as we did with the DA+DB systems analyzed in
Section~\ref{sec:DAB}. Once again, \Te\ and \logg\ for the DA
component and \Te\ for the DC component are the free parameters as we
fix \logg~=~8.0 for the DC star. The results of our fit are shown in
Figure~\ref{fg:fitDADC_0927} where we see that the fit to the
composite spectrum is nearly perfect. We obtain \Te~=~10,000~K and a
rather high surface gravity of \logg~=~8.91 for the DA component,
paired to a significantly cooler DC component with \Te~=~5600~K. More
importantly, in Figure~\ref{fg:plotDADC_0927} we compare the model
flux for the combined DA+DC system to our observed spectrum and we see
that the slopes are in excellent agreement. Consequently, we conclude
that the observed slope of 0927$-$173 can be explained by the
additional presence of an unresolved companion, a cool DC white
dwarf.

\section{GLOBAL PROPERTIES AND DISCUSSION}

\subsection{Adopted Atmospheric Parameters}

The atmospheric parameters we have adopted for all the stars in our
sample are summarized in Table~\ref{tab:MS99}. This includes all the
normal DA white dwarfs in our sample as well as all the stars we have
analyzed in detail in the preceding sections including the DAB and DAZ
stars, and the DA+DB and DA+dM binaries. We also include the nine
weakly magnetic white dwarfs from Section~\ref{sec:mag} and the three
peculiar objects from Section~\ref{sec:pec}. In all, a total of 1265
white dwarfs and their atmospheric parameters are listed in
Table~\ref{tab:MS99}. The entries in Table~\ref{tab:MS99} are ordered
by their WD numbers, and we list the values of \Te\ and \logg\ as well
as masses derived from the evolutionary models of \citet{wood95} with
thick hydrogen layers. Although the Montr\'eal group has computed
evolutionary models of their own \citep{fontaine01}, these are better
suited for the cooler end of the white dwarf cooling sequence. At
higher temperatures, the Wood and Fontaine et~al. models are quite
comparable and, in fact, the Wood models have the benefit of using
post-asymptotic giant branch (AGB) evolutionary sequences as a
starting point (M. A. Wood, 2010, private communication). Furthermore,
low-mass white dwarfs, below 0.46 \msun\ and \Te\ $<$ 50,000~K, are
likely helium core white dwarfs, and we rely instead on the
evolutionary models from \citet{althaus01} for those stars. For masses
higher than 1.3~\msun, we use the zero temperature calculations of
\citet{hamada61}. The values in parentheses represent the
uncertainties of each parameter, calculated by combining the internal
error, which is the dominant source of uncertainty, obtained from the
covariance matrix of the fitting algorithm with the external error,
obtained from multiple observations of the same object, estimated for
DA stars at 1.2\% in \Te\ and 0.038 dex in \logg\ (see LBH05 for
details). The effect on the uncertainties caused by different values
of S/N is included in the external error but is minimized by the fact
that the majority of our data have rather high S/N \citep[see Figure
12 of][]{gianninas05}. In addition, we list absolute visual magnitudes
determined using the photometric calibrations from
\citet{holberg06}. We also provide an updated spectroscopic
classification for each star, with temperature indices\footnote{See
  MS99 for a description of how the temperature index is calculated.}
based on our determinations of \Te.

By coupling the absolute magnitudes we have determined with the
apparent magnitudes available for each star in the literature, we
computed {\it spectroscopic distances} that we also list in
Table~\ref{tab:MS99}.  However, these determinations come with the
caveat that many of these apparent magnitudes are photographic or,
when $V$ was not available, we have substituted a $B$ magnitude, or a
$g$ magnitude from the SDSS, or even a Str\"omgren $y$
magnitude. Consequently, our distance estimates should be viewed with
caution. Additionally, we applied an extinction correction of
$A_V$=1.468\footnote{Taken from
  http://irsa.ipac.caltech.edu/applications/DUST/, see Section 4.7.2.}
when determining the distance to 0737$-$384. We note that our distance
of 142 pc for this star also places it outside NGC~2451
\citep[$\sim$180 pc;][]{platais01}.

Since there is a growing interest in identifying white dwarfs in the
local neighborhood, in particular those within 20 pc
\citep{holberg08}, we note that there is no strong evidence for new
white dwarfs in our sample within that distance, with the exception
perhaps of 0213+396 (16 pc). The distance to 1503$-$070 of 7 pc is
obviously underestimated since we fitted this star with non-magnetic
models (see Figure \ref{fg:fitsMAG}); the measured trigonometric
parallax actually suggests a distance of $\sim25$ pc. We also have
several objects with distances within 20 pc that are too cool
(\Te\ $<$ 6500~K) to derive meaningful atmospheric parameters with the
spectroscopic technique, in particular \logg. One must also remember
that the spectroscopic distances of cool DA stars located in the
temperature range where the high-\logg\ problem is encountered are
clearly underestimated since their luminosity is also underestimated
(smaller radii); this includes 0213+396, mentioned above, at
\Te~=~9370~K.

Finally, special notes have been added in Table~\ref{tab:MS99} to
designate objects that are known ZZ Ceti pulsators as well as stars
that are photometrically constant. We note that all the stars analyzed
in \citet{gianninas10} are also listed in Table~\ref{tab:MS99}. This
includes both the DAO and DA+BP stars, with a note for all stars that
exhibit the Balmer line problem. For the DA white dwarfs that have
been determined to belong to DA+DB, DA+DC, and DA+dM systems, we list
only the parameters of the DA component. Finally, we include a note
for the weakly magnetic stars that the listed parameters were
determined with non-magnetic models.

\subsubsection{Erratum: 0102$-$185}

One of the reported DA+BP stars from \citet{gianninas10}, 0102$-$185,
is actually a much cooler DA white dwarf with \Te\ = 23,410~K and
\logg\ = 7.78, in contrast to the parameters previously reported. An
error during data reduction associated the wrong spectrum to this
star. The spectrum that was analyzed as 0102$-$185 is actually a
spectrum of 2211$-$495 from the same observing run. This reduces the
number of DA+BP stars from 18 to 17.

\subsection{Comparison with SPY}

In establishing the reliability of the results of large surveys such
as the one presented here, we believe it is both interesting and
important to compare our results with other studies that include many
of the same objects. As mentioned earlier, recent analyses of large
numbers of white dwarfs were presented both by LBH05 and
\citet{limoges10}, and although virtually all of the white dwarfs in
those studies are included here, we are also using the exact same
observations and analysis techniques making any meaningful comparison
a rather futile endeavor. Instead, what we require is an independent
analysis, which is exactly what the SPY sample analyzed by
\citet{koester09} represents. The DA white dwarfs in the SPY sample
had initially been analyzed in \citet{voss06}, however
\citet{koester09} have re-analyzed the same stars with a new
generation of models presented in \citet{koester10}. We must stress
that these models did not yet include the improved Stark profiles of
TB09. Consequently, as previously discussed, we will be using the
results obtained with the \citet{lemke97} broadening profiles in order
to compare with the SPY sample. More importantly, the white dwarfs
from SPY have been observed and analyzed with models and techniques
completely independent from our own and thus represent a truly
distinct data set.

The sample presented in \citet{koester09} consists of a total of 615
DA white dwarfs of which 362 are in common with our sample. The reason
that there is not a greater overlap is two fold. First, many of the
stars selected for the SPY project were taken from the Hamburg-ESO
\citep{christlieb01} and Hamburg-Schmidt \citep{homeier98} surveys and
these stars were not yet listed in MS99. Second, as we mentioned
earlier, many of the white dwarfs for which we were unable to obtain
spectra were situated in the southern hemisphere, and this also
contributed to the smaller overlap. In order to be sure we are
comparing all stars on an equal footing, we choose to compare only
single DA white dwarfs; any known double-degenerate systems or
magnetic white dwarfs are excluded. Furthermore, we also omit stars
with \Te\ $>$ 50,000~K and \Te\ $<$ 8000~K for the same reasons
outlined in \citet{koester09}. Namely, the models used by
\citet{koester09} do not include effects due to departures from LTE at
higher temperatures, and the spectra become less sensitive to changes
in \logg\ at lower temperatures.  These last two criteria reduce the
number of viable white dwarfs to 330.

\begin{figure}[!t]
\includegraphics[scale=0.50,bb=20 42 592 749]{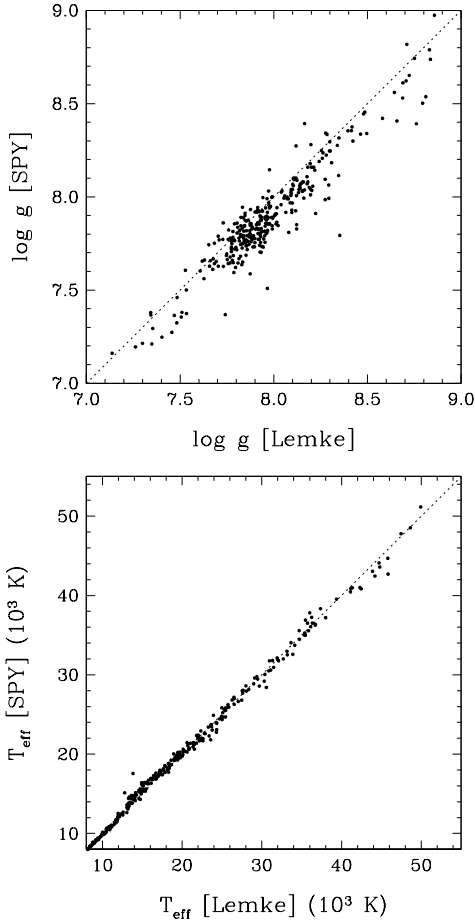}
\figcaption[f26.eps]{Comparison of \logg\ (top) and \Te\ (bottom)
  values between this work using the \citet{lemke97} Stark profiles
  and the results of \citet{koester09} for 330 DA white dwarfs common
  to both samples. The dotted lines represent the 1:1
  correlation. \label{fg:SPY}}
\end{figure}

The results of our comparison are presented in Figure~\ref{fg:SPY}. We
see that with the exception of a couple of outliers near \Te\ $\sim$
14,000~K, the agreement between our \Te\ values and those from SPY is
excellent. The two outliers (0318$-$021 and 0937$-$103) are stars
where both the cold and hot seeds lead to the same spectroscopic
solution (see LBH05 for a discussion about the hot and cold
solutions). Clearly, this is not the case in the SPY analysis of the
same objects as they derive larger values for \Te\ in both cases. In
contrast, the \logg\ determinations show a clear trend where our
measurements tend toward higher values than those from
SPY. \citet{koester09} had noted this problem (see their Figure~3) and
discussed several possibilities in an attempt to understand this
discrepancy. In particular, they point out that differences between
the ``Bergeron'' and ``Koester'' models could be the cause. Indeed,
LBH05 showed a comparison between their results and those of several
other studies. Three of those analyses used Koester models and all
three seemed to exhibit the same trend we see here. Another
possibility is that the apparent shift in \logg\ could be caused by
differences in the fitting procedures. Finally, \citet{koester09}
speculate that the very nature of the SPY data could be the origin of
the problem. The SPY spectra were obtained with the UV-Visual Echelle
Spectrograph at the Very Large Telescope, and the stitching together
of the various orders might have introduced artifacts in the
spectra. For the complete discussion, we refer the reader to
\citet{koester09}.

\begin{figure}[!t]
\includegraphics[scale=0.50,bb=20 42 592 749]{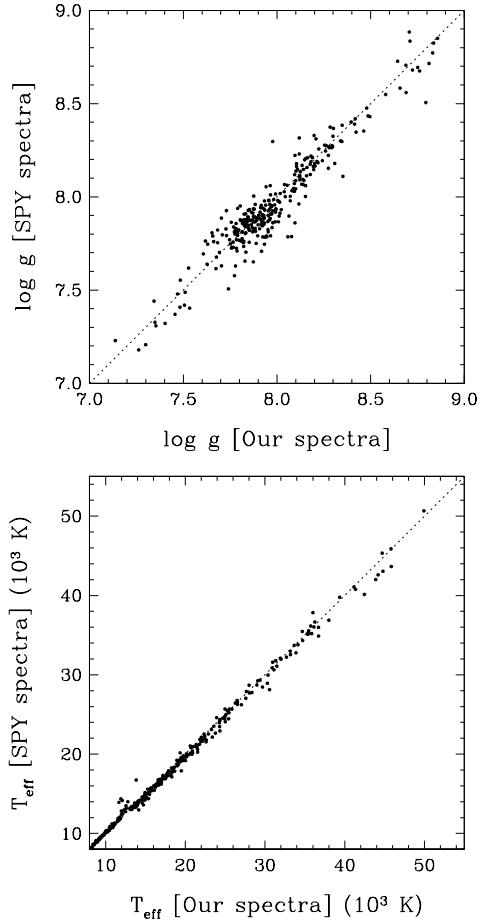}
\figcaption[f27.eps]{Comparison of \logg\ (top) and \Te\ (bottom)
  values between this work using the \citet{lemke97} Stark profiles
  and the results of our fits to the SPY spectra using the same
  models. The dotted lines represent the 1:1
  correlation. \label{fg:SPY2}}
\end{figure}

In an effort to better understand this shift in \logg\ values, we
decided to independently fit the SPY spectra, which were kindly
provided to us by D. Koester, using our models and fitting
procedures. In Figure~\ref{fg:SPY2}, we show the comparison between
these new values derived from the SPY spectra and those from our
spectra, all analyzed with the \citet{lemke97} Stark profiles. First,
we note that the agreement between the \Te\ determinations (bottom
panel) remains very good. Second, and more importantly, we do not
observe any systematic shift in the \logg\ values (top panel). There
is more scatter in the \logg\ plot but the overall agreement is
markedly better than was seen in Figure~\ref{fg:SPY}. Since we obtain
a satisfactory agreement between the atmospheric parameters measured
from both sets of spectra, we must conclude that there is no problem,
a priori, with the SPY spectra themselves. Differences between our
results and those of \citet{koester09} must necessarily arise from
differences in the models or the fitting techniques used. Nonetheless,
the fact that the \Te\ determinations agree so well in both of the
previous comparisons is encouraging. Furthermore, when using an
independent data set and relying on our own theoretical framework, we
succeed in getting both \Te\ and \logg\ values that match rather well.

\begin{figure}[!t]
\includegraphics[scale=0.50,bb=20 42 592 749]{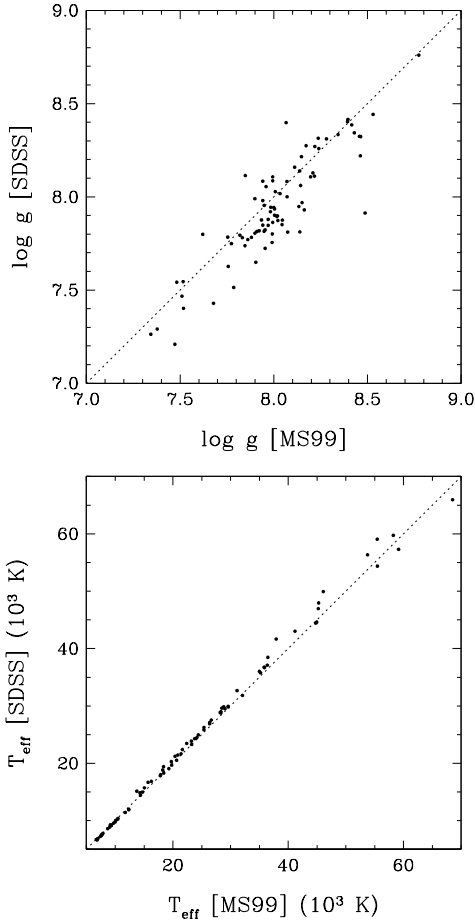}
\figcaption[f28.eps]{Comparison of \logg\ (top) and \Te\ (bottom)
  values derived from our observations and from fits to SDSS spectra
  of 83 DA white dwarfs common to both samples. The dotted lines
  represent the 1:1 correlation. \label{fg:SDSS}}
\end{figure}

\subsection{Comparison with SDSS}

Another independent source of spectra is the SDSS. Despite the
thousands of new white dwarfs discovered in the SDSS, very few are in
common with our sample. As we showed in Figure~\ref{fg:histoV}, the
much fainter sample of white dwarfs observed by the SDSS is the reason
that the two samples are so mutually exclusive. Nonetheless, we have
retained 83 DA white dwarfs in common with both samples.\footnote{We
  omit DA+dM binaries, DAO stars, and magnetic white dwarfs from this
  comparison.} We obtained the SDSS spectra for these stars from the
SDSS SkyServer\footnote{http://cas.sdss.org/dr7/en/} and proceeded to
fit them with the new models that include the TB09 Stark broadening
profiles. The comparison between our \Te\ and \logg\ determinations
and those obtained from the SDSS spectra is presented in
Figure~\ref{fg:SDSS}. As in the case of the comparison with SPY, we
note a very good agreement between both sets of \Te\ values (bottom
panel). However, we also note a second similarity with our SPY
comparison as the \logg\ values also seem to be systematically shifted
toward lower values from those determined from the SDSS spectra. This
concurs with the results of \citet{tremblay11a} who explained that any
differences must be due to the different spectra employed as they are
being analyzed within the same theoretical framework and the same
fitting method. \citet{tremblay11a} conclude that issues with the
reduction of the SDSS data still remain and are the reason for the
observed shift.

\begin{figure}[!t]
\includegraphics[scale=0.60,bb=70 207 592 604]{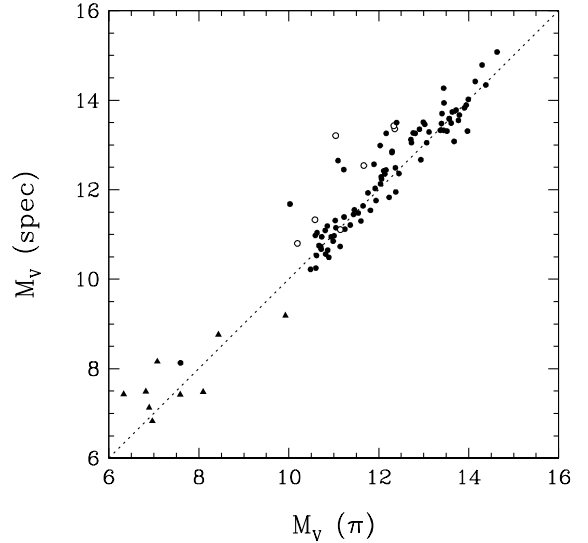}
\figcaption[f29.eps]{Comparison of absolute magnitudes derived from
  parallax measurements with the absolute magnitudes derived from our
  spectroscopic determinations of \Te\ and \logg\ for 101 white
  dwarfs. The open circles represent known or suspected
  double-degenerate binary systems while the triangles correspond to
  central stars of planetary nebulae. The dotted lines represent the
  1:1 correlation. \label{fg:compmv}}
\end{figure}

\subsection{Comparison with Parallax Measurements}

Yet another way of testing the reliability of our results is to
compare our derived values of absolute magnitudes with those obtained
from parallax measurements. The added incentive here is that the
absolute magnitudes obtained from parallaxes are completely model
independent. We have 92 DA white dwarfs in our sample with available
parallax measurements from either the Yale Parallax Catalogue
\citep{altena95}, or {\it Hipparcos} \citep{perryman97} and 9 central
stars of planetary nebulae with parallaxes taken from
\citet{harris07}. We present in Figure~\ref{fg:compmv} the comparison
between the $M_{V}$ values from parallax measurements and those
obtained through our spectroscopic analysis. We can see that the
overall agreement is quite good and this reinforces our confidence
that our results are accurate.

There are a number of obvious outliers in this figure that can be
easily explained as unresolved double degenerates (open circles), the
most important of which is L870-2 (0135$-$052), discovered by
\citet{saffer88}. But this is not always the case. For instance, we
find $M_V({\rm spec})=11.68$ for Ross 548 (0133$-$016, ZZ Ceti
itself), for a corresponding distance of 31 pc, while the distance
obtained from the trigonometric parallax measurement from the Yale
Parallax Catalogue is more than twice as large at 67 pc. We cannot
find an easy explanation for this discrepancy unless Ross 548 is also
an unresolved degenerate binary. As well, the absolute magnitudes of
cool DA stars found in the temperature range where the high-\logg\
problem is found are likely to be overestimated; this accounts for the
clump of objects away from the 1:1 correlation in the range
$12.5<M_V<13.5$.

We must finally mention that the observed agreement is very good
within the parallax uncertainties for {\it both} model grids. As TB09
pointed out, despite the fact that the new models yield higher values
of \Te\ and \logg\ (i.e., smaller radii), the two effects nearly
cancel each other out and the predicted luminosities (or $M_{V}$)
remain largely unchanged.

\begin{figure*}[!t]
\includegraphics[scale=0.65,angle=-90,bb=72 -9 521 734]{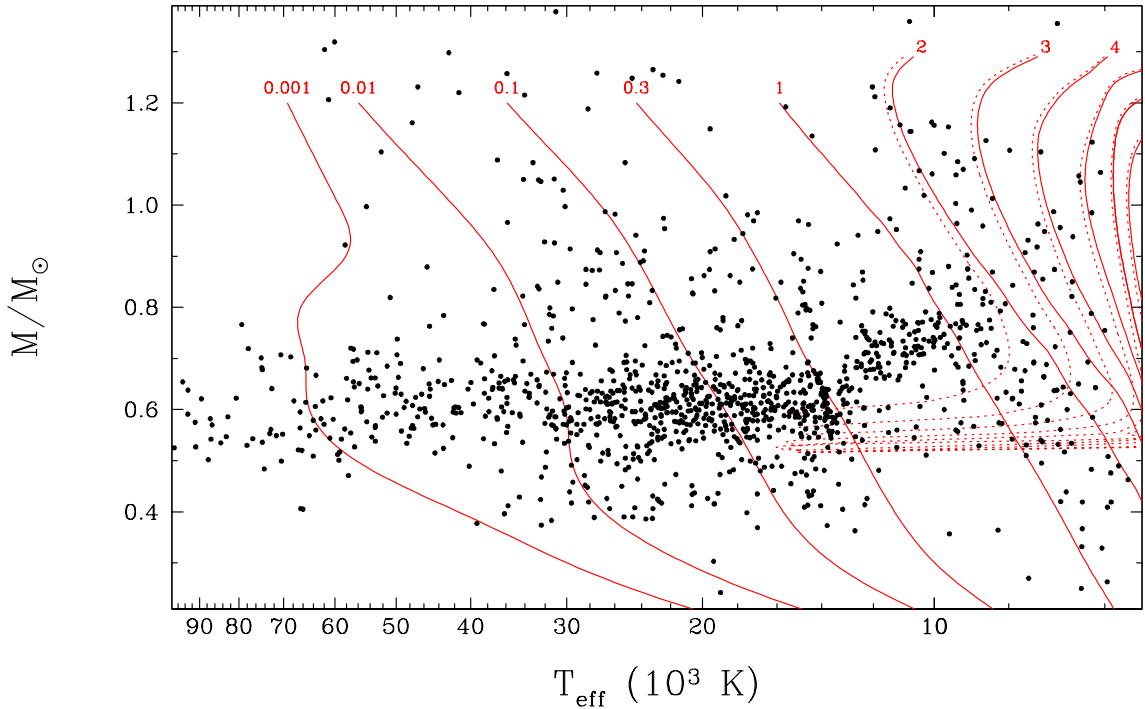}
\figcaption[f30.eps]{Mass distribution as a function of \Te\ for all
  the stars listed in Table~\ref{tab:MS99}. The solid red lines
  represent isochrones that take into account only the white dwarf
  cooling time whereas the dotted red lines include also the
  main-sequence lifetime. Each isochrone is labeled by its age in
  Gyr. Isochrones for $\tau <$ 1 Gyr are taken from \citet{wood95}
  whereas those for $\tau \geq$ 1 Gyr are taken from
  \citet{fontaine01}. \label{fg:mass_vs_t}}
\end{figure*}

\subsection{Mass Distribution}\label{sec:mass}

We display in Figure~\ref{fg:mass_vs_t}, the mass distribution as a
function of \Te\ for all the stars listed in Table~\ref{tab:MS99}. As
previously noted, the bulk of the distribution is centered around 0.6
\msun\ for \Te\ $>$ 13,000~K. Below this temperature, the well-known
``high-\logg\ problem'' manifests itself. Several solutions to this
problem have been explored over the years. First, \citet{bergeron90b}
showed how helium, spectroscopically invisible below 13,000~K, might
explain this problem. If analyzed with pure hydrogen models, DA white
dwarfs containing helium would appear to have higher \logg\ values
because the presence of helium increases the pressure in the
atmosphere, which is analogous to increasing the surface
gravity. Recently, \citet{tremblay10} attempted to detect the presence
of the \heif\ line in six DA white dwarfs using high-resolution
spectroscopy. They found no evidence of the helium line and concluded
that helium was not the solution. The reason helium was thought to be
the key ingredient in explaining the high-\logg\ problem is because it
was expected that convective mixing would bring it to the surface. In
addition, properly calibrating the mixing-length theory used to
describe convection was deemed crucial. \citet{bergeron95} addressed
this issue but even with the mixing length properly calibrated, the
high-\logg\ problem remained. The advent of the TB09 profiles was also
thought to represent a possible element of the solution. However, as
we have seen, the new profiles succeed in ``straightening out'' the
mass distribution, but do not seem to contribute in fixing the masses
at lower \Te. One of the few remaining avenues of research is the
computation of full three-dimensional hydrodynamic models of
convection as a replacement to the mixing-length theory that has
always been used, but ultimately constitutes an approximation. The
preliminary results from such computations, as presented by
\citet{tremblay11b}, show that these models produce corrections to the
values of \logg\ that are of the correct magnitude and direction when
compared to shifts deduced from the analysis of large samples of DA
white dwarfs. Hence, it seems that the solution to the high-\logg\
problem has finally been nailed down.

At the hotter end of the white dwarf sequence, we note in
Figure~\ref{fg:mass_vs_t} that the distribution remains essentially
continuous around 0.6 \msun\ unlike the previous determination shown
in Figure~3 of \citet{gianninas09} where there is a noticeable ``dip''
toward lower masses at higher \Te. Our new result is a consequence of
the new determinations of \Te\ and \logg\ for the DAO stars from
\citet{gianninas10}. As it was demonstrated, the use of the CNO models
to solve the Balmer line problem also resulted in higher masses for
the DAO stars, consistent with the remainder of the hydrogen-rich
white dwarf population.

\begin{figure}[!t]
\includegraphics[scale=0.50,bb=20 117 592 604]{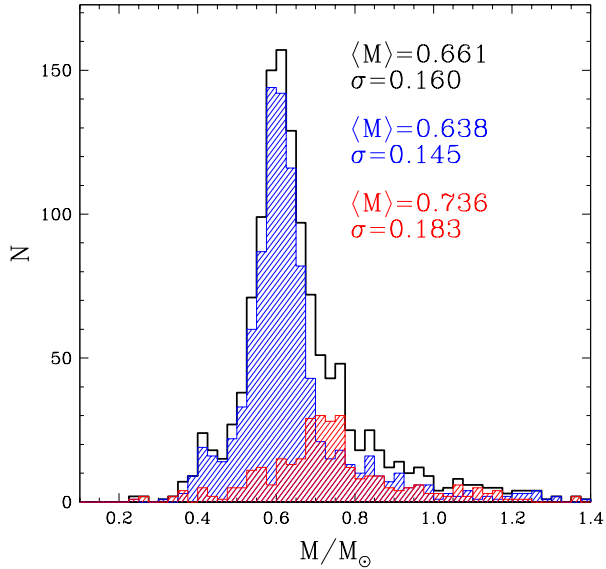}
\figcaption[f31.eps]{Mass distribution for all the stars listed in
  Table~\ref{tab:MS99} (solid line histogram). Also shown are the mass
  distributions for stars with \Te\ $>$ 13,000~K (hatched blue
  histogram) and for stars with \Te\ $<$ 13,000~K (hatched red
  histogram). The mean mass and dispersion for each distribution are
  indicated in the figure in units of solar
  masses. \label{fg:mass_dist}}
\end{figure}

We show in Figure~\ref{fg:mass_dist} the histogram that represents the
mass distribution of our sample regardless of effective
temperature. Besides the full distribution, we also plot the mass
distributions for stars above and below \Te\ = 13,000~K (in blue and
red, respectively). The reason for this split is that we want to
analyze the mass distribution for stars that are not affected by the
high-\logg\ problem. The overall shape of the distribution is in good
agreement with the latest determination of the SDSS DA mass
distribution presented by \citet{tremblay11a}, the only other large
scale analysis of DA white dwarfs employing the new TB09 Stark
profiles. In both cases, the mass distribution is strongly peaked near
0.6~\msun. The other notable features are the high-mass tail and the
low-mass component that peaks near 0.45~\msun. The mean mass of the
entire sample is $\langle M\rangle = 0.661$~\msun. However, if we
consider only stars with \Te~$\geq$~13,000~K, we get a lower mean mass
of $\langle M\rangle = 0.638$~\msun. This is higher than the value of
0.613~\msun\ obtained by \citet{tremblay11a} for the SDSS DR4
sample. However, as stated in Tremblay et al., the mean mass is
sensitive to outliers and as we will discuss below, the MS99 sample
contains a number of high-mass white dwarfs that are systematically
missed by many surveys. This fact is also evidenced in Figure~18 of
\citet{tremblay11a} where we note a dearth of stars for $M >
1.0$~\msun\ and \Te\ $\geq$~20,000~K.

We note in Figure~\ref{fg:mass_vs_t} a number of rather high mass
white dwarfs. In particular, if we restrict ourselves to stars with
\Te\ $>$ 13,000~K, thus avoiding the troublesome lower temperature
regime where the high-\logg\ problem reigns, we find that there are 20
white dwarfs with a mass $M \geq$ 1.1 \msun. Of these, 11 were
detected by {\it ROSAT} \citep{fleming96}. This trend is not at all
surprising. By definition, white dwarfs with a higher mass have a
smaller radius, which, in turn, means they are intrinsically fainter
at a given \Te. As such, magnitude-limited UV-excess surveys (PG,
KUV), and many of the other surveys looking for bright blue objects
(Montr\'eal-Cambridge-Tololo, Edinburgh-Cape) would have skipped these
stars because they are fainter. On the other hand, X-ray surveys like
the one conducted with {\it ROSAT}, catalog {\it all} sources,
regardless of the brightness of the object. Consequently, since ours
is not a magnitude-limited sample, these white dwarfs make their way
into our sample by virtue of being included in MS99.

At the other end of the mass spectrum, the presence of stars below the
dotted lines in Figure~\ref{fg:mass_vs_t} is significant. Those
isochrones include the main-sequence lifetime of the white dwarf's
progenitor. This represents a cutoff of sorts in that the possible
progenitor stars would not have had sufficient time to become white
dwarfs within the lifetime of the Galaxy, nor to have cooled down to
those temperatures. However, it is clear that there exists a
non-negligible low-mass component in the mass distribution. The
presence of these stars can be explained if we consider binary star
evolution under different guises. First, several of these stars are
known to be unresolved double-degenerate binary systems that have been
discovered through searches for radial velocity variations
\citep{maxted99,maxted00}. In such cases, the measured atmospheric
parameters are misleading if we assume that only a single star is
present. A second possibility involves the merger of two white
dwarfs. Indeed, the recent results presented by \citet[][and
references therein]{kilic11} seem to suggest that there are several
very low mass ($\sim$ 0.2 \msun) white dwarfs that are in close binary
systems. If these white dwarfs merge, they could potentially form a
white dwarf with a mass near 0.4 \msun. Finally, there is the
possibility of a helium core white dwarf \citep{althaus01}. Helium
core white dwarfs are formed when stars lose a significant amount of
their envelope's mass to a companion before the progenitor reaches the
tip of the red giant branch. Regardless which of these scenarios is
the most likely to account for the population of low-mass white
dwarfs, one thing is clear: these low-mass white dwarfs represent a
separate evolutionary channel distinct from the usual post-AGB models
that explain the evolution of the vast majority of white dwarfs.

\begin{figure}[!t]
\includegraphics[scale=0.50,bb=45 132 592 654]{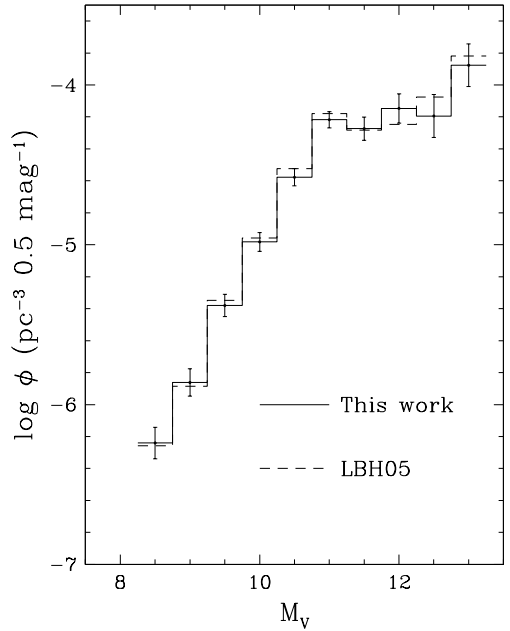}
\figcaption[f32.eps]{Luminosity function of all DA stars in the
  complete PG sample using the atmospheric parameters derived in this
  work (solid line) presented in half-magnitude bins, assuming a scale
  height for the Galaxy of $z_{0}$ = 250 pc. The dashed line
  represents the results of LBH05 shown here for
  comparison. \label{fg:PG}}
\end{figure}

\subsection{Revisiting the PG Luminosity Function}

Besides revising the atmospheric parameters of all the normal DA stars
from the complete sample with the use of the TB09 profiles, there is
also a number of white dwarfs whose atmospheric parameters are further
refined from the LBH05 analysis as they are members of certain
distinct classes of objects. In particular, there are 16 DA+dM
binaries, 10 DAO stars,\footnote{PG 1305$-$017 is included in this
  sample but, as outlined in \citet{gianninas10}, the parameters
  obtained by \citet{bergeron94} using their stratified models are
  adopted, as such, the parameters for this star remain unchanged.} 5
DA+BP stars, 7 magnetic white dwarfs and finally the DA component from
the 1115+166 DA+DB binary system. For the magnetic white dwarfs not
included in Table~\ref{tab:MS99}, we adopt the parameters listed in
Table~\ref{tab:mag}. With these up-to-date atmospheric parameters in
hand, we proceed to recompute the PG luminosity function in exactly
the same manner described in LBH05. We compare our new determination
of the PG luminosity function with that of LBH05 in
Figure~\ref{fg:PG}. We see some variations in the individual magnitude
bins with respect to the LBH05 result, but this is to be expected
considering the revised atmospheric parameters we have employed in
computing the luminosity function. We can also compare the space
densities, in other words, the total number of DA white dwarfs per
pc$^{-3}$. This is determined by integrating the area under the
luminosity function. LBH05 had determined a value\footnote{LBH05
  actually give a value of $5.0 \times 10^{-4}$~pc$^{-3}$ but this
  number is inaccurate for reasons unknown to one of the coauthors of
  both studies (P.B.)  who obtained the correct number provided here.}
of $3.07 \times 10^{-4}$~pc$^{-3}$ for $M_{V} < 12.75$. Our new
determination of the PG luminosity function yields a slightly lower
value of $2.92 \times 10^{-4}$~pc$^{-3}$ for the same magnitude range.
This represents only a 5.1\% difference with the value from LBH05.  At
first glance, one might think that the higher temperatures produced by
the new TB09 Stark profiles means that stars are intrinsically
brighter than we once thought, meaning they can potentially be further
away thus reducing the local density of white dwarfs. However, our
earlier comparison between the absolute magnitudes obtained from
spectroscopy and those determined from parallax measurements showed
that the intrinsic brightness of these white dwarfs has not changed in
an appreciable way since the stellar radii are also smaller (higher
masses). Under these circumstances, the luminosity function should
remain almost unaffected. However, closer inspection of the individual
values of \Te, \logg, and \mv\ reveals that small differences do cause
stars to jump from one magnitude bin to the next, which can ultimately
affect the determination of the space density of white dwarfs.

\begin{figure}[!t]
\includegraphics[scale=0.345,angle=-90,bb=62 36 546 759]{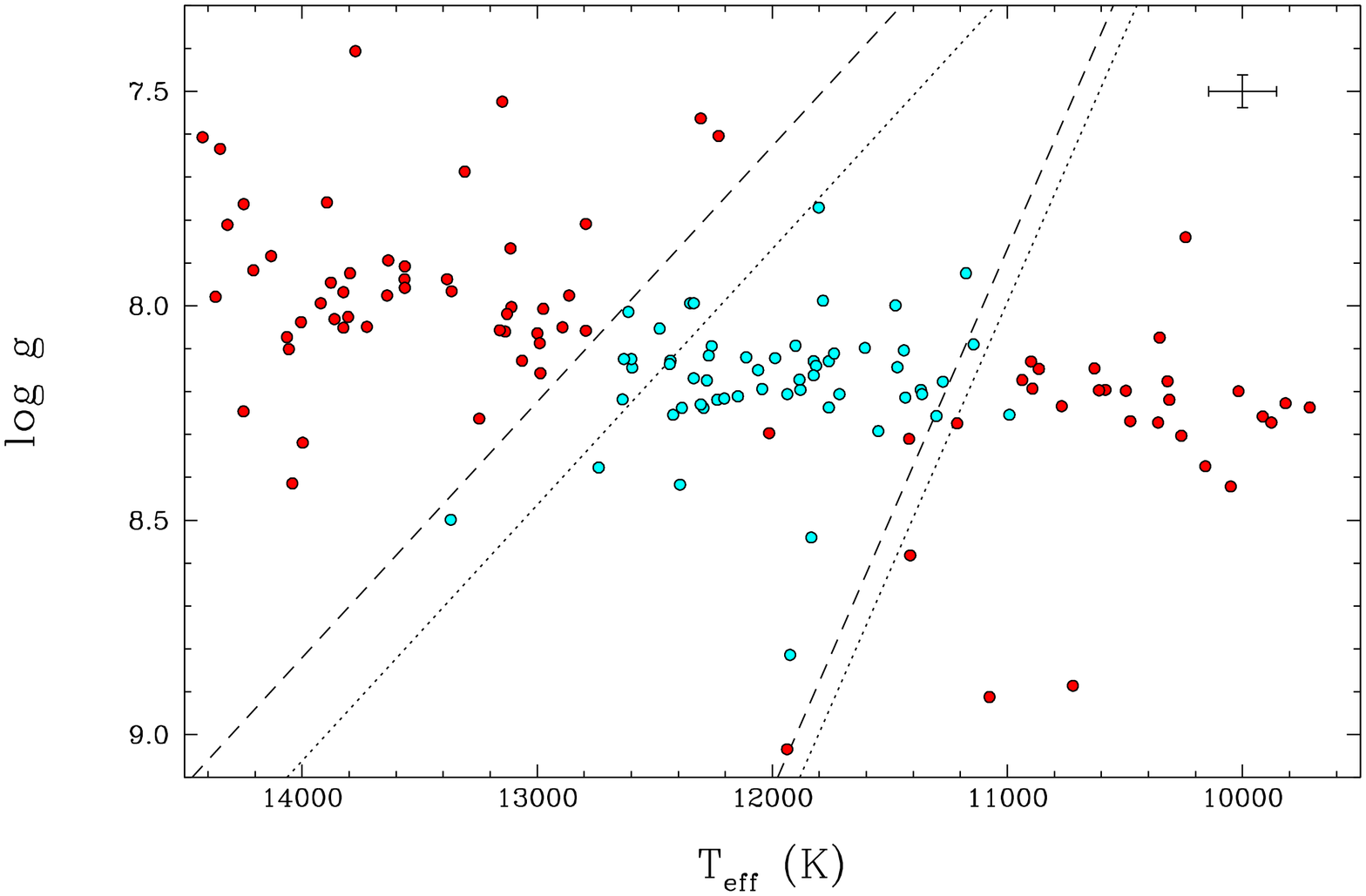}
\figcaption[f33.eps]{\Te$-$\logg\ distribution for DA white dwarfs
  with high-speed photometric measurements. The cyan circles represent
  the 56 ZZ Ceti stars in our sample and the red circles are the 145
  photometrically constant DA stars from Table~\ref{tab:MS99}. The
  dotted lines represent the empirical boundaries of the instability
  strip as established in \citet{gianninas06} whereas the dashed lines
  correspond to our new determinations. The error bars represent the
  average uncertainties of the spectroscopic method in the region of
  the ZZ Ceti instability strip. \label{fg:strip_phot}}
\includegraphics[scale=0.45,bb=20 267 592 604]{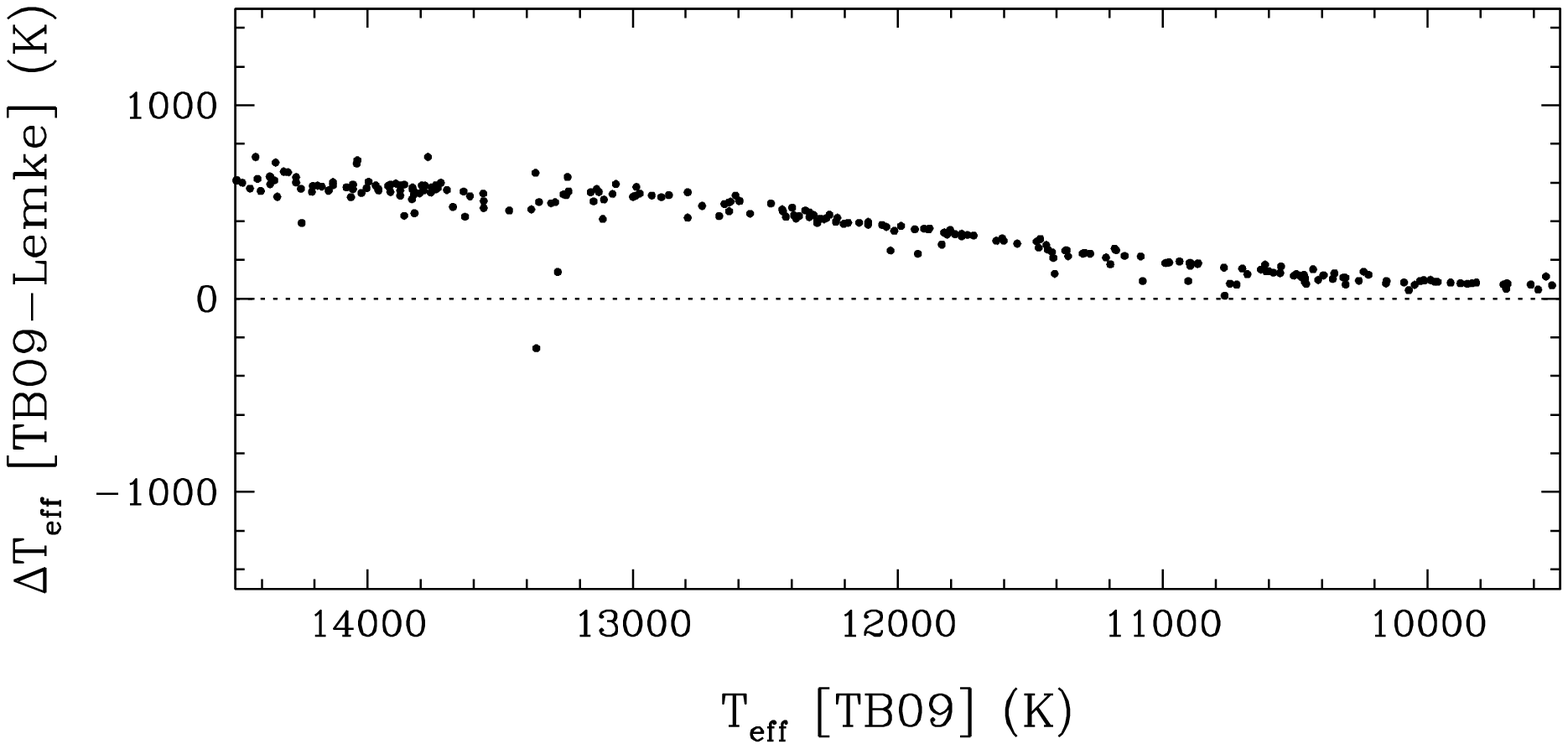}
\figcaption[f34.eps]{Differences in \Te\ determinations between the
  improved Stark profiles of TB09 and the profiles of \citet{lemke97}
  as a function of \Te\ for stars near the ZZ Ceti instability
  strip. The dotted line represents the 1:1
  correlation. \label{fg:deltgZZ}}
\end{figure}

\begin{figure}[!t]
\includegraphics[scale=0.345,angle=-90,bb=62 36 546 759]{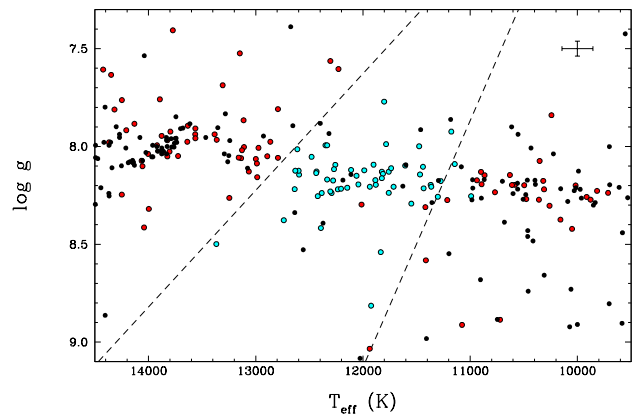}
\figcaption[f35.eps]{Same as Figure~\ref{fg:strip_phot} but for our
  entire sample of white dwarfs. The black circles represent the stars
  for which we do not have any high-speed photometric data. The dashed
  lines correspond to our new determination of the empirical
  boundaries of the ZZ Ceti instability strip. \label{fg:strip_surv}}
\end{figure}

\subsection{ZZ Ceti Instability Strip}

In Table~\ref{tab:MS99}, we list a number of stars that have been
observed in high-speed photometry in order to ascertain if they are
photometrically variable. In total, we identify 145 white dwarfs that
are photometrically constant and 56 stars that are known ZZ Ceti
pulsators. Using this photometric sample of 201 white dwarfs, we wish
to revisit the empirical determination of the boundaries of the ZZ
Ceti instability strip. In particular, we wish to see how the use of
the TB09 Stark broadening profiles has affected the location of the
instability strip as compared to our last result in
\citet{gianninas07}. We plot in Figure~\ref{fg:strip_phot} the
location of these 201 white dwarfs in the \Te$-$\logg\ plane. For the
purposes of comparison, we plot as dotted lines the empirical
boundaries of the instability strip as determined in
\citet{gianninas07}. Our new determination of these boundaries is
shown as dashed lines. First, we notice that the entire strip has
shifted by a few hundred Kelvins toward higher \Te. Second, we see
that the strip has also widened by several hundred Kelvins. These
changes are a direct result of the use of the new TB09 broadening
profiles. We plot in Figure~\ref{fg:deltgZZ} differences in \Te\
obtained by using the old \citet{lemke97} profiles versus the
calculations of TB09, focusing on the range of \Te\ where the ZZ Ceti
instability strip lies. We see that the new values of \Te\ are
systematically higher than the old values, this explains why the whole
instability strip is shifted toward higher effective
temperatures. However, we also notice that $\Delta$\Te\ increases as a
function of \Te\ from $\sim$ 200~K near \Te\ = 11,000~K, to $\sim$
600~K at \Te\ = 13,000~K. This is the reason that the instability
strip is now wider than previously established.

One aspect of the instability strip that remains unchanged is its
purity. With the exception of the two ``offending'' stars discussed
below (one pulsator and one non-variable), only pulsating white dwarfs
are to be found within the instability strip and only photometrically
constant white dwarfs without, within the uncertainties. We stress
that the significance of a pure instability strip is twofold. First,
if the strip is pure, one can predict the variability of white dwarfs
whose atmospheric parameters places them within the confines of the
instability strip. Second, if the strip is pure, then it represents an
evolutionary phase through which all hydrogen-rich white dwarfs must
pass. As a result, asteroseismological analyses of these stars can
yield important information about their internal structure, in
particular the thickness of the hydrogen layer. This in turn can be
applied to the entire population of hydrogen-rich white dwarf stars.

We also note that the blue edge seems to be quite clearly defined
whereas there is a certain level of ambiguity associated with the
location of the red edge. As far as the blue edge is concerned, the
sharp transition between non-variable to variable stars is probably a
consequence of very specific conditions being required for the driving
mechanism to ignite pulsations at the surface of the white dwarf. On
the other hand, the dying off of pulsations at the red edge is not so
well understood and it is speculated that the interaction between
convection and the pulsations is what ultimately signals the demise of
variability. It is conceivable that this process is not as clear cut
and thus the exact temperature at which a star will cease pulsating is
more ambiguous, producing a red edge whose precise location is more
difficult to nail down.

As mentioned above, there are two notable exceptions to the ``pure''
instability strip. The first of the two offending white dwarfs is
1959+059. This is a ZZ Ceti star that lies outside the instability
strip near $\sim$ 11,000~K. This star was first reported as being
variable by \citet{voss07}. However, our initial spectrum for this
object was one of the many provided to us by C. Moran (1999, private
communication). Since the discovery of its variability is more recent,
it is likely that the Moran spectrum was not averaged over several
pulsation cycles as is necessary in order to obtain a proper measure
of the atmospheric parameters for ZZ Ceti stars. Indeed,
\citet{voss07} report a period of 1350~s and an amplitude of 5.69~mma,
consistent with cooler ZZ Ceti stars that have longer periods and
larger amplitudes as compared to their hotter counterparts. These
large amplitudes signify that \Te\ can easily swing back and forth by
several hundred Kelvins resulting in the star moving in and out of the
instability strip \citep{fontaine08}. However, we finally secured a
new spectrum for 1959+059, averaged over several pulsation cycles,
during our last observing run in 2011 April. The new values do not
substantially change the position of the star in the \Te$-$\logg\
plane and it remains outside the empirical boundaries of the
instability strip. This is not altogether inconsistent with the notion
of a pure instability strip in the context of the difficulty of
precisely defining the red edge of the strip, as discussed above.

The second offending star is HS~1612+5528, the non-variable white
dwarf lying in the heart of the instability strip. This star was
reported as ``not observed to vary'' by \citet{voss2006}.
\citet{gianninas09} independently observed HS~1612+5528 and also
detected no variations down to a limit of 0.2\%. However, there are ZZ
Ceti stars with amplitudes as low as 0.05\%. It is possible that in
both cases the star was observed during a period of destructive
interference. Alternatively, HS~1612+5528 could represent the first ZZ
Ceti star whose pulsations are hidden from us due to geometric
considerations. In other words, since we cannot resolve the surface of
the star, the excited pulsation mode, or modes, might produce no
detectable variations if the combined effect of the alternating hot
and cold areas is nil. This would likely require a chance alignment
where, for example, the rotation axis of the star is perpendicular to
our line of sight. Whatever the case may be, we are planning to secure
new high-speed photometry for this star during the next trimester to
settle, once and for all, the issue of its photometric status.

Finally, in Figure~\ref{fg:strip_surv} we add all the remaining stars
from our survey whose photometric status is unknown. We can identify a
total of 14 ZZ Ceti candidates that lie within the empirical
boundaries of the instability strip. However, as \citet{gianninas06}
demonstrated, it is important and worthwhile to observe all white
dwarfs that are near either edge of the instability strip as well. It
could be argued that adding a few more ZZ Ceti stars to the mix is not
especially important since hundreds have now been identified through
the SDSS and follow-up work. However, the SDSS variables are, for the
most part, a few magnitudes fainter than the white dwarfs shown
here. Besides the discovery light curves, it is hard to conceive of
meaningful asteroseismological studies of the ZZ Ceti stars from SDSS.
In contrast, these brighter candidates could be studied on small to
medium-sized telescopes equipped with proper instrumentation.

The most intriguing of all the candidates is certainly 1659+662
(GD~518). With \logg~=~9.08, it would easily become the most massive
ZZ Ceti star ever discovered, surpassing even 1236$-$495 (BPM~37093,
LTT~4816) the current heavyweight of the group. High-speed photometric
measurements of this object are being planned as well.

\section{CONCLUSION}

We have presented the results of our spectroscopic survey of
hydrogen-rich white dwarfs from MS99. First, we discovered that many
stars listed in the catalog are still erroneously classified as white
dwarfs. In particular, we list a total of 68 stars that are
misclassified, 27 of which need to be reclassified as the result of
our observations. We then examined the spectroscopic content of our
survey. The majority of stars in our sample are simple DA white dwarfs
but a number of them show more than just the characteristic Balmer
lines of hydrogen. Indeed, our sample includes both DAB and DAZ white
dwarfs as well as DA+dM binaries and magnetic white dwarfs. In
addition, there are several white dwarfs that are cool enough that no
Balmer lines are present, at least not within the spectral range
covered by our observations. With the exclusion of the magnetic stars
and the cool, featureless white dwarfs, we have analyzed the spectra
of all the white dwarfs in our sample. In order to do so, and with the
goal of obtaining the most accurate atmospheric parameters possible,
we have endeavored to use the most modern, up-to-date, and appropriate
model atmospheres in our analyses of these white dwarfs.

First, for the analysis of all the hydrogen-rich white dwarfs we have
used models that include the new Stark broadening profiles from
TB09. The analysis of the hot DAO and DA+BP stars was presented in
\citet{gianninas10} and featured models to which CNO, at solar
abundances, was added in order to overcome the Balmer-line
problem. The results of that analysis were incorporated
here. Furthermore, we used updated helium atmosphere models to analyze
several DA+DB double-degenerate binary systems and we developed a
technique that uses M dwarf templates to correct for the contamination
in the composite spectra of DA+dM binary systems. We also analyzed
several DAZ stars by including calcium in the calculation of the
synthetic spectrum only. Two of these are newly identified DAZ stars.

In switching from the older Stark broadening profiles of
\citet{lemke97} to those of TB09, we also explored how our atmospheric
parameters have changed as a result. We observed that the values of
\Te\ are shifted toward higher temperatures while the values of \logg\
show a more obvious and significant trend toward higher values. With
that said, we also saw how the new profiles improved the overall shape
of the mass distribution as a function of \Te. The distribution now
remains largely constant around 0.60 \msun\ until $\sim$ 13,000~K
where the well-known high-\logg\ problem takes over. The new
three-dimensional hydrodynamic models of convection presented in
\citet{tremblay11b} have finally, and convincingly, solved this
problem by replacing the more approximate mixing-length theory used to
this day.

We also wanted to test the reliability of our results by comparing
them with those of an independent study. The \citet{koester09}
analysis of DA white dwarfs from the SPY sample was ideal for this
purpose since several hundred white dwarfs were in common with both
samples but were observed and analyzed in a completely independent
fashion. We showed that our \Te\ determinations are in excellent
agreement with those of \citet{koester09} despite a trend toward
higher \logg\ values in the Koester et al. analysis.

Our mass distribution yields a mean mass of 0.638 \msun\ when
considering only white dwarfs with \Te~$\geq$~13,000~K. The mean mass
is somewhat higher than other determinations, but it is understood to
be caused by high-mass white dwarfs that are missed in many surveys
but included here. We also note the presence of a low-mass component,
consistent with the results of most other large-scale studies of DA
white dwarfs. The low-mass stars are necessarily the product of binary
evolution either through mergers or a common envelope phase where a
significant fraction of their mass may have been lost.

We also recomputed the PG luminosity function and found a space
density which is only slightly lower than that determined by LBH05. It
would be easy to point the finger at the new TB09 Stark broadening
profiles since they yield higher temperatures, but they also yield
higher \logg\ measurements that translate to smaller radii. These
facts coupled together produce two very similar values of
$M_{V}$. However, even small differences can cause stars to switch
magnitude bins, which explains our marginally lower value. 

Finally, we re-examined the ZZ Ceti instability strip and found that
the TB09 Stark broadening profiles produce a wider instability strip
that is also systematically shifted toward higher \Te. Furthermore, we
have identified over a dozen new ZZ Ceti candidates among which might
be the most massive ZZ Ceti star ever discovered. High-speed
photometric observations are planned for all candidates in the very
near future.

We thank the director and staff of Steward Observatory and of the
Carnegie Observatories for the use of their facilities. We are also
grateful to D.~Koester for providing us with all of the SPY spectra,
M.-M.~Limoges for obtaining several of the optical spectra analyzed in
this paper, and R. Heller for inadvertant help with Table 5. This work
was supported in part by the NSERC Canada and by the Fund FQRNT
(Qu\'ebec). M.T.R. acknowledges support from FONDAP (15010003) and
Proyecto BASAL PB06 (CATA). P.B. is a Cottrell Scholar of Research
Corporation for Science Advancement.

\end{document}